\documentclass[12pt, a4paper]{article}
\usepackage[utf8]{inputenc}
\usepackage{hyperref}
\usepackage{geometry}
\usepackage{graphicx}
\usepackage{subcaption}
\usepackage[numbers]{natbib}
\usepackage{amssymb}                                                                                                                                                                                                                                                                                                                                                                                                                                                                                                                                                                                                                                                                                                                                                                                                                                                                                                                                                                                                                                                                                                                                                                                                                                                                                                                                                                                                                                                                                                                                                                                                                                                                                                                                                                                                                                                                                                                                                                                                                                                                                                                                                                                                                                                                                                                                                                                                                                                                                                                                                                                                                                                                                                                                                                                                                                                                                                                                                                                                                                                                                                                                                                                                                                                                                                                                                                                                                                                                                                                                                                                                                                                                                                                                                                                                                                                                                                                                                                                                                                                                                                                                                                                                                                          
\usepackage{amsmath}
\usepackage{slashed}
\usepackage{anyfontsize}
\usepackage{bm}
\usepackage{slashed}
\usepackage{multirow}
\usepackage{float}
\usepackage[font=small]{caption}
\usepackage{listings}
\newgeometry{lmargin=1.1in, rmargin=1.1in, tmargin=1in, bmargin=1in}
\title{Generalized Standard Model with higher-order derivatives under Rotor Mechanism and its Quantization}
\author{B.T.T.Wong\footnote{Corresponding author. The University of Hong Kong, Pokfulam road.  u3500478@connect.hku.hk. }}
\date{}

\begin{document}

\maketitle
\begin{abstract}
The Standard Model is the paradigm of particle physics which gives an accurate theory for fundamental particle interactions. However, the extension of Standard Model with higher-order derivatives is not a well-studied subject. This paper is a follow-up work of the previous study of the generalized Abelian gauge field theory and Yang-Mills theory under rotor mechanism of order $n$ of higher order derivatives, and we apply it to the Standard Model of particle physics. Rotor mechanism on scalar field and Dirac field is also studied. We will study the quantization of the rotored Standard Model using path integral approach. We also inherit the previous result from the path integral quantization of generalized Abelian gauge field and apply it to our non-Abelian case. Then we carry out the generalized BRST quantization and prove the existence of the Slavnov-Taylor Identities of the rotor model.   Finally, we discuss the possibility of rotor model on taming the infinities arise from the self-energy correction of the Higgs boson in high spacetime dimension, thus this provides a partial solution and new insights to the Hierarchy problem. 
\end{abstract}
Keywords: Standard Model; Quantum field theory ; Rotor mechanism ;  High-order derivative field theory;Hierarchy Problem

\section{Introduction}
The Standard Model (SM) of particle physics is a well-established theory which describes particle interaction with great precision, with quantum field theory as an underlying mathematical foundation. Yet, higher-order derivative SM with a great potential to tame UV divergence is not a well studied subject. Promising model includes a recent study in Lee-Wick Standard Model which stabilizes quadratic divergence \cite{LWSM}, of which arises from the generalization of Lee-Wick electrodynamics \cite{LW1,LW2}. 

The study of higher-order derivative quantum field theory is particular interesting because it can eliminate ultraviolet (UV) divergences in scattering amplitudes \cite{ho1,ho2,ho3,ho4,ho5}. There are studies on higher order derivative scalar field and gauge field theories, and these theories have contributions in quantum gravity and modified gravity \cite{h1,h2,h3,h4,h5,ho6b, ho6c,ho8}. Higher-order derivative theory also shows its appearance in string theory \cite{h9,h10,h11,h12,h13}. In our context of study, Yang-Mills theory with higher-order derivative has been studied in references \cite{ho8x, ho9x, ho10x}. Quantization of higher-order derivative quantum field theory using path integral approach has been studied in \cite{ho9, ho10, ho11,ho12,ho13}.

In our previous work, we have established the generalized, higher-order derivative Yang-Mills theory by rotor mechanism \cite{BW2}, which follows upon our further previous work in the abelian counterpart \cite{BW1}. The path integral quantization of generalized abelian gauge field theory under rotor model is conducted in our previous work \cite{BW3}. The Yang-Mills action in $D$ dimensional spacetime is given by \cite{YangMills}
\begin{equation} \label{eq:YM}
S_{\mathrm{YM}} = -\frac{1}{2} \int d^D x \mathrm{Tr}\,G_{\mu\nu} G^{\mu\nu} = -\frac{1}{4} \int d^D x \,G^a_{\mu\nu} G^{\mu\nu\,a} \,
\end{equation}
where 
\begin{equation}
G_{\mu\nu} = \partial_{\mu}T_{\nu} -\partial_{\nu}T_{\mu} -ig[T_{\mu}, T_{\nu} ] \,,
\end{equation}
for which $G_{\mu\nu}=G^a_{\mu\nu}t^a$ and $T_{\mu} = T_{\mu}^a t^a $ are matrices with $t^a$ the generators of $\mathrm{SU}(N)$ Lie group. Using the Lie algebra $[t^a , t^b] =if^{abc}t^c$, this gives the gauge field strength as,
\begin{equation}
G_{\mu\nu}^a = \partial_{\mu}T_{\nu}^a -\partial_{\nu}T_{\mu}^a +gf^{abc}T_{\mu}^b T_{\nu}^c\,.
\end{equation}
We have $D=4n+4$ for a renormalizable theory with unity gauge field dimension \cite{BW3}. By carrying our integration by parts on the kinetic term and expanding the remaining terms, equation(\ref{eq:YM}) explicitly is,
\begin{equation} \label{eq:basic}
S_{\mathrm{YM}} = \int d^D x  \bigg( T^{\mu a} \hat{R}_{\mu\nu} T^{\nu a} - \frac{g}{2} f^{abc}(\partial^{\mu}T^{\nu a} - \partial^{\nu}T^{\mu a})T_{\mu}^b T_{\nu}^c - \frac{g^2}{4} f^{abc}f^{ade} T_{\mu}^b T_{\nu}^c T^{\mu d} T^{\nu e} \bigg)\,,
\end{equation}
where $\hat{R}_{\mu\nu}= \frac{1}{2}(\Box\eta_{\mu\nu} - \partial_{\mu}\partial_{\nu}  )$ is defined as the projection tensor. Under the rotor mechanism we introduced in our previous work in \cite{BW1}, which is the successive action of the projection tensors on the original gauge field, 
\begin{equation} \label{eq:rotor}
T_{n\,\mu_n} = \hat{R}_{\mu_n \mu_{n-1}}\hat{R}^{\mu_{n-1} \mu_{n-2}} \cdots \hat{R}_{\mu_3 \mu_2} \hat{R}^{\mu_2 \mu_1} \hat{R}_{\mu_1 \mu_0} T^{\mu_0} = \frac{1}{2^{n-1}}  P_{\mu_n}^{\,\,\,\mu_{n-1}} P_{\mu_{n-1}}^{\,\,\,\mu_{n-2}} \cdots P_{\mu_3}^{\,\,\,\mu_{2}} P_{\mu_2}^{\,\,\,\mu_{1}} \hat{R}_{\mu_1 \mu_0} T^{\mu_0} \,,
\end{equation}
for which 
\begin{equation} \label{eq:propagator}
P_{\mu_j}^{\,\,\,\mu_{j-1}} = \Box \delta_{\mu_j}^{\,\,\,\mu_{j-1}}
\end{equation}
is defined as the propagator. This is known as the rotor transformation which generates high-order derivative gauge fields in the action \cite{BW1,BW3}. The $n$-th order Yang-Mills action after rotor transformation of gauge field under Lorentz gauge is \cite{BW3}
\begin{equation} \label{eq:general}
\begin{aligned}
S^{(n)}_{\mathrm{YM}}&=-\frac{1}{4} \int d^D x \,G^a_{n\,\mu\nu} G^{\mu\nu\,a}_n \\
&=\int d^D x  \bigg( T^{\mu a}_n \hat{R}_{\mu\nu} T^{\nu a}_n - \frac{g}{2} f^{abc}(\partial^{\mu}T^{\nu a}_n - \partial^{\nu}T^{\mu a}_n)T_{\mu}^b T_{n\,\nu}^c - \frac{g^2}{4} f^{abc}f^{ade} T_{n\,\mu}^b T_{n\,\nu}^c T^{\mu d}_n T^{\nu e}_n \bigg) \\
&=\int d^D x  \bigg( \frac{1}{4^n} \Box^n T^{\mu a} \hat{R}_{\mu\nu} \Box^n T^{\nu a} - \frac{1}{2\cdot 8^n}g f^{abc}(\partial^{\mu}\Box^n T^{\nu a} - \partial^{\nu} \Box^n T^{\mu a})\Box^n T_{\mu}^b \Box^n T_{\nu}^c \\
&\quad\quad \quad\quad\quad - \frac{g^2}{4\cdot 16^n} f^{abc}f^{ade} \Box^n T_{\mu}^b \Box^n T_{\nu}^c \Box^n T^{\mu d} \Box^n T^{\nu e} \bigg) 
\end{aligned}
\end{equation}
Therefore, under the rotor mechanism, the gauge field transforms as \cite{BW2},
\begin{equation}
T_{\mu} \rightarrow T_{n\,\mu} = \frac{1}{2^n} \Box^n T_{\mu} \,.
\end{equation}
And the gauge field strength becomes \cite{BW2}
\begin{equation}
G_{n\mu\nu}^a = \frac{1}{2^n} \partial_{\mu} \Box^n T_{\nu}^{ a} - \frac{1}{2^n} \partial_{\nu} \Box^n T_{\mu}^{ a} + \frac{1}{4^n} gf^{abc} \Box^n T^b_\mu  \Box^n T^c_\nu\,.
\end{equation}
The $n-$th ordered covariant derivative is given by \cite{BW2},
\begin{equation} \label{eq:covariant}
D_{n\,\mu} =  \partial_{\mu} - \frac{i}{2^n} g\Box^n T_{\mu}^c t^{c}  \,,
\end{equation}
The equation of motion of this generalized Yang-Mills theory is \citep{BW2},
\begin{equation}
D_{n \,\mu} G_n^{\mu\nu a} = \partial_{\mu} G^{\mu\nu a}_n + \frac{1}{2^n} gf^{abc}\Box^n T_{\mu}^b G^{\mu\nu c}_n = 0 \,.
\end{equation}
The Noether's current is \cite{BW2}
\begin{equation}
J^{\alpha} =\bigg(-\frac{1}{4^n}\Box^n \tilde{G}^{\alpha\beta \,k} -\frac{1}{2\cdot 8^n}g f^{kbc} ( \Box^n T^{\alpha b} \Box^n T^{\beta c} - \Box^n T^{\alpha c} \Box^n T^{\beta b}  )\bigg)\delta \Box^n T^k_{\beta} \,,
\end{equation}
and the associated Noether's charge is given by \cite{BW2}
\begin{equation}
Q=\int d^{D-1}x j^0 = \int d^{D-1}x\bigg(-\frac{1}{4^n}\Box^n \tilde{G}^{0\beta \,k} -\frac{1}{2\cdot 8^n}g f^{kbc} ( \Box^n T^{0 b} \Box^n T^{\beta c} - \Box^n T^{0 c} \Box^n T^{\beta b}  )\bigg)\delta \Box^n T^k_{\beta} \,,
\end{equation}
where $\tilde{G}^{\alpha\beta\,k}$ is the Maxwellian gauge field strength. It can be seen that when $n=0$ (no rotation), we get back the original Yang-Mills theory.

Under the unitary transformation by SU(N) representation, the spinor transforms as
\begin{equation}
\psi^\prime (x) = U(x) \psi(x) \,,
\end{equation}
where $U(x) = e^{i\alpha^a (x) t^a}$. The covariant derivative and field operator under rotor model transform, respectively as,
\begin{equation}
\begin{cases}
D_{n\,\mu}^\prime (x) &= U(x) D_{n\,\mu}(x) U^\dagger (x) \\
G_{n\,\mu\nu}^\prime (x) &= U(x) G_{n\,\mu\nu}(x) U^\dagger (x)\,.
\end{cases}
\end{equation}
It follows that the rotor gauge field transforms as,
\begin{equation} \label{eq:transform}
\Box^n T_{\mu}^\prime = U\Box^n T_{\mu} U^\dagger + \frac{i\cdot 2^n}{g} U \partial_{\mu} U^\dagger\,.
\end{equation}
Infinitesimally, the rotored gauge field transforms as
\begin{equation} \label{eq:17}
\Box^n T_{\mu}^{\prime a} =\Box^n T_{\mu}^{ a} + \frac{2^n}{g}\partial_{\mu}\alpha^a + f^{abc}\Box^n T_{\mu}^b \alpha^c \,.
\end{equation}
Therefore, the infinitesimal change in rotored gauge field is
\begin{equation} \label{eq:18}
\delta \Box^n T_{\mu}^a = \Box^n T_{\mu}^{\prime a} -\Box^n T_{\mu}^{ a} =\frac{2^n}{g}(D_{n\,\mu}\alpha)^a \,.
\end{equation} 
The $n$-th ordered gauge field strength can be defined through the commutator of the $n$-th ordered covariant derivative,
\begin{equation}
G_{n\,\mu\nu} = \frac{i}{g} [D_{n\,\mu} , D_{n\,\nu} ] \,.
\end{equation}
And infinitesimally it transforms as
\begin{equation}
G_{n\,\mu\nu}^\prime = G_{n\,\mu\nu}-f^{abc}\alpha^b G_{n\,\mu\nu}^c \,.
\end{equation}
It is noted that when $n=0$, this gives us back all the properties of the transformation rules of the original Yang-Mills theory.

We will proceed the quantization of the generalized Yang-Mills theory under rotor mechanism with Feynman path integral approach. The quantum amplitude can be computed as an integral of all possible field configurations over the exponential of the action \cite{Feyn1,Feyn2,Peskin}. Similarly to our previous work for the abelian case in \citep{BW3}, as in the generalized model it involves the transformation of field by $T^{\mu} \rightarrow \Box^n T^{\mu}$, therefore in the path integral we sum over all possible configurations of $\Box^n T^{\mu}$ instead of $T^{\mu}$, i.e. the integration measure changes as
\begin{equation}
\int \mathcal{D}T^{\mu}(x) \rightarrow \int \mathcal{D}\Box^n T^{\mu}(x) \,.
\end{equation}

\section{Generalized spin-0 scalar field  theory under rotor mechanism}
In this section, we will complete the study of generalized scalar field theory under rotor mechanism, this will generate scalar fields with higher-order derivatives. First consider the massless scalar field theory in $D$-dimension
\begin{equation}
S = \int d^D x \, \frac{1}{2} \partial_{\mu} \phi \partial^\mu \phi  = \int d^D x \, \phi \Big( -\frac{\Box}{2}\Big) \phi \,.
\end{equation}
According to the definition of rotor mechanism in (\ref{eq:rotor}), we define the rotor mechanism as the successive operations of the operator that couples to the gauge fields. For the gauge field case, the operator is the projection tensor $\hat{R}_{\mu\nu} = \frac{1}{2}(\Box \eta_{\mu\nu} - \partial_{\mu}\partial_{\nu} )$. For the scalar field case, we have the rotor operator as $\hat{\tilde{R}}= -\frac{\Box}{2}$. This scalar rotor operator can be recovered from tracing the the projection tensor, up to some scaling factor, 
\begin{equation}
\hat{R} = \eta^{\mu\nu} \hat{R}_{\mu\nu} = \hat{R}^{\mu}_{\,\,\,\mu} = \frac{D-1}{2} \Box \,.
\end{equation}
It follows that 
\begin{equation}
\hat{\tilde{R}} = -\frac{1}{D-1} \hat{R} \,.
\end{equation}
The rotor mechanism on scalar field is simply
\begin{equation}
\phi \rightarrow \prod_{j=1}^n \bigg(-\frac{1}{D-1}  \bigg)\hat{R}^{\mu_j}_{\,\,\,\mu_j} \phi =  \frac{(-1)^n}{2^n} \Box \cdots \Box \Box \phi = \frac{(-1)^n}{2^n} \Box^n \phi \,.
\end{equation}
The generalized massless scalar field theory is therefore
\begin{equation}
\begin{aligned}
S &= \int d^D x \, \frac{1}{2} \partial_{\mu} \bigg(\frac{(-1)^n}{2^n} \Box^n \phi  \bigg)\partial^{\mu} \bigg(\frac{(-1)^n}{2^n} \Box^n \phi  \bigg) =\int d^D x \, \frac{1}{2} \partial_{\mu} \bigg(\frac{1}{2^n} \Box^n \phi  \bigg)\partial^{\mu} \bigg(\frac{1}{2^n} \Box^n \phi  \bigg) \\
&=\frac{1}{4^n} \int d^D x \,  \Box^n \phi \Big(-\frac{\Box}{2} \Big) \Box^n \phi \,.
\end{aligned} 
\end{equation}
Notice that when $n=0$, this restores back to the original case. Therefore, we see that both the gauge field and scalar field transforms under the rotor mechanism by the same form,
\begin{equation}
T_{\mu} \rightarrow T_{n\,\mu} = \frac{1}{2^n} \Box^n T_{\mu} \quad \quad , \quad\quad \phi \rightarrow \phi_n = \frac{1}{2^n} \Box^n \phi \,.
\end{equation}
For the massive scalar field theory, the generalized action under rotor mechanism is 
\begin{equation}
S= \frac{1}{2 \cdot 4^n} \int d^D x \,\partial_{\mu} \Box^n \phi \partial^\mu \Box^n \phi - m^2 \Box^n \phi \Box^n \phi = \frac{1}{4^n} \int d^D x \,  \Box^n \phi \Big(-\frac{\Box + m^2}{2} \Big) \Box^n \phi \,.
\end{equation}
The Euler-Lagrangian equation is
\begin{equation}
\partial_\mu \frac{\partial \mathcal{L}}{\partial_{\mu} \Box^n \phi} = \frac{\partial \mathcal{L}}{\partial \Box^n \phi} \,.
\end{equation}
This gives the equation of motion as
\begin{equation}
\Box^{n+1}\phi + m^2 \Box^n \phi = 0 \,.
\end{equation}

Now consider the complex-scalar field theory.
First consider the action with two scalar fields,
\begin{equation} \label{eq:complex}
\begin{aligned}
S&= \frac{1}{2 \cdot 4^n} \int d^D x \,\sum_{i=1,2} (\partial_{\mu} \Box^n \phi_i \partial^\mu \Box^n \phi_i - m^2 \Box^n \phi_i \Box^n \phi_i ) \\
&= \frac{1}{4^n} \int d^D x \, \sum_{i=1,2} \Box^n \phi_i \Big(-\frac{\Box + m^2}{2} \Big) \Box^n \phi_i \,.
\end{aligned}
\end{equation}
Now define a n-rotored complex scalar field,
\begin{equation}
\Box^n \Phi = \frac{1}{\sqrt{2}} (\Box^n \phi_1+ i \Box^n \phi_2) \quad \text{and } \quad \Box^n \Phi^\dagger = \frac{1}{\sqrt{2}} (\Box^n \phi_1 -i \Box^n \phi_2) \,.
\end{equation}
Then the action in \ref{eq:complex} can be written as
\begin{equation}
S= \frac{1}{4^n} \int d^D x \, \partial_\mu \Box^n \Phi^\dagger \partial^\mu \Box^n \Phi - m^2 \Box^n \Phi^\dagger \Box^n \Phi = \frac{1}{4^n} \int d^D x\,\Box^n \Phi^\dagger (-\Box^n -m^2 ) \Box^n \Phi \,.
\end{equation}
The two Euler-Lagrangian equations are 
\begin{equation}
\partial_\mu \frac{\partial \mathcal{L}}{\partial_{\mu} \Box^n \Phi} = \frac{\partial \mathcal{L}}{\partial \Box^n \Phi} \quad \text{and} \quad \partial_\mu \frac{\partial \mathcal{L}}{\partial_{\mu} \Box^n \Phi^\dagger} = \frac{\partial \mathcal{L}}{\partial \Box^n \Phi^\dagger} \,. 
\end{equation}
This gives the two equations of motion as follows:
\begin{equation}
\Box^{n+1}\Phi^\dagger + m^2 \Box^n \Phi^\dagger = 0 \quad \text{and} \quad \Box^{n+1}\Phi + m^2 \Box^n \Phi = 0  \,.
\end{equation}
Similar to the argument in our previous paper \cite{BW1}, the scalar field action under rotor mechanism is renormalizable in $D=4n+4$ dimension with unity scalar field dimension.

\section{Generalized spin-1/2 Dirac field theory theory under rotor mechanism}
Next, we will investigate how the Dirac field transforms under rotor mechanism. First we consider the massless Dirac action in $D$-dimensional spacetime,
\begin{equation}
S = \int d^D x \, \bar{\psi} i \gamma^\mu \partial_{\mu} \psi = \int d^D x \, \bar{\psi} i \slashed{\partial} \psi     \equiv   \int d^D x\, \bar{\psi}_a i \gamma^\mu_{ab} \partial_{\mu} \psi_b \,.
\end{equation}
So in analogy to the case of gauge field and scalar field, we have the rotor operator as the matrix $\hat{R}_{ab} = i\gamma^\mu_{ab} \partial_{\mu} $. Hence under rotor transformation, the spinor field transforms as
\begin{equation}
\psi_{a_0} \rightarrow (i^n) \gamma^{\mu_{n}}_{a_n a_{n-1}}\gamma^{\mu_{n-1}}_{a_{n-1} a_{n-2}} \cdots \gamma^{\mu_1}_{a_1 a_0} \partial_{\mu_{n}}\partial_{\mu_{n-1}} \cdots \partial_{\mu_1} \psi_{a_0} \,. 
\end{equation}
Now let's see how the adjoint spinor transforms. Consider ,
\begin{equation}
\gamma^0_{a_0 b_0} \psi_{b_0} \rightarrow (i^n) \gamma^{\mu_{n}}_{a_n a_{n-1}}\gamma^{\mu_{n-1}}_{a_{n-1} a_{n-2}} \cdots \gamma^{\mu_1}_{a_1 a_0} \partial_{\mu_{n}}\partial_{\mu_{n-1}} \cdots \partial_{\mu_1} \gamma^0_{a_0 b_0} \psi_{b_0} \,.
\end{equation}
Then by taking the Hermitian conjugate
\begin{equation}
 (\gamma^0_{a_0 b_0} \psi_{b_0})^\dagger \rightarrow (-i)^n (\partial_{\mu_{n}}\partial_{\mu_{n-1}} \cdots \partial_{\mu_1}\psi_{b_0}^\dagger\gamma^{0\dagger}_{b_0 a_0})\gamma^{\mu_1 \dagger}_{a_0 a_1} \cdots \gamma^{\mu_{n-1} \dagger}_{a_{n-2}a_{n-1}}\gamma^{\mu \dagger}_{a_{n-1}a_n} \,.
\end{equation}
As $(\gamma^0_{a_0 b_0} \psi_{b_0})^\dagger =\psi_{b_0}^\dagger \gamma^{0\dagger}_{ b_0 a_0} $ and because $(\gamma^0)^\dagger = \gamma^0$, therefore $ \bar{\psi}_{a_0} =\psi_{b_0}^\dagger \gamma^{0}_{ b_0 a_0}  $. Hence the adjoint spinor transforms as follow:
\begin{equation}
\bar{\psi}_{a_0} \rightarrow (-i)^n (\partial_{\mu_{n}}\partial_{\mu_{n-1}} \cdots \partial_{\mu_1}\bar{\psi}_{a_0} ) \gamma^{\mu_1 \dagger}_{a_0 a_1} \cdots \gamma^{\mu_{n-1} \dagger}_{a_{n-2}a_{n-1}}\gamma^{\mu \dagger}_{a_{n-1}a_n} \,.
\end{equation}
Therefore, the generalized Dirac field theory under rotor mechanism is
\begin{equation} \label{eq:Dirac}
S=\int d^D x (\partial_{\nu_n}\partial_{\nu_{n-1}}\cdots\partial_{\nu_1} \bar{\psi} )\gamma^{\nu_1 \dagger}\gamma^{\nu_2 \dagger}\cdots \gamma^{\nu_n \dagger}\, i\slashed{\partial} \, \gamma^{\mu_n} \gamma^{\mu_{n-1}}\cdots \gamma^{\mu_1} \partial_{\mu_{n-1}}\partial_{\mu_{n-2}}\cdots\partial_{\mu_1} \psi  \,.
\end{equation}
Now, let's define short-hand tensor notation by
\begin{equation}
\gamma^{\alpha\beta\gamma\cdots} \equiv \gamma^\alpha \gamma^\beta \gamma^\gamma \cdots \quad \text{and} \quad \nabla_{\alpha\beta\gamma\cdots} \equiv \partial_{\alpha}\partial_{\beta}\partial_{\gamma}\cdots \,. 
\end{equation}
Then the action in \ref{eq:Dirac} can be formally written as
\begin{equation}
S=\int d^D x (\nabla_{\nu_n \cdots \nu_1} \bar{\psi}) \, \gamma^{\dagger \nu_1 \cdots \nu_n} i \slashed{\partial} \,\gamma^{\mu_n \cdots \mu_1} \nabla_{\mu_n \cdots \mu_1} \psi \,. 
\end{equation}
The Dirac action with the mass term is given by
\begin{equation}
S= \int d^D x \, \bar{\psi} i \gamma^\mu \partial_{\mu} \psi - m\bar{\psi}\psi \,.
\end{equation}
Under rotor mechanism, the whole action transforms as 
\begin{equation}
S=\int d^D x (\nabla_{\nu_n \cdots \nu_1} \bar{\psi}) \, \gamma^{\dagger \nu_1 \cdots \nu_n} i \slashed{\partial} \,\gamma^{\mu_n \cdots \mu_1} \nabla_{\mu_n \cdots \mu_1} \psi -m (\nabla_{\nu_n \cdots \nu_1} \bar{\psi}) \, \gamma^{\dagger \nu_1 \cdots \nu_n} \,\gamma^{\mu_n \cdots \mu_1} \nabla_{\mu_n \cdots \mu_1} \psi\,.
\end{equation}
The Euler-Lagrangian equation is given by
\begin{equation}
\partial_{\mu} \frac{\partial \mathcal{L}}{\partial_\mu (\nabla_{\nu_n \cdots \nu_1} \bar{\psi}) \, \gamma^{\dagger \nu_1 \cdots \nu_n}  } = \frac{\partial \mathcal{L}}{\partial (\nabla_{\nu_n \cdots \nu_1} \bar{\psi}) \, \gamma^{\dagger \nu_1 \cdots \nu_n}} \,.
\end{equation}
This gives the equation of motion as
\begin{equation}
i \slashed{\partial} \,\gamma^{\mu_n \cdots \mu_1} \nabla_{\mu_n \cdots \mu_1} \psi -m \gamma^{\mu_n \cdots \mu_1} \nabla_{\mu_n \cdots \mu_1} \psi =0\,.
\end{equation}
\subsection{Generalized Quantum electrodynamics under rotor mechanism}
Now we proceed to develop the theory of higher-order derivative quantum electrodynamics (QED) by rotor mechanism. From \cite{BW1}, the general Maxwell action under rotor mechanism is 
\begin{equation}
S = -\frac{1}{4^{n+1}} \int d^D x \, \Box^n G_{\mu\nu} \Box^n G^{\mu\nu} \,.
\end{equation}
From the Dirac action under rotor mechanism, we expect the interactive term can be achieved by replacing the ordinary partial derivative into covariant derivative $D_n$
\begin{equation}
S=\int d^D x (\nabla_{\nu_n \cdots \nu_1} \bar{\psi}) \, \gamma^{\dagger \nu_1 \cdots \nu_n} i \slashed{D}_n \,\gamma^{\mu_n \cdots \mu_1} \nabla_{\mu_n \cdots \mu_1} \psi \,, 
\end{equation}
where
\begin{equation}
D_{n\,\alpha} = \partial_{\alpha} + \frac{ie}{2^n}\Box^n T_{\alpha}
\end{equation}
under Lorentz gauge. 
Therefore, the full QED action with interaction under rotor mechanism is given by
\begin{equation} \label{eq:QED}
\begin{aligned}
S_{\mathrm{QED}} &= \int d^D x \, \bigg(-\frac{1}{4^{n+1}} \Box^n G_{\mu\nu} \Box^n G^{\mu\nu} +(\nabla_{\nu_n \cdots \nu_1} \bar{\psi}) \, \gamma^{\dagger \nu_1 \cdots \nu_n} (i \gamma^\alpha  \partial_\alpha -m) \,\gamma^{\mu_n \cdots \mu_1} \nabla_{\mu_n \cdots \mu_1} \psi \\
& \quad\quad\quad\quad\quad-\frac{e}{2^n}  (\nabla_{\nu_n \cdots \nu_1} \bar{\psi}) \, \gamma^{\dagger \nu_1 \cdots \nu_n} \gamma^\alpha \Box^n T_{\alpha}
 \gamma^{\mu_n \cdots \mu_1} \nabla_{\mu_n \cdots \mu_1} \psi \bigg)\,.
\end{aligned}
\end{equation}

\section{Path integral quantization under rotor mechanism}
\subsection{Generalized Yang-Mills Theory}
In this section, we will study the quantization of general Yang-Mills theory by path integral approach in detail. From now on, we take the transformed $\Box^n T^\mu$ field as the field variable. Simply speaking, the physics is changed by $T^{\mu}\rightarrow \Box^n T^{\mu}$. The quantum amplitude of the $\Box^n T^\mu$ field in the renormalizable $4n+4$ dimension is
\begin{equation}
\langle \Box^n T^{\mu}_f (t_f, \pmb{\mathrm{x}}) | e^{-i\hat{H}(t_f - t_i )} | \Box^n T^{\mu}_i ( t_i , \pmb{\mathrm{x}} ) \rangle = \int \mathcal{D} \Box^n T^{\mu}(x) \, \exp\Big(iS_{\mathrm{YM}}^{(n)}[\Box^n T^{\mu}]  \Big) \,,
\end{equation}
where $| \Box^n T^{\mu}_i ( t_i , \pmb{\mathrm{x}} ) \rangle$ is the field state at initial time $t_i$ and $| \Box^n T^{\mu}_i ( t_f , \pmb{\mathrm{x}} ) \rangle$ is the field state at final time $t_f$, and $\hat{H}$ is the Hamiltonian.
The sourced generating functional is a functional of 4-(covariant) transformed vector current $\Box^n J_{\mu}(x)$,
\begin{equation}
Z[\Box^n J_{\mu}(x)] = \int \mathcal{D}\Box^n T^{\mu}(x)\,\exp\bigg(iS_{\mathrm{YM}}^{(n)}[\Box^n T^{\mu}]  \,+\,2 i \int d^{4n+4} x \, \mathrm{Tr}\Big(\Box^n J_{\mu}(x) \Box^n T^{\mu}(x)\Big) \bigg) \,. 
\end{equation} \label{eq:FullPartitionFunctionalGaugeField}
The normalized generating functional for gauge field is
\begin{equation}
\mathcal{Z}[\Box^n J_{\mu}] = \frac{Z[\Box^n J_{\mu}]}{Z[0]} = \frac{\int \mathcal{D}\Box^n T^{\mu}(x)\,\exp\bigg(iS_{\mathrm{YM}}^{(n)}[\Box^n T^{\mu}]  \,+\,2 i \int d^{4n+4} x \mathrm{Tr}\Big(\Box^n J_{\mu}(x) \Box^n T^{\mu}(x)\Big) \bigg)}{ \int \mathcal{D} \Box^n T^{\mu}(x) \, \exp\Big(iS_{\mathrm{YM}}^{(n)}[\Box^n T^{\mu}]  \Big)} \,.
\end{equation}
The free $n$-point correlation function is given by
\begin{equation}
\langle 0 | \mathrm{T} \Box^n \hat{T}^{\nu_1}(x_1)  \Box^n\hat{T}^{\nu_2}(x_2) \cdots \Box^n \hat{T}^{\nu_n}(x_n) |0 \rangle = \frac{1}{i^n} \frac{\delta^n}{\delta\Box^n J_{\nu_1}(x_1) \delta \Box^n J_{\nu_2}(x_2) \cdots \delta \Box^nJ_{\nu_n}(x_n)} \mathcal{Z}[\Box^n J_{\mu}] \bigg\vert_{\Box^n J_{\mu} =0} \,,   
\end{equation}
where noting that each gauge field has different Lorentz indices and $\mathrm{T}$ means the time ordering operator. Then the path integral representation is
\begin{equation} \label{eq:GaugeFieldPathInteral}
\begin{aligned}
& \quad \langle 0 | \mathrm{T} \Box^n \hat{T}^{\nu_1}(x_1)   \cdots \Box^n \hat{T}^{\nu_n}(x_n) |0 \rangle \\
&= \frac{\int \mathcal{D}\Box^n T^{\mu}(x)\,\Box^n T^{\nu_1}(x_1)  \cdots \Box^n T^{\nu_n}(x_n)   \exp\Big(iS_{\mathrm{YM}}^{(n)}[\Box^n T^{\mu}] \Big)}{\int \mathcal{D}\Box^n T^{\mu}(x) \,\exp\Big(iS_{\mathrm{YM}}^{(n)}[\Box^n T^{\mu}]\Big)} \,.
\end{aligned}
\end{equation}
Since the projection tensor in the action $S_{\mathrm{YM}}^{(n)}$ is not invertible, we need to perform gauge fixing. We will use the Fadeev-Popov method. Recall the gauge field transforms as (\ref{eq:transform})
\begin{equation}
\Box^n T_{\mu}^\prime =\Box^n T_{\mu}^U = U\Box^n T_{\mu} U^\dagger + \frac{i\cdot 2^n}{g} U \partial_{\mu} U^\dagger\,.
\end{equation}
Now we need to choose a gauge fixing functional,
\begin{equation} \label{eq:gaugecondition}
G(\Box^n T^{\mu}) = \partial_{\mu} \Box^n T^\mu (x) - w(x) \,,
\end{equation}
where $w(x)$ is some arbitrary matrix ($w(x) = w^a (x) t^a $ ) . Now we use the identity,
\begin{equation}
1 =\int \mathcal{D}U \delta[G(\Box^n T^{\mu U} )  ] \det \bigg( \frac{\delta G[\Box^n T^{\mu U} (x)]}{\delta U(y)} \bigg)
\end{equation}
Next we insert this into (\ref{eq:FullPartitionFunctionalGaugeField}),
\begin{equation}
\begin{aligned}
Z[\Box^n J_{\mu}(x)] & = \int \mathcal{D}U \mathcal{D}\Box^n T^{\mu}(x)\, \delta[G(\Box^n T^{U\mu } )]\det \bigg( \frac{\delta G[\Box^n T^{U\mu } (x)]}{\delta U(y)} \bigg) \\
&\quad\quad\quad\quad\quad\quad \times\exp\bigg(iS_{\mathrm{YM}}^{(n)}[\Box^n T^{\mu}]  \,+\,2 i \int d^{4n+4} x \, \mathrm{Tr}\Big(\Box^n J_{\mu}(x) \Box^n T^{\mu}(x)\Big) \bigg) \,. 
\end{aligned}
\end{equation} 
But since the action is gauge invariant, $S_{\mathrm{YM}}^{(n)}[\Box^n T^{U\mu }] =S_{\mathrm{YM}}^{(n)}[\Box^n T^{\mu}] $, also the integration measure remains unchanged, $\mathcal{D}\Box^n T^{\mu U} = \mathcal{D}\Box^n T^{\mu} $, therefore we can write
\begin{equation}
\begin{aligned}
Z[\Box^n J_{\mu}(x)] & = \int \mathcal{D}U \mathcal{D}\Box^n T^{U\mu }(x)\, \delta[G(\Box^n T^{U\mu } )]\det \bigg( \frac{\delta G[\Box^n T^{U\mu } (x)]}{\delta U(y)} \bigg) \\
&\quad\quad\quad\quad\quad\quad \times\exp\bigg(iS_{\mathrm{YM}}^{(n)}[\Box^n T^{U\mu }]  \,+\,2 i \int d^{4n+4} x \, \mathrm{Tr}\Big(\Box^n J_{\mu}(x) \Box^n T^{\mu}(x)\Big) \bigg) \,. 
\end{aligned}
\end{equation}
It is remarked that the source term is not gauge invariant unless $D^{\mu} \Box^n T_{\mu} = 0$. Now we can relabel the transformed, rotored gauge field variable to $\Box^n T^{U\mu} = \Box^n T^{\prime \mu} $, 
\begin{equation} \label{eq:intermediate1}
\begin{aligned}
Z[\Box^n J_{\mu}(x)] & = \int \mathcal{D}U \mathcal{D}\Box^n T^{\prime\mu }(x)\, \delta[G(\Box^n T^{\prime\mu } )]\det \bigg( \frac{\delta G[\Box^n T^{\prime\mu } (x)]}{\delta U(y)} \bigg) \\
&\quad\quad\quad\quad\quad\quad \times\exp\bigg(iS_{\mathrm{YM}}^{(n)}[\Box^n T^{\prime\mu }]  \,+\,2 i \int d^{4n+4} x \, \mathrm{Tr}\Big(\Box^n J_{\mu}(x) \Box^n T^{\mu}(x)\Big) \bigg) \,. 
\end{aligned}
\end{equation}
Now we choose some normalization $N(\xi)$ which is dependent on the gauge fixing parameter $\xi$ such that
\begin{equation} \label{eq:Gaussian}
1 = N(\xi) \int \mathcal{D}w(x) \exp\bigg( -i\int d^{4n+4} x \,\mathrm{Tr} \Big( \frac{w^2 (x)}{2\xi} \Big)   \bigg) \,.
\end{equation} 
Then we relate the rotored field variable $\Box^n T^{\prime\mu } (x)$ to $\Box^n T^{\mu } (x)$. And inserting equation(\ref{eq:Gaussian}) to (\ref{eq:intermediate1}), now we obtain,
\begin{equation}
\begin{aligned}
Z[\Box^n J_{\mu}(x)] & = \Big(N(\xi)\int \mathcal{D}U \Big) \int \mathcal{D}w(x) \mathcal{D}\Box^n T^{\mu }(x)\, \delta[G(\Box^n T^{\mu } )]\det \bigg( \frac{\delta G[\Box^n T^{\mu } (x)]}{\delta U(y)} \bigg) \\
& \times\exp\bigg(iS_{\mathrm{YM}}^{(n)}[\Box^n T^{\mu }]  \,+\,2 i \int d^{4n+4} x \, \mathrm{Tr}\Big(\Box^n J_{\mu}(x) \Box^n T^{\mu}(x)\Big) -i\int d^{4n+4} x \,\mathrm{Tr} \Big( \frac{w^2 (x)}{\xi}     \Big) \bigg) \,,
\end{aligned}
\end{equation}
where $N(\xi)\int \mathcal{D}U$ is some constant and can be integrated out. Now employing the gauge condition in \ref{eq:gaugecondition}, in which the Dirac delta function picks out the term of $\partial_{\mu} \Box^n T^\mu (x)$, we obtain,
\begin{equation} \label{eq:intermediate2}
\begin{aligned}
Z[\Box^n J_{\mu}(x)] & \propto \int \mathcal{D}\Box^n T^{\mu }(x)\,\det \bigg( \frac{\delta G[\Box^n T^{\mu } (x)]}{\delta U(y)} \bigg) \\
& \times\exp\bigg(iS_{\mathrm{YM}}^{(n)}[\Box^n T^{\mu }]  \,+\,2 i \int d^{4n+4} x \, \mathrm{Tr}\Big(\Box^n J_{\mu}(x) \Box^n T^{\mu}(x)\Big) \\
& \quad\quad\quad\quad - \frac{i}{\xi}\int d^{4n+4} x \,\mathrm{Tr} \Big( \partial_{\mu} \Box^n T^\mu (x)\partial^{\mu} \Box^n T^\mu (x)    \Big) \bigg) \,,
\end{aligned}
\end{equation}
where the last term in equation (\ref{eq:intermediate2}) is the gauge fixing term. We define the gauge-fixed action as
\begin{equation}
S_{\xi ,\mathrm{YM}}^{(n)} = \int d^{4n+4}x \bigg( -\frac{1}{2}\mathrm{Tr}\, G_{n\,\mu\nu}G_n^{\mu\nu} - \frac{1}{\xi}\mathrm{Tr}\,( \partial_{\mu} \Box^n T^\mu (x)\partial^{\mu} \Box^n T^\mu (x)    )     \bigg) \,.
\end{equation}
Now we proceed to calculate the terms in the determinant. First notice that
\begin{equation}
\delta \Box^n T^{U}_{\mu} = -\frac{1}{g} \big( \partial_{\mu} \alpha + ig \big[ \frac{1}{2^n} \Box^n T^U_\mu , \alpha \big] \big) = -\frac{1}{g} D_{n\,\mu} \alpha \,.
\end{equation}
For $G[\Box^n T_\mu (x)] = g\partial^{\mu} \Box^n T_{\mu} (x)$, then
\begin{equation}
\delta G[\Box^n T_\mu (x)] =- \partial^\mu D_{n\,\mu} \alpha \,.
\end{equation}
Since $i\alpha = \delta U \cdot U^{-1}$, we have 
\begin{equation}
\frac{\delta G[\Box^n T^\mu (x)]}{\delta U(y)} = i \delta^{4n+4}(x-y)\partial^\mu D_{n\,\mu} \cdot U(y).
\end{equation}
Since the determinant of $U(y)$ is unity, it follows that
\begin{equation}
\det\bigg(\frac{\delta G[\Box^n T^\mu (x)]}{\delta U(y)}   \bigg) = \det \bigg( i \delta^{4n+4}(x-y)\partial^\mu D_{n\,\mu}  \bigg) \,.
\end{equation} 
Now we can implement the Grassmann action as follow
\begin{equation}
\det (i\partial^\mu D_{n\,\mu}) =\int \mathcal{D} \Box^n c^* \mathcal{D}\Box^n c \exp\bigg( -2i \int d^{4n+4} x \, \mathrm{Tr} (\Box^n c^* \partial^\mu D_{n\,\mu} \Box^n c ) \bigg) \,,
\end{equation}
where $\Box^n c^*, \Box^n c$ are rotored transformed Grassmann fields, known as the Faddeev-Popov ghost fields. Therefore, the full sourced partition functional now gives
\begin{equation} \label{eq:72}
\begin{aligned}
Z[\Box^n J_{\mu}(x)] & \propto \int \mathcal{D}\Box^n T^{\mu }(x) \mathcal{D}\Box^n c^*(x) \mathcal{D} \Box^n c(x) \\
& \times\exp\bigg(iS_{\mathrm{YM}}^{(n)}[\Box^n T^{\mu }]  \,+\,2 i \int d^{4n+4} x \, \mathrm{Tr}\Big(\Box^n J_{\mu}(x) \Box^n T^{\mu}(x)\Big) \\
& \quad\quad\quad\quad - \frac{i}{\xi}\int d^{4n+4} x \,\mathrm{Tr} \Big( \partial_{\mu} \Box^n T^\mu (x)\partial^{\mu} \Box^n T^\mu (x)    \Big) \bigg) \\
&\quad\quad\quad\quad -2i \int d^{4n+4} x \, \mathrm{Tr} \big(\Box^n c^*(x) \partial^\mu D_{n\,\mu} \Box^n c(x) \big) \bigg) \,.
\end{aligned}
\end{equation} 

\subsection{Generalized Scalar field theory}
The quantization of generalized scalar field theory follows similarly to that of the abelian gauge field case in our previous paper \cite{BW2}, where we will not repeat the steps here. Generally speaking, the quantum amplitude follows the form of
\begin{equation}
\langle \Box^n \phi_f (t_f, \pmb{\mathrm{x}}) | e^{-i\hat{H}(t_f - t_i )} | \Box^n \phi_i ( t_i , \pmb{\mathrm{x}} ) \rangle = \int \mathcal{D} \Box^n \phi(x) \, \exp\Big( \frac{i}{4^n} \int d^{4n+4} x \,  \Box^n \phi \Big(-\frac{\Box + m^2}{2} \Big) \Box^n \phi \Big) \,,
\end{equation}
The sourced partition functional is then given by
\begin{equation}
Z[\Box^n J(x)] = \int \mathcal{D} \Box^n \phi(x) \, \exp\Big( \frac{i}{4^n} \int d^{4n+4} x \,  \Box^n \phi \Big(-\frac{\Box + m^2}{2} \Big) \Box^n \phi + i\int d^{4n+4}x \, \Box^n J \Box^n \phi \Big) \,,
\end{equation}
The normalized generating functional for the scalar field is
\begin{equation}
\mathcal{Z}[\Box^n J] = \frac{Z[\Box^n J]}{Z[0]} =\frac{\int \mathcal{D} \Box^n \phi(x) \, \exp\Big( \frac{i}{4^n} \int d^{4n+4} x \,  \Box^n \phi \Big(-\frac{\Box + m^2}{2} \Big) \Box^n \phi + i\int d^{4n+4}x \, \Box^n J \Box^n \phi \Big) }{\int \mathcal{D} \Box^n \phi(x) \, \exp\Big( \frac{i}{4^n} \int d^{4n+4} x \,  \Box^n \phi \Big(-\frac{\Box + m^2}{2} \Big) \Box^n \phi \Big) } \,.
\end{equation}
The free $n$-point correlation function is given by
\begin{equation}
\langle 0 | \mathrm{T} \Box^n \phi(x_1)  \Box^n \phi(x_2) \cdots \Box^n \phi(x_n) |0 \rangle = \frac{1}{i^n} \frac{\delta^n}{\delta\Box^n J(x_1) \delta \Box^n J(x_2) \cdots \delta \Box^n J(x_n)} \mathcal{Z}[\Box^n J] \bigg\vert_{\Box^n J =0} \,,   
\end{equation}
which is evaluated to be
\begin{equation}
\begin{aligned}
&\quad \langle 0 | \mathrm{T} \Box^n \phi(x_1)  \Box^n \phi(x_2) \cdots \Box^n \phi(x_n) |0 \rangle \\
&= \frac{\int \mathcal{D}\Box^n (x) \Box^n \phi(x_1)\Box^n \phi(x_2 ) \cdots \Box^n \phi(x_n) \exp\Big( \frac{i}{4^n} \int d^{4n+4} x \,  \Box^n \phi \Big(-\frac{\Box + m^2}{2} \Big) \Box^n \phi \Big)}{\int \mathcal{D}\Box^n \phi(x)\exp\Big( \frac{i}{4^n} \int d^{4n+4} x \,  \Box^n \phi \Big(-\frac{\Box + m^2}{2} \Big) \Box^n \phi \Big)} \,.
\end{aligned}
\end{equation}
Using the same technique in \cite{BW3} similar to the case of gauge field, one finds the Feynman propagator as
\begin{equation} \label{eq:83}
\langle 0 | \mathrm{T} \Box^n \phi(x) \Box^n \phi (y)|0\rangle = \int \frac{d^{4n+4}p}{(2\pi)^{4n+4}} \frac{i\cdot 4^n}{p^{4n}(p^2 - m^2)} e^{-ip\cdot (x-y)} \,.
\end{equation}
Or in the most generalized case, for the case of inhomogeneous order, we have
\begin{equation}
\langle 0 | \mathrm{T} \Box^n \phi(x) \Box^k \phi (y)|0\rangle = \int \frac{d^{2n+2k+4}p}{(2\pi)^{2n+2k+4}} \frac{i\cdot 2^{n+k}}{p^{2n+2k}(p^2 - m^2)} e^{-ip\cdot (x-y)} \,.
\end{equation}
From \ref{eq:83}, when $n=0$ which is the unrotored case this restores us back to the original two point correlation and the Feynman propagator of the scalar fields, 
\begin{equation}
\langle 0 | \mathrm{T}  \phi(x) \phi (y)|0\rangle = \int \frac{d^{4}p}{(2\pi)^{4}} \frac{i}{(p^2 - m^2)} e^{-ip\cdot (x-y)} \,.
\end{equation}

\subsection{Generalized Dirac field and Electrodynamics}
For the quantization of Dirac field under rotor mechanism, first we consider the quantum amplitude as follow:
\begin{equation}
\begin{aligned}
&\quad\langle  \gamma^{\mu_n \cdots \mu_1} \nabla_{\mu_n \cdots \mu_1} \psi(t_f ,\pmb{\mathrm{x}}) | e^{-i\hat{H}(t_f - t_i )} | \nabla_{\nu_n \cdots \nu_1} \bar{\psi}( t_i , \pmb{\mathrm{x}} )  \, \gamma^{\dagger \nu_1 \cdots \nu_n}   \rangle\\
& = \int \mathcal{D}[(\nabla_{\nu_n \cdots \nu_1} \bar{\psi}) \, \gamma^{\dagger \nu_1 \cdots \nu_n} ] \mathcal{D}[ \gamma^{\mu_n \cdots \mu_1} \nabla_{\mu_n \cdots \mu_1} \psi ] \\
&\quad\quad\quad\quad \times\exp\bigg( i\int d^D x \,(\nabla_{\nu_n \cdots \nu_1} \bar{\psi}) \, \gamma^{\dagger \nu_1 \cdots \nu_n} (i \slashed{\partial}-m) \,\gamma^{\mu_n \cdots \mu_1} \nabla_{\mu_n \cdots \mu_1} \psi  \bigg) \,.
\end{aligned}
\end{equation}
The sourced partition functional is then given by
\begin{equation} \label{eq:sourcedP}
\begin{aligned}
&\quad Z[ \nabla_{\beta_n \cdots \beta_1} \bar{J}( x )  \, \gamma^{\dagger \beta_1 \cdots \beta_n}\,,\,\gamma^{\alpha_n \cdots \alpha_1} \nabla_{\alpha_n \cdots \alpha_1} J(x)] \\
&=\int \mathcal{D}[(\nabla_{\nu_n \cdots \nu_1} \bar{\psi}) \, \gamma^{\dagger \nu_1 \cdots \nu_n} ] \mathcal{D}[ \gamma^{\mu_n \cdots \mu_1} \nabla_{\mu_n \cdots \mu_1} \psi ] \\
&\quad\quad\times\exp\bigg( i\int d^D x \,(\nabla_{\nu_n \cdots \nu_1} \bar{\psi}) \, \gamma^{\dagger \nu_1 \cdots \nu_n} (i \slashed{\partial}-m) \,\gamma^{\mu_n \cdots \mu_1} \nabla_{\mu_n \cdots \mu_1} \psi \\
&\quad\quad\quad+i\int d^D x (\nabla_{\nu_n \cdots \nu_1} \bar{\psi}) \, \gamma^{\dagger \nu_1 \cdots \nu_n} \gamma^{\alpha_n \cdots \alpha_1} \nabla_{\alpha_n \cdots \alpha_1} J(x) \\
&\quad\quad\quad+i\int d^D x \nabla_{\beta_n \cdots \beta_1} \bar{J}( x )  \, \gamma^{\dagger \beta_1 \cdots \beta_n} \gamma^{\mu_n \cdots \mu_1} \nabla_{\mu_n \cdots \mu_1} \psi \bigg) \,.
\end{aligned}
\end{equation}
And it is remarked that
\begin{equation} \label{eq:86}
\begin{aligned}
&\quad iS[(\nabla_{\nu_n \cdots \nu_1} \bar{\psi}) \, \gamma^{\dagger \nu_1 \cdots \nu_n} , \gamma^{\mu_n \cdots \mu_1} \nabla_{\mu_n \cdots \mu_1} \psi   ] \\
&= \iint d^D x  d^D y (\nabla_{\nu_n \cdots \nu_1} \bar{\psi}(x)) \, \gamma^{\dagger \nu_1 \cdots \nu_n} i\delta^D (x-y)(i\slashed{\partial} -m) \gamma^{\mu_n \cdots \mu_1} \nabla_{\mu_n \cdots \mu_1} \psi (y)
\end{aligned} \,.
\end{equation}
Now perform fourier transform,
\begin{equation}
\psi(x) = \int \frac{d^D p}{(2\pi)^D} \tilde\psi(p) e^{-ip\dot x} \quad \text{and}\quad \bar{\psi}(x) =  \int \frac{d^D p}{(2\pi)^D} \bar{\tilde{\psi}}(p) e^{+ip\dot x} \,.
\end{equation}
Equation \ref{eq:86} yields
\begin{equation}
\begin{aligned}
&\quad iS[(\nabla_{\nu_n \cdots \nu_1} \bar{\psi}) \, \gamma^{\dagger \nu_1 \cdots \nu_n} , \gamma^{\mu_n \cdots \mu_1} \nabla_{\mu_n \cdots \mu_1} \psi   ] \\
&=\iint d^D x d^D y \iint \frac{d^D p}{(2\pi)^D} \frac{d^D q}{(2\pi)^D}(i^n)(-i)^n p_{\nu_n}\cdots p_{\nu_1} \bar{\tilde{\psi}}(p)\gamma^{\dagger \nu_1 \cdots \nu_n}i\delta^D (x-y)(\slashed{q} -m)\gamma^{\mu_n \cdots \mu_1} \\
&\quad\quad\quad\quad\quad\quad\quad\quad\quad\quad\quad\quad\quad\quad\quad\quad\quad\quad\quad\quad \times q_{\mu_n}\cdots q_{\mu_1} \tilde{\psi}(q) e^{-ip\cdot x +iq\cdot y} \\
&=\int d^D x \iint \frac{d^D p}{(2\pi)^D} \frac{d^D q}{(2\pi)^D} p_{\nu_n}\cdots p_{\nu_1} \bar{\tilde{\psi}}(p)\gamma^{\nu_1\dagger} \cdots \gamma^{\nu_n\dagger}i(\slashed{q} -m)\gamma^{\mu_n} \cdots \gamma^{\mu_1} q_{\mu_n}\cdots q_{\mu_1} \tilde{\psi}(q) e^{-i(p-q)\cdot x } \\
&= \iint \frac{d^D p}{(2\pi)^D} \frac{d^D q}{(2\pi)^D} (\slashed{p}^\dagger )^n \bar{\tilde{\psi}}(p)i(2\pi)^D \delta^D (p-q)(\slashed{q} -m)\slashed{q}^n \tilde{\psi}(q)  \\
&=\int \frac{d^D p}{(2\pi)^D} \bar{\tilde{\psi}}(p) (\slashed{p}^\dagger )^n  i(\slashed{p} -m) (\slashed{p} )^n \tilde{\psi}(p) \,.
\end{aligned}
\end{equation}
Therefore, we have the matrix element as
\begin{equation}
\tilde{M} (p,q) = -i (2\pi)^D \delta^D (p-q)  (\slashed{p}^\dagger )^n (\slashed{q} - m) (\slashed{q})^n \,.
\end{equation}
Therefore, the momentum green's function for Dirac field under rotor mechanism is
\begin{equation}
\tilde{M}^{-1} (p,q) = \frac{i}{(\slashed{p}^\dagger )^n (\slashed{q} - m) (\slashed{q})^n} (2\pi )^D \delta ^D (p-q)\,.
\end{equation}
Hence, the Feynman propagator in momentum space is
\begin{equation}
S_F^{(n)}(p) = \frac{i}{(\slashed{p}^\dagger )^n (\slashed{p} - m) (\slashed{p})^n} \,.											
\end{equation}
Now, let's evaluate the sourceless partition functional, which is
\begin{equation}
\begin{aligned}
&\quad Z[0,0]=\int \mathcal{D}[(\nabla_{\nu_n \cdots \nu_1} \bar{\psi}) \, \gamma^{\dagger \nu_1 \cdots \nu_n} ] \mathcal{D}[ \gamma^{\mu_n \cdots \mu_1} \nabla_{\mu_n \cdots \mu_1} \psi ] \\
&\quad\quad\quad\quad\quad\quad\quad \times \exp\bigg( i\int d^D x \,(\nabla_{\nu_n \cdots \nu_1} \bar{\psi}) \, \gamma^{\dagger \nu_1 \cdots \nu_n} (i \slashed{\partial}-m) \,\gamma^{\mu_n \cdots \mu_1} \nabla_{\mu_n \cdots \mu_1} \psi \bigg) \,,
\end{aligned}
\end{equation}
which is 
\begin{equation}
\begin{aligned}
&\quad Z[0,0]=\int \mathcal{D}[(\nabla_{\nu_n \cdots \nu_1} \bar{\psi}) \, \gamma^{\dagger \nu_1 \cdots \nu_n} ] \mathcal{D}[ \gamma^{\mu_n \cdots \mu_1} \nabla_{\mu_n \cdots \mu_1} \psi ] \\
&\quad\quad \times \exp\bigg( i\iint d^D x d^D y \,(\nabla_{\nu_n \cdots \nu_1} \bar{\psi}(x)) \, \gamma^{\dagger \nu_1 \cdots \nu_n} \delta^D (x-y)(i \slashed{\partial}-m) \,\gamma^{\mu_n \cdots \mu_1} \nabla_{\mu_n \cdots \mu_1} \psi(y) \bigg) \,,
\end{aligned}
\end{equation}
They by using the identity from Gaussian integral,
\begin{equation}
\int \bigg(\prod_{j=1}^n d\theta_j^* d\theta_j\bigg)  \exp( -\theta^{*T} M\theta) = \det(M)  \,, 
\end{equation}
we obtain
\begin{equation}
Z[0,0] = \det[\delta^D (x-y)(i\slashed{\partial} -m ) ] \,.
\end{equation}
Now, using the usual fact from Gaussian integral with sourced term,
\begin{equation}
\int \bigg(\prod_{j=1}^n d\theta_j^* d\theta_j\bigg)  \exp( -\theta^{*T} M\theta + \eta^{*T}\theta + \theta^{*T}\eta   ) = \det(M) \exp(  \eta^{*T} M^{-1} \eta ) \,, 
\end{equation}
and with equations (\ref{eq:sourcedP}) and (\ref{eq:86}), the sourced partition functional for the Dirac field is evaluated to be 
\begin{equation}
\begin{aligned}
&\quad Z[ \nabla_{\beta_n \cdots \beta_1} \bar{J}( x )  \, \gamma^{\dagger \beta_1 \cdots \beta_n}\,,\,\gamma^{\alpha_n \cdots \alpha_1} \nabla_{\alpha_n \cdots \alpha_1} J(x)] \\
&=  \det[\delta^D (x-y)(i\slashed{\partial} -m ) ] \exp\bigg(\iint d^D x d^D y \nabla_{\beta_n \cdots \beta_1} \bar{J}( x )  \, \gamma^{\dagger \beta_1 \cdots \beta_n} M^{-1}(x,y) \gamma^{\alpha_n \cdots \alpha_1} \nabla_{\alpha_n \cdots \alpha_1} J(y)  \bigg) \\
&=Z[0,0] \exp\bigg(\iint d^D x d^D y \nabla_{\beta_n \cdots \beta_1} \bar{J}( x )  \, \gamma^{\dagger \beta_1 \cdots \beta_n} M^{-1}(x,y) \gamma^{\alpha_n \cdots \alpha_1} \nabla_{\alpha_n \cdots \alpha_1} J(y) \bigg) \,.
\end{aligned}
\end{equation}
The normalized generating functional for the Dirac field is
\begin{equation}
\begin{aligned}
\mathcal{Z}[  \nabla_{\beta_n \cdots \beta_1} \bar{J}( x )  \, \gamma^{\dagger \beta_1 \cdots \beta_n}\,,\,\gamma^{\alpha_n \cdots \alpha_1} \nabla_{\alpha_n \cdots \alpha_1} J(x)] = \frac{Z[  \nabla_{\beta_n \cdots \beta_1} \bar{J}( x )  \, \gamma^{\dagger \beta_1 \cdots \beta_n}\,,\,\gamma^{\alpha_n \cdots \alpha_1} \nabla_{\alpha_n \cdots \alpha_1} J(x)]}{Z[0,0]} \,.
\end{aligned}
\end{equation}
The two point correlation function is given by
\begin{equation}
\begin{aligned}
&\quad\langle 0| \mathrm{T}[\gamma^{\mu_n \cdots \mu_1} \nabla_{\mu_n \cdots \mu_1} \psi(x_1 )][ (\nabla_{\nu_n \cdots \nu_1} \bar{\psi}(x_2 )) \, \gamma^{\dagger \nu_1 \cdots \nu_n} ]|0\rangle \\
&=\frac{1}{i^2} \frac{\delta^2 \mathcal{Z}[  \nabla_{\beta_n \cdots \beta_1} \bar{J}( x )  \, \gamma^{\dagger \beta_1 \cdots \beta_n}\,,\,\gamma^{\alpha_n \cdots \alpha_1} \nabla_{\alpha_n \cdots \alpha_1} J(x)]}{\delta[\gamma^{\mu_n \cdots \mu_1} \nabla_{\mu_n \cdots \mu_1} J(x_1 )]\delta[ (\nabla_{\nu_n \cdots \nu_1} \bar{J}(x_2 )) \, \gamma^{\dagger \nu_1 \cdots \nu_n} ] } \bigg\vert_{\gamma^{\mu_n \cdots \mu_1} \nabla_{\mu_n \cdots \mu_1} J(x_1 ) = (\nabla_{\nu_n \cdots \nu_1} \bar{J}(x_2 )) \, \gamma^{\dagger \nu_1 \cdots \nu_n} =0} \,.
\end{aligned}
\end{equation}
The variational principle follows that
\begin{equation}
\frac{\delta \gamma^{\alpha_n \cdots \alpha_1} \nabla_{\alpha_n \cdots \alpha_1} J(x)}{\delta \gamma^{\mu_n \cdots \mu_1} \nabla_{\mu_n \cdots \mu_1} J(y )} =\delta^D (x-y) 
\end{equation}
and
\begin{equation}
\frac{\delta \nabla_{\beta_n \cdots \beta_1} \bar{\psi}(x ) \, \gamma^{\dagger \beta_1 \cdots \beta_n} }{\delta \nabla_{\nu_n \cdots \nu_1} \bar{\psi}(y ) \, \gamma^{\dagger \nu_1 \cdots \nu_n} } =\delta^D (x-y) \,.
\end{equation}
This would give 
\begin{equation}\label{eq:104}
\begin{aligned}
&\quad\langle 0| \mathrm{T}[\gamma^{\mu_n \cdots \mu_1} \nabla_{\mu_n \cdots \mu_1} \psi(x_1 )][ (\nabla_{\nu_n \cdots \nu_1} \bar{\psi}(x_2 )) \, \gamma^{\dagger \nu_1 \cdots \nu_n} ]|0\rangle = M^{-1}(x,y) \\
& =\int \frac{d^D p }{(2\pi)^D} \frac{i}{(\slashed{p}^\dagger )^n (\slashed{p} - m) (\slashed{p})^n}    e^{-ip\cdot (x-y)}\,.
\end{aligned}
\end{equation}
For the inhomogeneous case, we will have,
\begin{equation} \langle 0| \mathrm{T}[\gamma^{\mu_n \cdots \mu_1} \nabla_{\mu_n \cdots \mu_1} \psi(x_1 )][ (\nabla_{\nu_k \cdots \nu_1} \bar{\psi}(x_2 )) \, \gamma^{\dagger \nu_1 \cdots \nu_k} ]|0\rangle 
=\int \frac{d^D p }{(2\pi)^D} \frac{i}{(\slashed{p}^\dagger )^n (\slashed{p} - m) (\slashed{p})^k}    e^{-ip\cdot (x-y)}\,.
\end{equation}
From equation (\ref{eq:104}), when $n=0$, this restores us back the fermion two-point correlation function and Feynman propagator
\begin{equation}
\langle 0| \mathrm{T}\psi(x)\bar{\psi}(y)|0\rangle =\int \frac{d^D p }{(2\pi)^D} \frac{i}{\slashed{p}-m} e^{-ip\cdot (x-y)}\,.
\end{equation}

To quantize the generalized electrodynamics under rotor mechanism, we first consider the sourceless functional as follow:
\begin{equation}
Z[0,0,0]=\int \mathcal{D}[(\nabla_{\nu_n \cdots \nu_1} \bar{\psi}) \, \gamma^{\dagger \nu_1 \cdots \nu_n} ] \mathcal{D}[ \gamma^{\mu_n \cdots \mu_1} \nabla_{\mu_n \cdots \mu_1} \psi ] \mathcal{D}\Box^n T^\mu \exp\big( iS_{\mathrm{QED}} \big) \,,
\end{equation}
where $ S_{\mathrm{QED}}$ is the quantum electrodynamic action in \ref{eq:QED}. The normalized generating functional for QED is
\begin{equation}
\begin{aligned}
&\quad\mathcal{Z}_{\mathrm{QED}}[  \nabla_{\beta_n \cdots \beta_1} \bar{J}( x )  \, \gamma^{\dagger \beta_1 \cdots \beta_n}\,,\,\gamma^{\alpha_n \cdots \alpha_1} \nabla_{\alpha_n \cdots \alpha_1} J(x), \Box^n J^{\mu}] \\
&= \frac{ \exp{(i \int d^D z  \mathcal{L}_{\mathrm{int}}[\mathrm{source}] )} Z[  \nabla_{\beta_n \cdots \beta_1} \bar{J}( x )  \, \gamma^{\dagger \beta_1 \cdots \beta_n}\,,\,\gamma^{\alpha_n \cdots \alpha_1} \nabla_{\alpha_n \cdots \alpha_1} J(x), \Box^n J^{\mu}]}{ \{\exp{(i \int d^D z  \mathcal{L}_{\mathrm{int}}[\mathrm{source}] )} Z[  \nabla_{\beta_n \cdots \beta_1} \bar{J}( x )  \, \gamma^{\dagger \beta_1 \cdots \beta_n}\,,\,\gamma^{\alpha_n \cdots \alpha_1} \nabla_{\alpha_n \cdots \alpha_1} J(x), \Box^n J^{\mu}] \}|_{\mathrm{sourced\,\,terms = 0}}} \,,
\end{aligned}
\end{equation}
where
\begin{equation}
\begin{aligned}
&\quad\exp{\bigg(i \int d^4 z \,. \mathcal{L}_{\mathrm{int}}[\mathrm{source}] \bigg)} \\ 
&= \exp\bigg( i \int d^D z  \frac{1}{i \delta \nabla_{\beta_n \cdots \beta_1} \bar{J}( z )  \, \gamma^{\dagger \beta_1 \cdots \beta_n} }\frac{1}{i\delta \gamma^{\alpha_n \cdots \alpha_1} \nabla_{\alpha_n \cdots \alpha_1} J(y)}\frac{1}{i\delta \Box^n J^\mu (z)}\bigg)
\end{aligned} \\\,.
\end{equation}
Then the full physical two-point correlation function is 
\begin{equation}
\langle \Omega| \mathrm{T}\psi(x)\bar{\psi}(y)|\Omega\rangle = \frac{1}{i^2} \frac{\delta^2 \mathcal{Z}_{\mathrm{QED}}}{\delta\nabla_{\beta_n \cdots \beta_1} \bar{J}( x )  \, \gamma^{\dagger \beta_1 \cdots \beta_n} \delta\gamma^{\alpha_n \cdots \alpha_1} \nabla_{\alpha_n \cdots \alpha_1} J(y) }\bigg\vert_{\mathrm{source\,\,terms =0}}
\end{equation}
with $|\Omega\rangle $ the physical vacuum.

\subsection{A summary of Feynman propagators under rotor mechanism}
Using the path integral quantization technique approach, we can obtain the Feynman propagators under rotor mechanism ( n-rotors of $\Box^n$ operators) as follow:
\subsubsection*{Scalar spin-0 boson:}
\begin{equation} \label{eq:106}
\Delta_F^{(n)} (x-y) = \int \frac{d^{D}p}{(2\pi)^{D}} \frac{i\cdot 4^n}{p^{4n}(p^2 - m^2)} e^{-ip\cdot (x-y)} \,.
\end{equation}
\subsubsection*{Massless spin-1 gauge boson:}
\begin{equation}
D_{\mu\nu}^{(n)} (x-y) = \int \frac{d^{D}p}{(2\pi)^{D}} \frac{-i\cdot 4^n g_{\mu\nu}}{p^{2+4n}} e^{-ip\cdot (x-y)} \,.
\end{equation}
\subsubsection*{Massless spin-1 gauge boson in Lorentz gauge:}
\begin{equation}
D_{\mu\nu}^{(n)} (x-y) = \int \frac{d^{D}p}{(2\pi)^{D}} \frac{-i\cdot 4^n }{p^{2+4n}}\bigg(g_{\mu\nu} - (1-\xi)\frac{p_{\mu}p_{\nu}}{p^2} \bigg) e^{-ip\cdot (x-y)} \,.
\end{equation}
\subsubsection*{Massive spin-1 gauge boson:}
\begin{equation}
D_{\mu\nu}^{(n)} (x-y) = \int \frac{d^{D}p}{(2\pi)^{D}} \frac{-i\cdot 4^n}{p^{4n}(p^2 - M^2)}\bigg( g_{\mu\nu} - \frac{p_{\mu}p_\nu}{M^2}\bigg) e^{-ip\cdot (x-y)} \,.
\end{equation}
\subsubsection*{Massive spin-1 gauge boson in Lorentz gauge}
\begin{equation}
D_{\mu\nu}^{(n)} (x-y) = \int \frac{d^{D}p}{(2\pi)^{D}} \frac{-i\cdot 4^n}{p^{4n}}\bigg(  \frac{g_{\mu\nu}}{p^2 - M^2} - \frac{1}{p^2 - M^2}\bigg(1-\frac{1}{\xi}\bigg)\bigg( \frac{1}{M^2 - \frac{p^2}{\xi}} \bigg)p_{\mu}p_{\nu}   \bigg) e^{-ip\cdot (x-y)} \,.
\end{equation}
\subsubsection*{Dirac spin-1/2 fermion:}
\begin{equation}
S_F^{(n)} ( x-y) =\int \frac{d^D p }{(2\pi)^D} \frac{i}{(\slashed{p}^\dagger )^n (\slashed{p} - m) (\slashed{p})^n}    e^{-ip\cdot (x-y)}\,.
\end{equation}
\subsubsection*{Massless spin-1 gluon:}
\begin{equation}
D^{ab\,(n)}_{\mu\nu}(x-y) = \delta^{ab}D^{(n)}_{\mu\nu}(x-y)=\int \frac{d^{D}p}{(2\pi)^{D}} \frac{-i\cdot 4^n \delta^{ab}g_{\mu\nu}}{p^{2+4n}}\bigg(g_{\mu\nu} - (1-\xi)\frac{p_{\mu}p_{\nu}}{p^2} \bigg) e^{-ip\cdot (x-y)} \,.
\end{equation}
\subsubsection*{Ghost field:}
\begin{equation}
G^{ab\,(n)}(x-y) = \int\frac{d^D p}{(2\pi)^D} \frac{i \delta^{ab}}{p^{2 + 4n}} e^{-ip\cdot (x-y)} \,.
\end{equation}
The corresponding vertices in rotored quantum field theory is
\begin{figure}[H]
\centering
\includegraphics[trim=0cm 0cm 0cm 0cm, clip, scale=0.5]{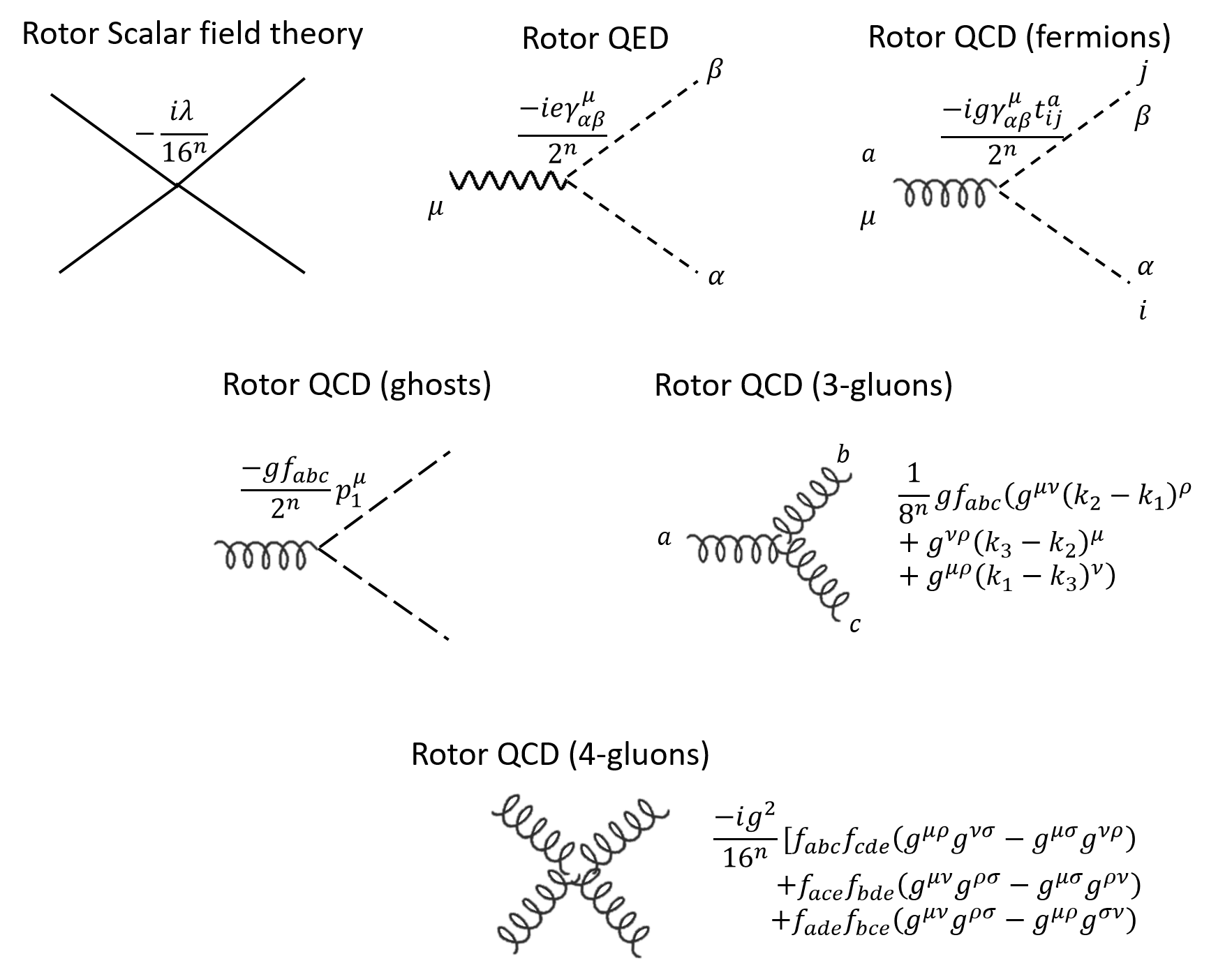}
\caption[]
{ Vertices in momentum space under rotor quantum field theory. \label{fig:vertices}}
\end{figure}

\section{BRST Symmetry of the rotor model }
It is known that the Faddeev-Popov ghost field arises from the gauge fixing procedure of non-abelian gauge field, and so do the rotored ghost field $\Box^n c$ appears in the generalized action (\ref{eq:72}) of the sourced partition function. It is noted that the rotor propagators contains both physical transverse polarized degree of freedom and the unpolarized time and longitudinal-polarized degree of freedom, this potentially leads to problems in loop diagrams, whicn contains non-physical degree of freedom. In order to construct a valid quantum rotored Yang-Mills gauge field theory of general $n$-th order, we need to check that if the Yang-Mills theory under rotor model possess BRST (Becchi-Rouet-Stora-Tyutin) symmetry, such that the contributions from the non-physical degree of freedom are cancelled by the rotored ghosts. Since the cancellation extends to the full theory, the most convenient way is to check if the rotored non-abelian gauge field and rotored ghost field has BRST symmetry. 

To see the BRST symmetry in the rotor mechanism, we need to slightly rewrite the action of a set of rotored fields
\begin{equation}
S_{\xi}[\Box^n T_\mu, \Box^n c^*, \Box^n c, \nabla_{\nu_n \cdots \nu_1} \bar{\psi} \, \gamma^{\dagger \nu_1 \cdots \nu_n}\, , \,\gamma^{\mu_n \cdots \mu_1} \nabla_{\mu_n \cdots \mu_1} \psi   ]
\end{equation}
by  using a Lagrange multiplier. We also introduce a set of new commuting rotor auxiliary fields $ \Box^n B = \Box^n B^a t^a \,,$ which has no independent dynamics. Then
\begin{equation} \label{eq:115}
\begin{aligned}
&\quad S_{\xi}[\Box^n T_\mu, \Box^n c^*, \Box^n c, \nabla_{\nu_n \cdots \nu_1} \bar{\psi} \, \gamma^{\dagger \nu_1 \cdots \nu_n}\, , \,\gamma^{\mu_n \cdots \mu_1} \nabla_{\mu_n \cdots \mu_1} \psi \,, \Box^n B    ] \\
&= \int d^{4n+4}x \bigg(-\frac{1}{2}\mathrm{Tr}G_{n\,\mu\nu}G^{\mu\nu}_n + (\nabla_{\nu_n \cdots \nu_1} \bar{\psi}) \, \gamma^{\dagger \nu_1 \cdots \nu_n} (i \gamma^\alpha  D_{n\,\alpha} -m) \,\gamma^{\mu_n \cdots \mu_1} \nabla_{\mu_n \cdots \mu_1} \psi \\
&\quad\quad\quad\quad\quad\quad\quad+ 2\mathrm{Tr}(\partial^\mu \Box^n c^*)(D_{n\,\mu}\Box^n c) + \frac{1}{4^n}\xi \mathrm{Tr}\,(\Box^n B \Box^n B)  + \frac{2}{4^n}\mathrm{Tr}(\Box^n B\partial_{\mu}\Box^n T^\mu)
 \bigg) \,.
\end{aligned}
\end{equation}
By ignoring an overall constant factor, the sourced partition function is given by
\begin{equation}
\begin{aligned}
Z[\mathrm{source}] &=\int\mathcal{D}\Box^n T_{\mu}\mathcal{D}[(\nabla_{\nu_n \cdots \nu_1} \bar{\psi})\gamma^{\dagger \nu_1 \cdots \nu_n}]\mathcal{D}[\gamma^{\mu_n \cdots \mu_1} \nabla_{\mu_n \cdots \mu_1} \psi]\mathcal{D}\Box^n c^* \mathcal{D}\Box^n c \mathcal{D}\Box^n B \\
&\quad\quad\quad\times \exp(iS_\xi + \text{source terms}) \,.
\end{aligned}
\end{equation}
We find that the action in (\ref{eq:115}) has a rotored BRST symmetry by the following infinitesimal transformations. Let  $\epsilon$ be a constant Grassmann number,
\begin{equation} \label{eq:117}
\begin{aligned}
\delta \Box^n T_{\mu} &= 2^n \epsilon D_{n\,\mu}\Box^n c\\
\delta (  \gamma^{\mu_n \cdots \mu_1} \nabla_{\mu_n \cdots \mu_1} \psi ) &= ig\epsilon c \gamma^{\mu_n \cdots \mu_1} \nabla_{\mu_n \cdots \mu_1} \psi \\
\delta [(\nabla_{\nu_n \cdots \nu_1} \bar{\psi}) \, \gamma^{\dagger \nu_1 \cdots \nu_n} ] &= -ig\epsilon c (\nabla_{\nu_n \cdots \nu_1} \bar{\psi}) \, \gamma^{\dagger \nu_1 \cdots \nu_n} \\
\delta\Box^n c &= ig\epsilon \Box^n c \Box^n c \\
\delta \Box^n c^* &= \frac{1}{2^n}\epsilon \Box^n B \\
\delta \Box^n B &= 0 \,.
\end{aligned}
\end{equation}
Now we check it is a symmetry. If the action in \ref{eq:115} satisfies the BRST symmetry, it satisfies $\delta S_{\xi} =0$ under the infinitesimal transformation in \ref{eq:117}.  First notice that $\delta \Box^n T_{\mu} = 2^n D_{n\,\mu}(\epsilon \Box^n c)$ and $\delta (  \gamma^{\mu_n \cdots \mu_1} \nabla_{\mu_n \cdots \mu_1} \psi ) = ig(\epsilon c) \gamma^{\mu_n \cdots \mu_1} \nabla_{\mu_n \cdots \mu_1} \psi$, this is just as if the gauge parameter $\Box^n \alpha= \epsilon \Box^n c$ where 
$\Box^n \alpha \equiv \Box^n \alpha^a t^a$ is the phase matrix parameter in the unitary transformation $U =\exp (i\Box^n \alpha)$. It follows that
\begin{equation}
 -\frac{1}{2}\delta(\mathrm{Tr}G_{n\,\mu\nu}G^{\mu\nu}_n)=0 \quad \text{and}\quad \delta[ (\nabla_{\nu_n \cdots \nu_1} \bar{\psi}) \, \gamma^{\dagger \nu_1 \cdots \nu_n} (i \gamma^\alpha  D_{n\,\alpha} -m) \,\gamma^{\mu_n \cdots \mu_1} \nabla_{\mu_n \cdots \mu_1} \psi ] =0 \,.
\end{equation}
Next we check 
\begin{equation}
\begin{aligned}
\delta(D_{n\,\mu} \Box^n c) &= \delta \Big(\partial_{\mu}\Box^n c - ig\Big[\frac{1}{2}\Box^n T_{\mu}\,,\, \Box^n c  \Big]  \Big)\\
&=\partial_{\mu}\delta\Box^n c -ig\Big[ \frac{1}{2^n}\Box^n T_{\mu}\, ,\,\delta\Box^n c \Big] -ig\Big[ \frac{1}{2}\delta\Box^n T_{\mu}\, ,\, \Box^n c \Big]\\
&= D_{n\,\mu} \delta\Box^n c -ig[\epsilon D_{n\,\mu} \Box^n c , \Box^n c ] \\
&= D_{n\,\mu} (ig\epsilon\Box^n c\Box^n c )-ig\epsilon (D_{n\,\mu}\Box^n c) \Box^n c +ig\Box^n c(\epsilon D_{n\,\mu}\Box^n c) \\
&= ig\epsilon (D_{n\,\mu} \Box^n c)\Box^n c + ig\epsilon\Box^n c (D_{n\,\mu}\Box^n c) -ig\epsilon (D_{n\,\mu} \Box^n c) \Box^n c +ig\Box^n c \epsilon (D_{n\,\mu} \Box^n c) \\
&= -ig\Box^n c \epsilon (D_{n\,\mu}\Box^nc) + ig \Box^n c \epsilon (D_{n\,\mu}\Box^n c) \\
&=0 \,.
\end{aligned}
\end{equation}
And since $\delta\Box^n B = 0$, then the remaining variation gives
\begin{equation}
\begin{aligned}
&\quad\delta\Big(2\mathrm{Tr}(\partial^\mu \Box^n c^*)(D_{n\,\mu}\Box^n c) + \frac{1}{4^n}\xi \mathrm{Tr}\,(\Box^n B \Box^n B)  + \frac{2}{4^n}\mathrm{Tr}(\Box^n B\partial_{\mu}\Box^n T^\mu)
 \Big)\\
&=2 \mathrm{Tr}(\partial^\mu \delta\Box^n  c^*)(D_{n\,\mu}\Box^n c) +\mathrm{Tr}(\partial^\mu \Box^n  c^*)(\delta D_{n\,\mu}\Box^n c) + \frac{2}{4^n}\mathrm{Tr}(\Box^n B\partial_{\mu}\delta \Box^n T^\mu) \\
&= -2\mathrm{Tr}(\delta\Box^n c^* \partial^\mu D_{n\,\mu} \Box^n c) +\frac{2}{4^n} \mathrm{Tr} \Big( \Box^n B \partial^\mu (2^n \epsilon D_{n\,\mu}\Box^n c) \Big) \\
&= -2\mathrm{Tr}\Big(\frac{1}{2^n}\epsilon \Box^n B \partial_{\mu}D_{n\,\mu}\Box^n c \Big) + \frac{2}{2^n}\mathrm{Tr} \Big( \Box^n B \partial^\mu (2^n \epsilon D_{n\,\mu}\Box^n c) \Big) \\
&=0 \,.
\end{aligned}
\end{equation}
Therefore,  it satisfies $\delta S_{\xi} =0$. Thus, there exists a BRST symmetry for the rotor mechanism.

Next, we can use a notation $Q$ to generalize the BRST symmetry,
\begin{equation}
\delta_{\epsilon} \Box^n T_{\mu} = \epsilon Q\Box^n T_{\mu}\,,
\end{equation}
and
\begin{equation}
\delta_{\epsilon}(\gamma^{\mu_n \cdots \mu_1} \nabla_{\mu_n \cdots \mu_1} \psi ) = \epsilon Q \gamma^{\mu_n \cdots \mu_1} \nabla_{\mu_n \cdots \mu_1} \psi \,.
\end{equation}
A key property is that 
\begin{equation}
Q^2 = 0
\end{equation}
so that BRST symmetry under rotor mechanism is nilpotent. For example
\begin{equation}
\epsilon Q \epsilon^\prime Q \Box^n T_{\mu} = \delta_{\epsilon}(\epsilon^\prime D_{n\,\mu}\Box^n c) = \epsilon^\prime \delta_{\epsilon}(D_{n\,\mu} \Box^n c ) = 0 \,.
\end{equation}
And
\begin{equation}
\begin{aligned}
\epsilon Q \epsilon^\prime Q \Box^n c &= \delta_{\epsilon} (ig\epsilon^\prime \Box^n c \Box^n c)\\
&= ig\epsilon^\prime [(\delta_{\epsilon}\Box^n c) \Box^n c + \Box^n c (\delta_{\epsilon} \Box^n c)]\\
&= ig\epsilon^\prime[ig\epsilon(\Box^n c)^2 \Box^n c + \Box^n c ig\epsilon (\Box^n c)^2]\\
&= -g^2 \epsilon^\prime [\epsilon (\Box^n c)^3 - \epsilon (\Box^n c)^3]\\
&=0 \,.
\end{aligned}
\end{equation}
Therefore in general, we have
\begin{equation}
Q^2 (\Box^n (\mathrm{field})) = 0\,.
\end{equation}
\subsection{Slavnov-Taylor Identities for the Rotor Model }
The Ward identity imposes the condition of gauge symmetry in quantum electrodynamics for abelian gauge fields. The generalization of Ward identity to non-abelian gauge field is manifested in Slavnov-Taylor identities \cite{Slavnov1,Taylor,Slavnov2}, which is far less trivial. And the generalization to rotor model is complicated, but it is essential to check for gauge symmetry of rotored Yang Mills theory. First consider the gauge-fixed sourced generating functional,
\begin{equation} \label{eq:127}
\begin{aligned}
Z[\Box^n J] &= \int \mathcal{D}\Box^n T^{\mu} \mathcal{D}\Box^n c^* \mathcal{D} \Box^n c \exp\bigg(i\int d^{4n+4}x \big(\mathcal{L}^{(n)}_{\mathrm{YM}} + \Box^n J_{\mu}^a \Box^n B^{\mu\,a} \\
&\quad\quad\quad\quad+ \frac{1}{2\cdot 4^n \xi}(\partial_{\mu}\Box^n T^\mu)^2 + \Box^n c^{*a}\partial^\mu D_{n\mu}^{ab} \Box^n c^{b}     \big) \bigg) \\
&=\int \mathcal{D}\Box^n T^{\mu} \mathcal{D}\Box^n c^* \mathcal{D} \Box^n c \exp\bigg(i\int d^{4n+4}x \big(\mathcal{L}^{(n)}_{\mathrm{YM}} + \Box^n J_{\mu}^a \Box^n B^{\mu\,a} \\
&\quad\quad\quad\quad+ \frac{1}{2\cdot 4^n \xi}(\partial_{\mu}\Box^n T^\mu)^2 + \Box^n c^{*a}\Box^{n+1}c^a + \frac{g}{2^n}f^{abc}\Box^n c^{*a}\Box^n c^{b} \partial^\mu \Box^n T_\mu^c  \big) \bigg) \,,
\end{aligned}
\end{equation}  
where $\frac{1}{2\cdot 4^n \xi}(\partial_{\mu}\Box^n T^\mu)^2 $ is the gauge-fixing term of rotored non-abelian gauge field. And we define the matrix element 
\begin{equation}
M^{ab} = \partial^{\mu}D_{n\mu}^{ab} \,.
\end{equation}
To obtain the generalized Slavnov-Taylor identities of the rotor model, the integral in (\ref{eq:127}) we make an infinitesimal rotored gauge transformation of the field variables as in equation (\ref{eq:17}) and (\ref{eq:18}), where the phase satisfies
\begin{equation}
M^{ab} \alpha^b (x) = \chi^a (x)
\end{equation}
for some arbitrary function $\chi^a (x)$. The inverse of $M^{ab}$, $M^{-1\,ab}$ is the Green's function of the rotor ghost-field under external rotor gauge field $\Box^n T^\mu$. By carrying out variation for the integral (\ref{eq:127}) and equating the coefficient $\chi^a (x)$ to zero, we obtain 
\begin{equation} \label{eq:130}
\begin{aligned}
&\int \mathcal{D}\Box^n T^{\mu} \mathcal{D}\Box^n c^* \mathcal{D} \Box^n c \exp\bigg(i\int d^{4n+4}x \big(\mathcal{L}^{(n)}_{\mathrm{YM}} + \Box^n J_{\mu}^a \Box^n B^{\mu\,a}\frac{1}{2\cdot 4^n \xi}(\partial_{\mu}\Box^n T^\mu)^2 + \Box^n c^{*a}) \bigg) \\
&\times\bigg\{ -\frac{1}{2^n\xi} \partial_\mu \Box^n T_{\mu}^a (y) + \int d^{4n+4} z \Box^n c^{*a}(y) \Box^n J^{\mu\,b} (z) [D_{n\mu}\Box^n c(z)]^b     \bigg\} =0 \,.
\end{aligned}
\end{equation}
Equation (\ref{eq:130}) is the generalized Slavnov-Taylor identity for the rotor model. By differentiating (\ref{eq:130}) with respect to external source $\Box^n J_{\mu}$ and putting $\Box^n J_{\mu}=0$, we obtain the following identity
\begin{equation}
\frac{1}{\alpha}\langle 0 | \mathrm{T} \partial^\mu \Box^n A_{\mu}^a (x) \Box^n A^b_{\nu}(y) = -i \langle 0 |\mathrm{T} \Box^n c^{*a}(x) (D_{n\nu}\Box^n c)^b (y) |0\rangle\,.
\end{equation}

\section{Generalized Higgs mechanism under rotor mechanism}
The Higgs mechanism is responsible for the explanation of mass acquisition of gauge boson through the process of spontaneous symmetry breaking, it also explains how the fermions couple with the Higgs field to main mass \cite{Higgs1, Higgs2, Englert, Weinberg, Salam, Kibble}. In addition, it also predicts all possible Higgs interactions and decays. 

Under rotor mechanism, the potential term is \begin{equation}
V=  \frac{\mu^2}{4^n} \Box^n \phi^\dagger   \Box^n \phi + \frac{\lambda}{16^n} (\Box^n \phi^\dagger   \Box^n \phi)^2 \,, 
\end{equation}
where $\mu^2 < 0$ and $\lambda > 0$ for spontaneous symmetry breaking. 
The Higgs action is
\begin{equation}
S=\int d^D x \bigg( -\frac{1}{4^{n+1}} \Box^n G_{\mu\nu} \Box^n G^{\mu\nu} + \frac{1}{4^n} (D_{n\,\mu}\Box^n \phi)^\dagger (D_{n\,\mu}\Box^n \phi) - \frac{\mu^2}{4^n} \Box^n \phi^\dagger   \Box^n \phi - \frac{\lambda}{16^n} (\Box^n \phi^\dagger   \Box^n \phi)^2  \bigg) \,,
\end{equation}
This Lagrangian is invariant under U(1) transformation,
\begin{equation}
(\Box^n \phi) \rightarrow (\Box^n \phi)^\prime = e^{i\theta}(\Box^n \phi) \,.
\end{equation}
When expressed in terms of two real rotored scalar fields $\Box^n \phi_1$ and $\Box^n \phi_2$, action reads
\begin{equation}
V(\Box^n \phi_1 , \Box^n \phi_2) = \frac{\mu^2}{2 \cdot 4^n} \big( (\Box^n \phi_1)^2  +    (\Box^n \phi_2)^2  \big) + \frac{\lambda}{4 \cdot 16^n}\big( (\Box^n \phi_1)^2  +    (\Box^n \phi_2)^2  \big)^2 \,.
\end{equation}
The potential has an infinite minimum when
\begin{equation} \label{eq:124}
 (\Box^n \phi_1)^2  +    (\Box^n \phi_2)^2   = \frac{-4^n \cdot \mu^2}{\lambda} = v_n^2 \,.
\end{equation}
Upon the elimination of Goldstone boson by Unitary gauge and by perturbation of the non-zero vacuum expectation value, we have
\begin{equation}
\Box^n \phi (x) = \frac{1}{\sqrt{2}} \big(v_n + \Box^n h (x)\big) \,.
\end{equation}
Then after spontaneously symmetry breaking, we have
\begin{equation} \label{eq:lag}
\begin{aligned}
S&=\int d^D x \bigg( -\frac{1}{4^{n+1}} \Box^n G_{\mu\nu} \Box^n G^{\mu\nu} \\
&\,\,\,\,\,\,\,\,+\frac{1}{2\cdot 4^n}\Big(\partial_{\mu} - \frac{i}{2^n}g\Box^n T_{\mu}(v_n + \Box^n h)   \Big)\Big(\partial^{\mu} + \frac{i}{2^n}g\Box^n T_{\mu}(v_n + \Box^n h)   \Big) \\
&\,\,\,\,\,\,\,\, - \frac{\mu^2}{2\cdot 4^n}(v_n + \Box^n h)^2 - \frac{\lambda}{4\cdot 16^n}(v_n + \Box^n h)^4 \bigg) \\
&=\int d^D x \bigg( -\frac{1}{4^{n+1}} \Box^n G_{\mu\nu} \Box^n G^{\mu\nu} + \frac{1}{2\cdot 4^n}g^2 v_n^2 \Box^n T_\mu \Box^n T^\mu \\
&\,\,\,\,\,\,\,\,\quad\quad\quad\quad+ \frac{1}{2 \cdot 4^n}\partial_\mu \Box^n h \partial^\mu \Box^n h - \frac{1}{4^n}\lambda v_n^2 \Box^n h \Box^n h \\
&\,\,\,\,\,\,\,\,\quad\quad\quad\quad+ \frac
{1}{8^n} g^2 v_n \Box^n T_{\mu} \Box^n T^{\mu} \Box^n h + \frac{1}{2\cdot 8^n}g^2  \Box^n T_{\mu}\Box^n T^{\mu} \Box^n h \Box^n h \\
&\,\,\,\,\,\,\,\,\quad\quad\quad\quad - \frac{1}{8^n}\lambda v_n \Box^n h \Box^n h \Box^n h - \frac{1}{4\cdot 16^n}\lambda \Box^n h \Box^n h \Box^n h \Box^n h \bigg)\,.
\end{aligned}
\end{equation}
The mass term of the rotored massive gauge boson is identified as $M_n = gv_n$, and the mass term of the rotored Higgs boson is $m_n=\sqrt{2\lambda} v_n$. It is noted that when $n=0$, the action (\ref{eq:lag}) will restore back to the normal Higgs action with spontaneous symmetry breaking. From the third line in (\ref{eq:lag}), this gives the interactions between the rotored Higgs boson and rotored gauge field. We can rewrite them in terms of the metric,
\begin{equation}
\frac{1}{8^n}g M_n g_{\mu\nu}\Box^n T^\mu \Box^n T^\nu \Box^n h\,,\quad \frac{1}{2\cdot 8^n} g^2 g_{\mu\nu} \Box^n T^\mu \Box^n T^\nu \Box^n h \Box^n h \,.
\end{equation}
In terms of vertex diagram, we have the following feynman rules,
\begin{figure}[H]
\centering
\includegraphics[trim=0cm 0cm 0cm 0cm, clip, scale=0.6]{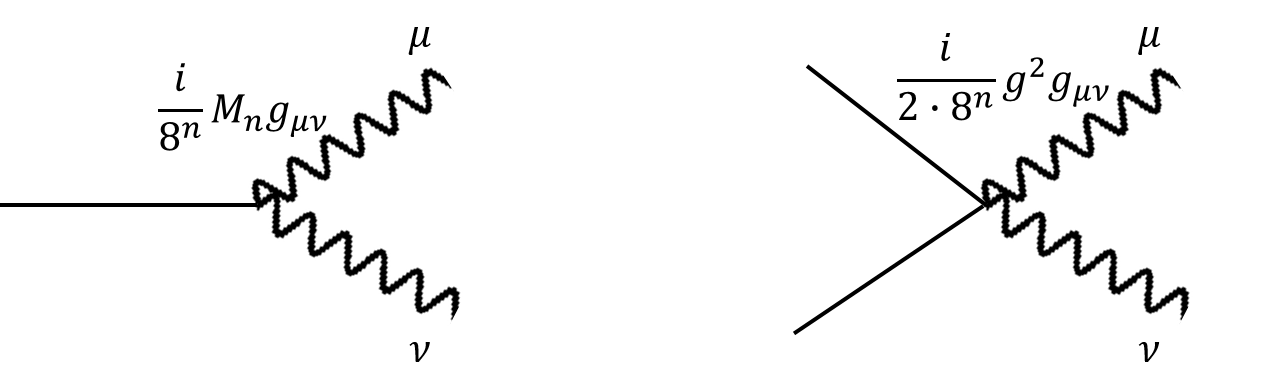}
\caption[]
{ Feyman rules for rotored Higgs boson decaying into two rotored bosons (left) and rotored scattering process (right).  \label{fig:vertex1}}
\end{figure}
From the fourth line in \ref{eq:lag}, this gives the rotored Higgs boson self interactions. The Feynman rules are as follow:
\begin{figure}[H]
\centering
\includegraphics[trim=0cm 0cm 0cm 0cm, clip, scale=0.6]{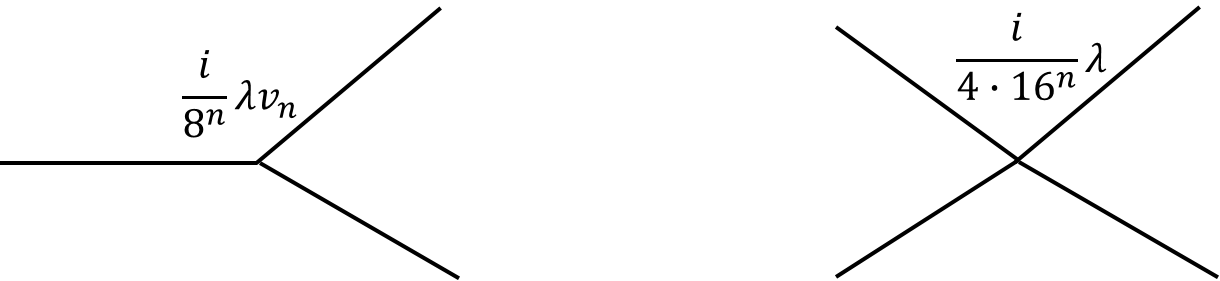}
\caption[]
{ Feyman rules for third order and fourth order rotored Higgs boson interactions.  \label{fig:vertex2}}
\end{figure}
\section{A discussion of Hierarchy Problem and the partial solution provided by the rotor mechanism}
It has been known that high-order derivative quantum field theory makes a good job in eliminating UV divergence \cite{ho1,ho2,ho3,ho4,ho5}. In this section, we will investigate how the high-order-derivative quantum field theory by the rotor mechanism can suppress UV divergence in the quantum loop processes. Although the standard model is renormalizable, i.e. the divergences in the Lagrangian can be treated by adding finite number of counter terms, the Hierarchy Problem which involves fine tunning of the Higgs mass remains a long-lasting problem in high-energy particle physics. Recently, reference \cite{LWSM} generalizes Lee-Wick electrodynamics with high-order field derivatives to the Standard Model and offers a solution to the Hierarchy Problem to tame UV divergences in one-loop level. Here, we will see how high-order derivative field theory by rotor mechanism can eliminate the UV divergence in amplitudes.  

Let us have a revisit to the Hierarchy problem.  For example, taking the 4-fields self-interaction of the Higgs boson the one-particle irreducible (1\,PI) function can be easily calculated by integrating through the loop momentum $k$,
\begin{equation}
\tilde{\Gamma}(p)=\frac{-i\lambda^2}{8}\int \frac{d^4 k}{(2\pi)^4}\frac{i}{k^2-m_{\rm H}^2}=\frac{-i\lambda^2}{128 \pi^2}\big[\Lambda^2-2m_{\rm H}^2\ln\left(\frac{\Lambda}{m_{\rm H}}\right)+\mathcal{O}\left(\frac{m_{\rm H}^2}{\Lambda^4}\right)\big]+\mathcal{O}(\lambda^4),
\end{equation}
which has a dominant quadratic divergence of high energy scale $\Lambda$. 
Hence, the renormalization of the bare Higgs boson's mass for the 4-field self interaction is,
\begin{equation}
\tilde{m}_{\rm H}^2=m_{\rm H}^2+i\tilde{\Gamma}(p)=m_{\rm H}^2+\frac{\lambda^2}{128 \pi^2}\big[\Lambda^2-2m_{\rm H}^2\ln\left(\frac{\Lambda}{m_{\rm H}}\right)+\mathcal{O}\left(\frac{m_{\rm H}^2}{\Lambda^2}\right)\big]+\mathcal{O}(\lambda^4).
\end{equation}
It is noted that $m_{\rm H}=\mu$ above is the bare mass, while $\tilde{m}_{\rm H}$ is the physical, observed mass-the renormalized mass. 
The full calculation of all quantum loop corrections from electroweak and Yukawa interaction to the $n^{\rm th}$-order loop takes the general form \citep{Fred},
\begin{equation}
{\rm \tilde{m}_{H}^2}=m_{\rm H}^{2}+\frac{\Lambda^{2}}{16 \pi^2}C_{n}(\mu^{2}=m_{\rm H}^2)\,, 
\end{equation}
where $C_{n}$ is the a polynomial expansion of the bare Higgs mass scale and it is a function of the Higgs self-coupling $\lambda$, electroweak couplings $g_{W}, g' $ and the Yukawa coupling $Y^{f}_{ij}$. To the first-order loop calculation \citep{Fred},
\begin{equation} \label{eq:3.5}
\mathrm{ \tilde{m}_{H}^2= m_{\rm H}^2}+\frac{\Lambda^2}{32 \pi^{2}}(4\lambda^2 + 3g'^{2}+9g_{W}^{2}-24Y^{f\,2}_{ij})+\mathcal{O}(\lambda^2,g^{\prime 4},g_{W}^4,Y^{f\,4}_{ij})\,.
\end{equation}
The physical Higgs boson mass is measured to be $\tilde{m}_{\rm H}\simeq 126\,\rm GeV$ and hence $\tilde{m}^2_{\rm H}\sim 10^4,\rm GeV^2$, while the quantum corrections lead to the quadratic divergence of the Planck scale $\Lambda\sim10^{19}\,\rm GeV$. From equation (\ref{eq:3.5}) we can see that the physical Higgs mass depends strictly on the strength of each of the coupling. The nature requires an extremely precise fine-tuning cancellation of the couplings and the bare Higgs mass to each $n^{\rm th}$ order of quantum loop divergence by $10^{17}$\,GeV, so as to obtain the low-energy, EW-scale Standard Model we observe today \citep{Martin}. This is known as the Hierarchy Problem. 

The Hierarchy Problem gives a strong motivation to search for New Physics (NP) particles with mass scales $\Lambda_{\rm NP}>\Lambda_{\rm EW}$ such that they can cancel these loop divergences. One important scheme is supersymmetry, which introduces new particles between bosonic and fermonic symmetry, such that each divergent bosonic loop has a fermionic counterpart, vice versa. Since fermonic amplitude has an extra factor of minus sign, this can cancels the bosonic loop.

Using the rotor mechanism, we will see that the divergent loop processes are suppressed and this will lead to convergent result. The generic problematic 1-loop divergent process includes, for example, the self-correction of the Higgs boson by the W-boson loop. The Feynman diagram is illustrated below.

\begin{figure}[H]
\centering
\includegraphics[trim=0cm 0cm 0cm 0cm, clip, scale=0.4]{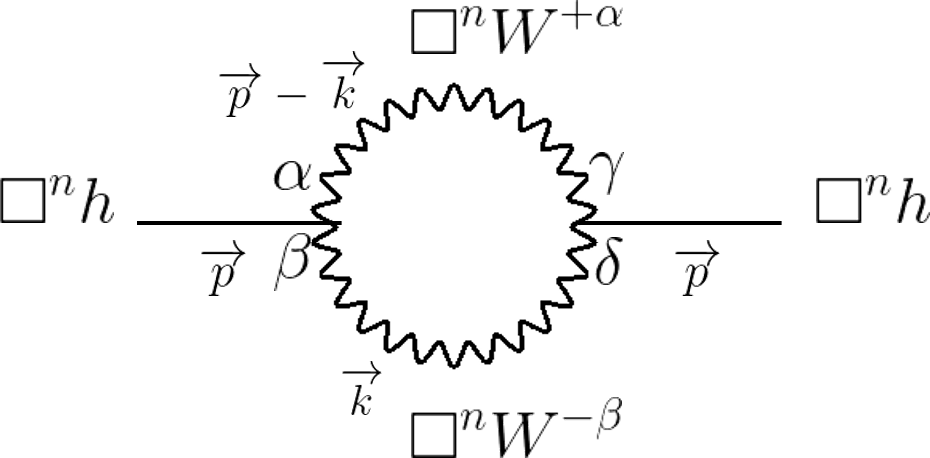}
\caption[]
{ Feyman diagram of self-energy correction of the Higgs boson by a W-boson loop.  \label{fig:loop1}}
\end{figure}

The $n$-rotored amplitude is given by
\begin{equation}
\mathcal{M}^{(n)}=\frac{g_W^2 (16)^n M_W^2}{64^n (2\pi)^D} \int d^D k \, g^{\alpha\beta}g^{\delta\gamma} \frac{g_{\beta\delta} - k_\beta k_\delta /M_W^2}{k^{4n} (k^2 - M_W^2)} \frac{g_{\gamma\alpha} -(p-k)_\gamma (p-k)_\alpha ) / M_W^2 }{(p-k)^{4n}((p-k)^2 - M_W^2 )} \,.
\end{equation}
By expanding, we obtain
\begin{equation} \label{eq:132}
\begin{aligned}
\mathcal{M}^{(n)} &=\frac{g_W^2  M_W^2}{4^n (2\pi)^D} \bigg( \int d^D k \frac{g^{\alpha\beta}g_{\beta\delta}g^{\delta\gamma}g_{\gamma\alpha} }{ k^{4n} (p-k)^{4n} (k^2-M_W^2 )( (p-k)^2 -M_W^2  )} \\ 
& \quad\quad\quad\quad\quad - \frac{1}{M_W^2}\int d^D k \frac{(p-k)^\alpha (p-k)_\alpha}{k^{4n} (p-k)^{4n} (k^2-M_W^2 )( (p-k)^2 -M_W^2  )} \\
&\quad\quad\quad\quad\quad - \frac{1}{M_W^2}\int d^D k \frac{k^\alpha k_\alpha}{k^{4n} (p-k)^{4n} (k^2-M_W^2 )( (p-k)^2 -M_W^2  )}\\
&\quad\quad\quad\quad\quad + \frac{1}{M_W^4}\int d^D k \frac{k_\delta (p-k)^\delta k^\alpha(p-k)_\alpha} {k^{4n} (p-k)^{4n} (k^2-M_W^2 )( (p-k)^2 -M_W^2  )} \bigg) \,.
\end{aligned}
\end{equation}
First, we can give a rough estimate of the amplitude by the leading term,
\begin{equation} \label{eq:132}
\begin{aligned}
\mathcal{M}^{(n)}_1 &\sim\frac{g_W^2 M_W^2}{4^n (2\pi)^D} \int d^D k \frac{g^{\alpha\beta}g_{\beta\delta}g^{\delta\gamma}g_{\gamma\alpha} }{ k^{4n} (p-k)^{4n} (k^2-M_W^2 )( (p-k)^2 -M_W^2  )} \\
&\sim \frac{2^{1-D} D g_W^2 M_W^2}{4^n \sqrt{\pi^D}\Gamma\big(\frac{D}{2}\big)}   \int^\Lambda  \frac{k^{D-1} dk}{k^{8n+4}} \,,
\end{aligned}
\end{equation}
where have have used the fact that of the integration of D-dimensional solid angle is given by $\int d\Omega_D = \frac{2\pi^{D/2}}{\Gamma(D/2)}$ with $\Gamma(x)$ the Gamma function. When $n=0$ and in four spacetime dimension $D=4$ which is the unrotored case in 4D spacetime (which is just the normal case), we have logarithmic divergence 
\begin{equation}
\mathcal{M}^{(0)}_1 \sim \frac{ g_W^2 M_W^2}{2\pi^2 }   \ln \Lambda \,.
\end{equation}
Hence in the normal case, the self-correction of the Higgs boson by the W boson is logarithmic divergent. But under rotor mechanism, by the second line of \ref{eq:132}.
we have
\begin{equation}
\mathcal{M}^{(n)}_1 \sim \frac{2^{1-D} D g_W^2 M_W^2}{4^n \sqrt{\pi^D}\Gamma\big(\frac{D}{2}\big)(D-8n-4)}  \Lambda^{D-8n-4}\,,
\end{equation}
In our 4D universe, we simply have 
\begin{equation}
\mathcal{M}^{(n)}_1 \sim -\frac{g_W^2 M^2_W}{ 2^{2n+4}\pi^2 n} \frac{1}{\Lambda^{8n}} \,.
\end{equation}
As $\Lambda$ is a very large value, which can be up to Planck scale $10^{19}$ GeV, even for $n=1$ $\mathcal{M}^{(1)}_1 \rightarrow 0$. For the second term,
\begin{equation}
\begin{aligned}
\mathcal{M}^{(n)}_2 &= \frac{g_W^2 }{4^n (2\pi)^D} \int d^D k  \frac{1}{k^{4n} (p-k)^{4n-2} (k^2-M_W^2 )( (p-k)^2 -M_W^2  )} \\
&\sim \frac{2^{1-D}g_W^2 }{4^n \sqrt{\pi^D} \Gamma\big(\frac{D}{2}\big)} \int^\Lambda \frac{k^{D-1} dk}{k^{8n+2}} \\
&=\frac{2^{1-D}g_W^2 }{4^n \sqrt{\pi^D} \Gamma\big(\frac{D}{2}\big)(D-8n-2)} \Lambda^{D-8n-2} \,.
\end{aligned}
\end{equation}
For the unrotored case $n=0$ and in $D=4$, 
\begin{equation}
\mathcal{M}^{(0)}_2 \sim \frac{ g_W^2}{16 \pi^2 }  \Lambda^2 \,,
\end{equation}
which contributes to a quadratic divergence. But under rotor mechanism,
\begin{equation}
\mathcal{M}^{(n)}_2 \sim \frac{g_W^2}{2^{(2n+4)} \pi^2 (1-4n)}\Lambda^{2-8n}\,.
\end{equation} 
For example for $n=1$, $\mathcal{M}^{(1)}_2 \propto \frac{1}{\Lambda^6}$ which is convergent. For the third term,
\begin{equation}
\begin{aligned}
\mathcal{M}^{(n)}_3 &= \frac{g_W^2 }{4^n (2\pi)^D} \int d^D k  \frac{1}{k^{4n-2} (p-k)^{4n} (k^2-M_W^2 )( (p-k)^2 -M_W^2  )} \\
&\sim \frac{2^{1-D}g_W^2 }{4^n \sqrt{\pi^D} \Gamma\big(\frac{D}{2}\big)} \int^\Lambda \frac{k^{D-1} dk}{k^{8n+2}} \\
&=\frac{2^{1-D}g_W^2 }{4^n \sqrt{\pi^D} \Gamma\big(\frac{D}{2}\big)(D-8n-2)} \Lambda^{D-8n-2} \,,
\end{aligned}
\end{equation}
which has the same result as the $\mathcal{M}_2^{(n)}$ case. For the last term, 
\begin{equation} \label{eq:132}
\begin{aligned}
\mathcal{M}^{(n)}_4 &\sim\frac{g_W^2}{4^n (2\pi)^D M_W^2} \int d^D k \frac{(k\cdot p - k^2)^2 }{ k^{4n} (p-k)^{4n} (k^2-M_W^2 )( (p-k)^2 -M_W^2  )} \\
&=\frac{g_W^2}{4^n (2\pi)^D M_W^2} \int d^D k \frac{(k\cdot p)^2 -2(k\cdot p)k^2 + k^4 }{ k^{4n} (p-k)^{4n} (k^2-M_W^2 )( (p-k)^2 -M_W^2  )} \\
&\sim \frac{g_W^2}{4^n (2\pi)^D M_W^2} \int d^D k \frac{k^2 p^2 -2(k\cdot p)k^2 + k^4}{[k^2 (p-k)^2]^{2n} (k^2-M_W^2 )( (p-k)^2 -M_W^2  )}\\
&= \frac{g_W^2}{4^n (2\pi)^D M_W^2} \int d^D k \frac{1}{[k^2 (p-k)^2]^{2n-1} (k^2-M_W^2 )( (p-k)^2 -M_W^2  )} \,.
\end{aligned}
\end{equation}
From the second line to the third line, we have made the order approximation. Notice that $(k\cdot p)^2 = k_\mu p^\mu k_\alpha p^\alpha = (k_\mu p^\alpha) p^\mu k_\alpha$. Then notice that $k^2 p^2 = k_{\mu}k^{\mu}p_{\alpha}p^{\alpha}=(k_{\mu}p^\alpha )k^{\mu}p_{\alpha} $. They are obviously different but they have the same order in $k$ and $p$. As here we are interested in order calculation, to simplify the analysis we make the above approximation. Then we have
\begin{equation}
\mathcal{M}^{(n)}_4 \sim  \frac{2^{1-D} g_W^2}{4^n \sqrt{\pi^D} \Gamma\big(\frac{D}{2}\big) M_W^2}  \int^\Lambda \frac{k^{D-1} dk}{k^{8n}} =   \frac{2^{1-D}g_W^2}{4^n \sqrt{\pi^D}  M_W^2 \Gamma\big(\frac{D}{2}\big)(D-8n)}\Lambda^{D-8n} \,.
\end{equation}
For the unrotored case $n=0$ and in $D=4$,
\begin{equation}
\mathcal{M}^{(0)}_4 \sim \frac{g_W^2}{32 \pi^2}\Lambda^4 \,,
\end{equation}
which is a seriously 4-th order divergent term. But under rotor mechanism,
\begin{equation}
\mathcal{M}^{(n)}_4 \sim \frac{g_W^2}{2^{(2n+5)}\pi^2 M_W^2 (1-2n)}\Lambda^{4-8n}\,.
\end{equation}
For example for $n=1$, $\mathcal{M}^{(1)}_2 \propto \frac{1}{\Lambda^4}$ which is convergent. Therefore, the amplitude is
\begin{equation} 
\mathcal{M}^{(n)} = \mathcal{M}^{(n)}_1 + \mathcal{M}^{(n)}_2 + \mathcal{M}^{(n)}_3 + \mathcal{M}^{(n)}_4 \sim \frac{g_W^2}{4^{n+2} \pi^2 \Lambda^{8n}} \bigg( \frac{M_W^2}{n} + \frac{\Lambda^2}{1-4n} +\frac{\Lambda^4}{4 M_W^2 (1-2n)}\bigg) 
\end{equation}
in $D=4$ spacetime dimension. We can see that when $n=1$ under rotor mechanism, the amplitude is already convergent in the high-energy regime. Therefore, we see how the rotor mechanism can remove infinities in loop diagrams at high energies.

Next, consider the self-correction of Higgs Boson by a fermion loop. The Feynman diagram is as follow:
\begin{figure}[H]
\centering
\includegraphics[trim=0cm 0cm 0cm 0cm, clip, scale=0.4]{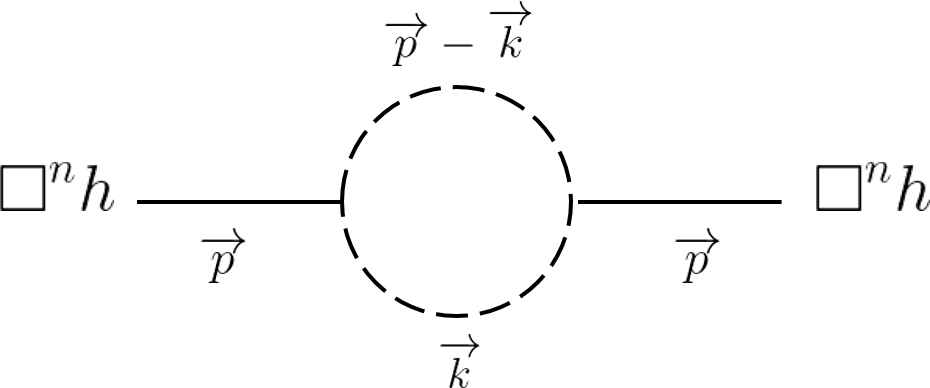}
\caption[]
{ Feyman diagram of self-energy correction of the Higgs boson by a fermionic loop.  \label{fig:loop2}}
\end{figure}
Since the vertex factor is just $-i\frac{m_f}{v_n}$, the $n$-rotored amplitude is given by
\begin{equation} \label{eq:136}
\mathcal{M}^{(n)} =-\frac{m_f^2}{v_n^2 (2\pi)^D} \int d^D k \,\mathrm{Tr}\bigg( \frac{1}{(\slashed{k}^\dagger)^n (\slashed{k} - m_f)(\slashed{k})^n } \frac{1}{(\slashed{p}^\dagger-\slashed{k}^\dagger)^n (\slashed{p}-\slashed{k} - m_f)(\slashed{p}-\slashed{k})^n} \bigg)\,.
\end{equation}
The minus sign is due to the Feynman rule for the fermion loop. Now using the following fact that
\begin{equation}
\begin{aligned}
\gamma_{\mu} \gamma^\mu &= 4I_D \\
(\gamma_{\mu} \gamma^\mu)^\dagger &= 4I_D^\dagger \\
\gamma^{\mu\dagger} \gamma_{\mu}^\dagger &= 4I_D \,.
\end{aligned}
\end{equation}
Equivalently, the amplitude in (\ref{eq:136}) reads
\begin{equation} \label{eq:143}
\mathcal{M}^{(n)} =-\frac{m_f^2}{v_n^2 (2\pi)^D} \int d^D k \,\mathrm{Tr} \bigg( \frac{(\slashed{k})^n (\slashed{k}+m_f )(\slashed{k}^\dagger )^n }{k^{4n}(k^2 -m_f^2 )} \frac{ (\slashed{p}-\slashed{k} )^n (\slashed{p}-\slashed{k} + m_f ) (\slashed{p}^\dagger-\slashed{k}^\dagger )^n }{(p-k)^{4n} ((p-k)^2 - m_f^2 )} \bigg) \,.
\end{equation}
The rough estimation of (\ref{eq:136}) gives 
\begin{equation}
\mathcal{M}^{(n)} \sim -\frac{m_f^2}{v_n^2 (2\pi)^D} \int d^D k \,\mathrm{Tr}\bigg(\frac{1}{\slashed{k}^{2+4n}}\bigg) \,.
\end{equation}
For the unrotored case, $n=0$ and in 4D spacetime, we have 
\begin{equation}
\mathcal{M}^{(n)} \sim -\frac{m_f^2}{v^2 (2\pi)^D} \Lambda^2
\end{equation}
which has a quadratic divergence. But under rotor mechanism we have 
\begin{equation}
\mathcal{M}^{(n)} \sim -\frac{m_f^2}{v_n^2 (2\pi)^D} \Lambda^{2-4n} \,.
\end{equation}
When $n=1$, we can see that $\mathcal{M}^{(n)}\propto \frac{1}{\Lambda^2}$, which is convergent. Thus the divergence is eliminated. 

Next, we consider the self-energy correction of the Higgs boson by the 3rd order self-interaction. The vertex factor is $\frac{i}{2\cdot 8^n} \frac{m_\mathrm{H}^2}{v_n}$. The Feynman diagram is
\begin{figure}[H]
\centering
\includegraphics[trim=0cm 0cm 0cm 0cm, clip, scale=0.4]{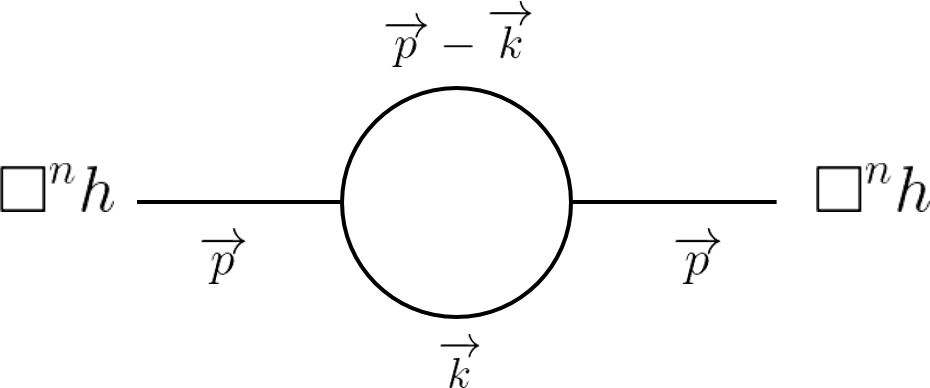}
\caption[]
{ Feyman diagram of self-energy correction of the Higgs boson by a 3rd-order self interaction.  \label{fig:loop2}}
\end{figure}
The amplitude for such process is
\begin{equation} \label{eq:152}
\mathcal{M}^{(n)} =\frac{(16)^n m_{\mathrm{H}}^4}{4\cdot 64^n (2\pi)^D v_n^2} \int d^D k \frac{1}{k^{4n}(k^2 - m_{\mathrm{H}}^2)}\frac{1}{(p-k)^{4n} ((p-k)^2 -m_{\mathrm{H}}^2 )} \,.
\end{equation}
By rough estimation, we have
\begin{equation}
\begin{aligned}
\mathcal{M}^{(n)} &\sim \frac{m_{\mathrm{H}}^4}{2^{D+2n+1} \sqrt{\pi^D} v_n^2} \int^\Lambda  \frac{k^{D-1} dk}{k^{4+8n}} \\
&\sim  \frac{m_{\mathrm{H}}^4}{2^{D+2n+1} \sqrt{\pi^D}(D-8n-4 ) v_n^2} \Lambda^{D-8n-4} \,.
\end{aligned}
\end{equation}
For the unrotored case in 4D spacetime, $n=0$, we have logarithmic divergence
\begin{equation}
\mathcal{M}^{(0)} \sim \frac{m_{\mathrm{H}}^4}{32 \pi^2 v_n^2} \ln \Lambda
\end{equation}
For the general case, if we consider $D=4$ for our universe, we have
\begin{equation}
\mathcal{M}^{(n)}\sim -\frac{m_{\mathrm{H}}^4}{2^{D+4n+4} \sqrt{\pi^D}n v_n^2}\frac{1}{\Lambda^{8n}} \,.
\end{equation}
For $n=1$, the matrix element goes for $\mathcal{M}^{(n)}\propto \frac{1}{\Lambda^8}$, which drops very quickly to zero for large $\Lambda$. Therefore, the problem of divergence vanishes in the rotor mechanism.

In general, the above method of power counting used by the above example applies to other diagrams with higher number of loops. The generic form takes the following:
\begin{equation}
\mathcal{M}^{(n)} \sim \int\cdots\int \frac{d^D k_1 d^D k_2 \cdots d^D k_L}{k_i^{4n} \cdots k_l^{4n} k^2_i \cdots k^2_l (\slashed{k}_j^\dagger)^n (\slashed{k}_j)^{n+1} \cdots (\slashed{k}_p^\dagger)^n (\slashed{k}_p)^{n+1}  } \,.
\end{equation}
Let $S$ be the superficial degree of divergence. And let $L$ be the number of loops, $N_{\mathrm{B}}$ be the number of internal spin-1 gauge boson propagator(s), $N_\mathrm{H}$ be the number of internal Higgs boson propagator(s) and $N_{f}$ be the number of internal fermion propagators. Then we have
\begin{equation}
\begin{aligned} \label{eq:PC}
S&=\text{(power of momentum in numerator)- (power of momentum in demoninator )} \\
&= DL -(2N_{\mathrm{B}} + 4N_{\mathrm{B}}n) -(2N_{\mathrm{H}} + 4N_{\mathrm{H}}n) -(N_f + 2 N_f n  )\\
&= DL- (2N_{\mathrm{B}} +  2N_{\mathrm{H}} + N_f) -2n(2N_{\mathrm{B}} +2 N_{\mathrm{H}} + N_f ) \\
&= DL - (2n+1)(2N_{\mathrm{B}} +  2N_{\mathrm{H}} + N_f) \,.
\end{aligned}
\end{equation}
Naively, we expect a diagram to have a divergence proportional to $\Lambda^S$. We expect logarithmic divergence $\log \Lambda $ when $S=0$, and no divergence when $S <0$. This takes place when the following inequality holds,
\begin{equation}
\begin{aligned}
 DL - (2n+1)(2N_{\mathrm{B}} +  2N_{\mathrm{H}} + N_f) & <0 \\
 n & > \frac{DL}{(2N_{\mathrm{B}} +  2N_{\mathrm{H}} + N_f)} - \frac{1}{2} \,,
\end{aligned}
\end{equation}
and also we demand $n>0$.
In addition, as the number of loops satisfy the following equation,
\begin{equation}
L= I - V +1 \,,
\end{equation}
where $I$ is the number of internal lines, $V$ is the number of vertices, it follows that
\begin{equation}
L = N_{\mathrm{B}} + N_{\mathrm{H}} + N_{f} - V +1  \,.
\end{equation}
Let us illustrate by an example. Consider the following 3rd-order self-energy correction of the Higgs boson,
\begin{figure}[H]
\centering
\includegraphics[trim=0cm 0cm 0cm 0cm, clip, scale=0.4]{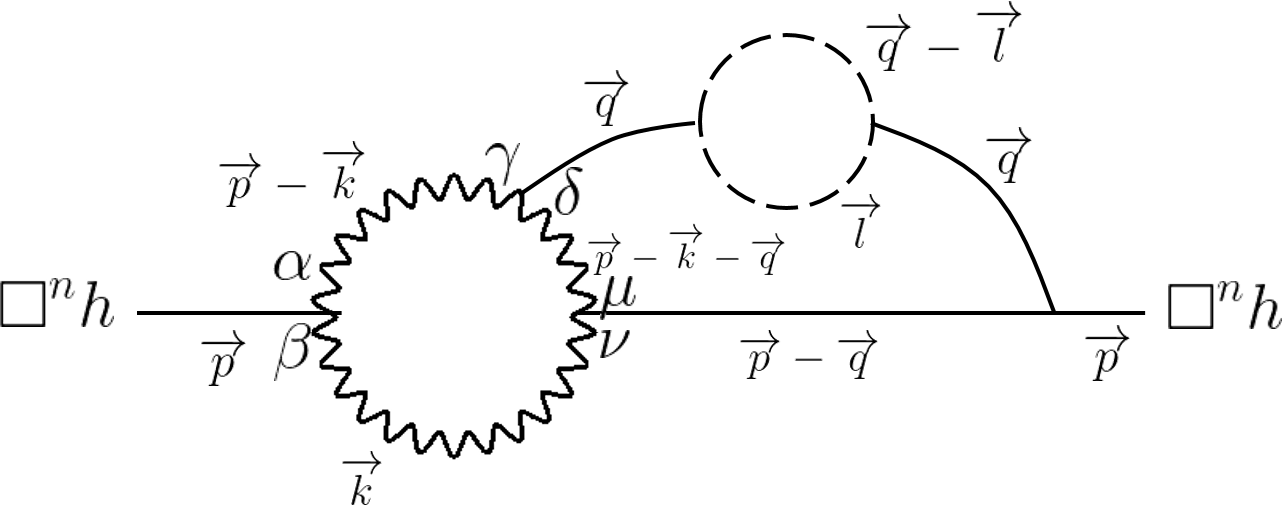}
\caption[]
{ Feyman diagram of self-energy correction of the Higgs boson by a 3-loop interaction.  \label{fig:loop4}}
\end{figure}
The Feynman amplitude can be easily written down as
\begin{equation}
\begin{aligned}
\mathcal{M}^{(n)} &= -\frac{M_W^3 m_f^2 m_{\mathrm{H}}^2}{2 (2\pi)^{3D} v_n^3} \iiint d^D k \,d^D q \,d^D l \,g^{\beta\alpha}g^{\gamma\delta}g^{\mu\nu} \\
&\,\,\,\,\times \frac{g_{\alpha\gamma} -(p-k)_\alpha (p-k)_\gamma  /M_W^2}{(p-k)^{4n}((p-k)^2 - M_W^2)}\frac{g_{\delta\mu} -(p-k-q)_\delta (p-k-q)_\mu /M_W^2}{(p-k-q)^{4n}((p-k-q)^2 - m_W^2)} \frac{g_{\nu\beta}-k_\nu k_\beta  /M_W^2}{k^{4n}(k^2 - M_W^2)} \\
&\,\,\,\,\times \bigg(\frac{1}{q^{4n}(q^2 -m_{\mathrm{H}}^2 )}\bigg)^2 \frac{1}{(p-q  )^{4n}(p-q)^2 -m_\mathrm{H}^2} \\
&\,\,\,\,\times\mathrm{Tr}\bigg(\frac{1}{(\slashed{l}^*)^n (\slashed{l}-m) (\slashed{l}^n)} \frac{1}{(\slashed{q}^*-\slashed{l}^*)^n ( \slashed{q} -\slashed{l} -m  ) (\slashed{q}-\slashed{l})^n}   \bigg) \,.
\end{aligned}
\end{equation}
To compute the superficial degree of UV divergence, we utilize the power counting formula in (\ref{eq:PC}). First notice that $N_{\mathrm{B}}=3,\,\, N_{\mathrm{H}}=3,\,\, N_f=2 ,\,\, V=6$, then $L = 3$. Hence it follows that $S= 3D-14(2n+1)= 3D-28n-14$. We can see that when $n=0$ for the unrotored case and $D=4$ spacetime $S=-2$, thus this diagram is convergent. We can also see that in $D=4$ spacetime $S=-2 -28n <0$, so there is no UV divergence.

Therefore, in this section, we have seen how infinities arise from high-energy ends of loop diagrams are tamed by higher-order-derivatives fields under rotor mechanism. In such way, the UV divergences of from the 1-loop self-correction of Higgs boson propagator are eliminated. 

\section{Explicit calculation of 1-loop self-correction of Higgs boson under rotor mechanism and a discussion on the Hierarchy Problem }
\subsection{Correction by the 3rd order Higgs vertex}
In this section, we will demonstrate the explicit calculation of 1-loop self correction of Higgs boson by third order Higgs vertex under rotor mechanism. We will show that although the rotor mechanism can tame ultraviolet UV divergence at high energies, it introduces Infra-red IR divergence at low energies in 4-dimensional spacetime. However, we will demonstrate that the rotor mechanism can remove all infinities in higher dimension using the example of 1-loop self-correction of Higgs Boson by third order Higgs vertex.  According to equation (\ref{eq:152}), the n-th order rotored amplitude is proportional to the following integral,
\begin{equation}
\mathcal{M}^{(n)} \propto \int d^D k \frac{1}{k^{4n}(k^2 - m_{\mathrm{H}}^2)}\frac{1}{(p-k)^{4n} ((p-k)^2 -m_{\mathrm{H}}^2 )} 
\end{equation}
Define the integral as
\begin{equation}
I^{(n)} = \int d^D k \frac{1}{k^{4n}(k^2 - m_{\mathrm{H}}^2)}\frac{1}{(p-k)^{4n} ((p-k)^2 -m_{\mathrm{H}}^2 )} 
\end{equation}
We will first demonstrate the calculation of $n=1$ case, then the general $n$ rotored case. 
Using the Feynman parameter method given in \cite{Peskin}
\begin{equation} \label{eq:164}
\frac{1}{A_1 A_2 \cdots A_n} = \int_0^1 \cdots \int_0^1 dx_1 dx_2 \cdots dx_n \,\delta \bigg(\sum_{i=1}^n  x_i -1 \bigg) \frac{(n-1)!}{(x_1 A_1 + x_2 A_2 +\cdots x_n A_n )^n} \,.
\end{equation}
For the $n=1$ rotored case, we can rewrite the Feynman integral using equation (\ref{eq:164})
\begin{equation}
\begin{aligned}
&\frac{1}{k^4 (p-k)^4 (k^2 - m_{\mathrm{H}}^2) ( (p-k)^2 -  m_{\mathrm{H}}^2   )} \\
& = \int_0^1 \int_0^1 \int_0^1 \int_0^1 \int_0^1 \int_0^1 dx dy dr ds du dv \delta(x+y+r+s+u+v-1)\\
&\quad\times
\frac{5!}{[xk^2 + yk^2 + r(p-k)^2 + s(p-k)^2 + u(k^2 - m_{\mathrm{H}}^2 ) + v((p-k)^2 - m_{\mathrm{H}}^2)]^6  }
\end{aligned}
\end{equation}
Now let us evaluate the terms in the denominator. 
\begin{equation}
\begin{aligned}
&\quad xk^2 + yk^2 + r(p-k)^2 + s(p-k)^2 + u(k^2 - m_{\mathrm{H}}^2 ) + v((p-k)^2 - m_{\mathrm{H}}^2) \\
&= xk^2 + yk^2 + r (p^2 - 2 p \cdot k + k^2) + s(p^2 - 2 p \cdot k + k^2) + u(k^2 - m_{\mathrm{H}}^2 ) \\
&\quad\quad+ v( p^2 - 2 p \cdot k + k^2  - m_{\mathrm{H}}^2 )\\
&= (x+y +r +s +u+v) k^2 + (r+s+v)p^2 -2(r+s+v)(p\cdot k) - (u+v)m_{\mathrm{H}}^2 \\
&= k^2 -2(r+s+v) +(r+s+v)p^2 - (u+v)m_{\mathrm{H}}^2 \\
&= (k-(r+s+v)p )^2 - (u+v)m_{\mathrm{H}}^2 \,, \\
\end{aligned}
\end{equation}
where in the fourth line we used the fact that $x+y+r+s+u+v =1$. We define new variable $l = k -(r+s+v)p$ and the effective mass $\Delta = (u+v)m_{\mathrm{H}}^2$. Also it is clearly that $d^D k = d^D l$. Therefore, now the integral reads
\begin{equation}
\begin{aligned}
I^{(1)}&= \int d^D k \int_{0}^1 \int_{0}^1\int_{0}^1\int_{0}^1\int_{0}^1\int_{0}^1 
dx dy dr ds du dv \\
& \quad\times\delta(x+y+r+s+u+v-1)\frac{120}{[ (k-(r+s+v)p)^2  ) - (u+v)m_{\mathrm{H}}^2  ]^6} \,.
\end{aligned}
\end{equation}
Then we have
\begin{equation}
I^{(1)}=\int d^D l  \int_{0}^1 \int_{0}^1\int_{0}^1\int_{0}^1\int_{0}^1\int_{0}^1 
dx dy dr ds du dv  \delta(x+y+r+s+u+v-1)\frac{120}{(l^2 - \Delta)^6} \,.
\end{equation}
Now we carry out Wick's rotation. The Wick's rotation simply amount to substitute $l^0 = i l_E^0$ and $\pmb{l} = \pmb{l}_E$. Then the integral now reads
\begin{equation}
\begin{aligned}
I^{(1)} &= 120 i(-1)^6 \int d\Omega_4 \int_0^\infty dl_E \int_{0}^1 \int_{0}^1\int_{0}^1\int_{0}^1\int_{0}^1\int_{0}^1 dx dy dr ds du dv \\
&\quad \times \delta(x+y+r+s+u+v-1) \frac{l_E^3}{(l_E^2 + \Delta )^6} \\
&= \frac{120 i (2\pi^2)}{\Gamma(2)}\int_0^\infty dl_E\int_{0}^1 \int_{0}^1\int_{0}^1\int_{0}^1\int_{0}^1\int_{0}^1 dx dy dr ds du dv\\
&\quad\times \delta(x+y+r+s+u+v-1) \frac{l_E^3}{(l_E^2 + \Delta )^6} \,.
\end{aligned}
\end{equation}
Next using the integral fact that 
\begin{equation}
\int_0^\infty \frac{x^3}{(x^2 +a)^6} dx = -\frac{a+5x^2}{40 (a+x^2)^5}\bigg\vert^\infty_0 = \frac{1}{40a^4} \,.
\end{equation}
Therefore, now the integral reads
\begin{equation}
\begin{aligned}
I^{(1)} &= \frac{240i\pi^2}{40 m_{\mathrm{H}}^8}\int_{0}^1 \int_{0}^1\int_{0}^1\int_{0}^1\int_{0}^1\int_{0}^1 dx dy dr ds du dv  \delta(x+y+r+s+u+v-1) \frac{1}{(u+v)^4}
\\
&= \frac{6 i \pi^2}{m_{\mathrm{H}}^8} \int_0^1 du \int_0^{1-u} dv \int_0^{1-u-v}ds \int_0^{1-s-u-v} dr \int_0^{1-r-s-u-v} dy \frac{1}{(u+v)^4} \\
&=\frac{6 i \pi^2}{m_{\mathrm{H}}^8}\lim_{u\rightarrow 0}\bigg( -\frac{1}{4} + \frac{1}{36u^2} - \frac{1}{4u} - \frac{1}{2}\ln u \bigg) \\
&=\frac{-3i \pi^2}{2 m_{\mathrm{H}}^8} + \mathrm{infinity} \,.
\end{aligned}
\end{equation}
We can see that the integral diverges when $u \rightarrow 0$, this is referred as the infra-red IR divergence. Finally, we obtain the amplitude as
\begin{equation}
\mathcal{M}^{(1)} = \frac{3i}{512 \pi^2 m_{\mathrm{H}}^4 v_n^2} + \mathrm{infinity}\,.
\end{equation}
Therefore, we can see that although the rotor mechanism resolves the high-energy infinity problem, it introduces infra-red IR divergences at low energy in $D=4$ spacetime.

\subsubsection*{The General $n$ case}
To calculate the amplitude of general $n$-rotored case, we need to use the following Feynman integral formula given in \cite{Peskin}
\begin{equation}
\begin{aligned}
\frac{1}{A_1^{m_1}A_2^{m_2}\cdots A_n^{m_n}} &= \int_0^1 \cdots  \int_0^1 \bigg(\prod_{i=1}^n dx_i \bigg)\delta\bigg(\sum_{i=1}^n x_i -1 \bigg)\frac{\prod_{i=1}^n x_i^{m_i -1} }{\big(\sum_{i=1}^n x_i A_i \big)^{\sum_{i=1}^n m_i}} \\
&\quad\quad\quad\quad\quad\quad\quad\quad\quad\times \frac{\Gamma(m_1 + m_2 +\cdots + m_n)}{\Gamma({m_1})\Gamma({m_2})\cdots \Gamma(m_n)} \,.
\end{aligned}
\end{equation} 
The n-rotored integral is calculated to be
\begin{equation}
\begin{aligned}
I^{(n)}&= \int d^D k \frac{1}{(k^{2})^{2n}(k^2 - m_{\mathrm{H}}^2)}\frac{1}{[(p-k)^{2}]^{2n} ((p-k)^2 -m_{\mathrm{H}}^2 )} \\
&= \int d^D k\int_0^1 \int_0^1 \int_0^1 \int_0^1 dx dy dr ds\delta(x+y+r+s-1)  \\
&\quad\times \frac{x^{2n-1} y^{2n-1}r^0 s^0 }{[xk^2 + y(p-k)^2 + r(k^2 - m_{\mathrm{H}}^2) +s((p-k)^2 - m_{\mathrm{H}}^2 ) ]^{4n+2}} \frac{\Gamma(4n+2)}{\Gamma(2n)\Gamma(2n)\Gamma(1)\Gamma(1)} \\
&= \int d^D k\int_0^1 \int_0^1 \int_0^1 \int_0^1 \delta(x+y+r+s-1) dx dy dr ds \\
&\quad\times \frac{x^{2n-1} y^{2n-1} }{[ (k-(y+s)p)^2 -(r+s)m_{\mathrm{H}}^2 ]^{4n+2}} \frac{\Gamma(4n+2)}{\Gamma^2(2n)} 
\end{aligned} \,.
\end{equation}
This time take $l = k - (y+s)p$ and the effective mass $\Delta(r+s)= m_{\mathrm{H}}^2$. Then we have
\begin{equation}
\begin{aligned}
I^{(n)} &= \frac{\Gamma(4n+2)}{\Gamma^2(2n)} \int d^D l \int_0^1 \int_0^1 \int_0^1 \int_0^1 dxdydrds\delta(x+y+r+s-1)  \frac{x^{2n-1} y^{2n-1}}{(l^2 - \Delta)^{4n+2}} \\
&=\frac{i(-1)^{4n+2}\Gamma(4n+2)}{\Gamma^{2}(2n)} \int_0^\infty dl_E \int d\Omega_D \int_0^1 \int_0^1 \int_0^1 \int_0^1 dxdydrds\delta(x+y+r+s-1)  \\
&\quad\quad\quad\quad\quad\quad\quad\quad\quad\quad\quad\quad\quad\quad\quad\quad\quad\quad\quad\quad\quad \times x^{2n-1}y^{2n-1} \frac{l_E^{D-1}}{(l_E^2 + \Delta)^{4n+2}}\\
&=\frac{2i(-1)^{4n+2}\pi^{D/2}\Gamma(4n+2)}{\Gamma^2(2n)\Gamma\big(\frac{D}{2}\big)}\int_0^\infty dl_E  \int_{0}^1 \int_{0}^1 \int_{0}^1 \int_{0}^1 dxdydrds \delta(x+y+r+s-1) \\
& \quad\quad\quad\quad\quad\quad\quad\quad\quad\quad\quad\quad\quad\quad\quad\quad\quad\quad\quad\quad\quad \times x^{2n-1}y^{2n-1} \frac{l_E^{D-1}}{(l_E^2 + \Delta)^{4n+2}} \,.\\
\end{aligned}
\end{equation}
Next we calculate the $l_E$ integral
\begin{equation}
\int_0^\infty dl_E \frac{l_E^{D-1}}{(l_E^2 + \Delta)^{4n+2}} = \frac{\Gamma\big(\frac{D}{2} \big) \Gamma\big( 4n+2 -\frac{D}{2} \big)}{2\Delta^{4n+2 -\frac{D}{2}}\Gamma(4n+2)} \,\,\,\,\text{if}\,\, 0< D < 8n+4 \,.
\end{equation}
Note that $D$ is out of the range, the integral does not converge. Therefore, now the $I^{(n)}$ integral is\
\begin{equation}
\begin{aligned}
I^{(n)}&=\frac{i(-1)^{4n+2} \pi^{D/2} \Gamma\big( 4n+2 -\frac{D}{2} \big)}{\Gamma^2 (2n)} \int_0^1 \int_0^1 \int_0^1 \int_0^1 dxdydrds \delta(x+y+r+s-1) \frac{x^{2n-1} y^{2n-1}}{\Delta^{4n+2 -\frac{D}{2}}} \\
&=\frac{i(-1)^{4n+2} \pi^{D/2} \Gamma\big( 4n+2 -\frac{D}{2} \big)}{\Gamma^2 (2n) m_{\mathrm{H}}^{8n+4 -D}} \int_0^1 \int_0^1 \int_0^1 \int_0^1 dxdydrds \delta(x+y+r+s-1) \frac{x^{2n-1} y^{2n-1}}{(r+s)^{4n+2 -\frac{D}{2}}} \\
&=\frac{i (-1)^{4n+2}\pi^{D/2} \Gamma\big( 4n+2 -\frac{D}{2} \big)}{\Gamma^2 (2n) m_{\mathrm{H}}^{8n+4 -D}} \int_0^1 dr \int_0^{1-r} ds  \int_0^{1-r-s} dy\frac{(1-y-r-s)^{2n-1} y^{2n-1}}{(r+s)^{4n+2 -\frac{D}{2}}} \\
&= \frac{i (-1)^{4n+2}\pi^{D/2} \Gamma\big( 4n+2 -\frac{D}{2} \big)}{\Gamma^2 (2n) m_{\mathrm{H}}^{8n+4 -D}} \frac{\Gamma^2 (2n)}{\Gamma(4n)} \int_0^1 dr \int_0^{1-r} ds\frac{(1-r-s)^{4n-1}}{(r+s)^{4n+2 -\frac{D}{2}}} \\
&= \frac{i(-1)^{4n+2} \pi^{D/2}\Gamma\big( 4n+2 -\frac{D}{2} \big)}{(-1)^{4n}\Gamma(4n)m_{\mathrm{H}}^{8n+4 -D}} \int_0^1 dr B \bigg( 1-\frac{1}{r}; 4n , 2-\frac{D}{2}\bigg) \,,
\end{aligned}
\end{equation}
where
\begin{equation}
B \bigg( 1-\frac{1}{r}; 4n , 2-\frac{D}{2}\bigg) = \int_0^{1-\frac{1}{r}} t^{4n-1} (1-t)^{1-\frac{D}{2}} dt
\end{equation}
is the incomplete Beta function. Hence, we need to compute the following integral
\begin{equation}
I^{(n)}(n,D) = \frac{i \pi^{D/2}\Gamma\big( 4n+2 -\frac{D}{2} \big)}{\Gamma(4n)m_{\mathrm{H}}^{8n+4 -D}} \int_0^1 dr \int^{1-\frac{1}{r}}_0  t^{4n-1} (1-t)^{1-\frac{D}{2}} dt \,.
\end{equation}
Let us define the integral
\begin{equation}
J(n,D) = \int_0^1 dr \int^{1-\frac{1}{r}}_0  t^{4n-1} (1-t)^{1-\frac{D}{2}} dt \,.
\end{equation}
First we analyse the case when $n=1/4$, which is not an integer. Although this corresponds to an unphysical rotor number, it is worth to study this case. It is found that, when $n=1/4$, the integral is computed to be
\begin{equation}
J\bigg(\frac{1}{4} ,D\bigg) = \frac{2}{D-4} \bigg(\frac{2}{D-2} -1 \bigg) \,\,\mathrm{if}\,\,D>2 \,.
\end{equation}
The plot is as follow:
\begin{figure}[H]
\centering
\includegraphics[trim=0cm 0cm 0cm 0cm, clip, scale=0.9]{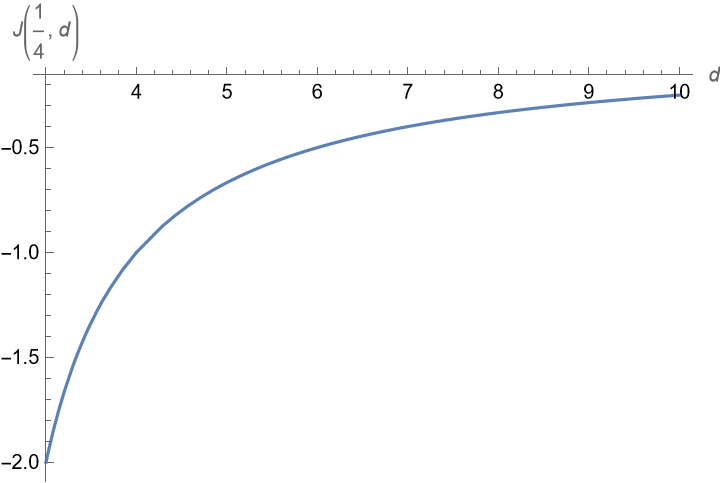}
\caption[]
{The plot of $J\Big(\frac{1}{4},D\Big)$ against spacetime dimension $D$.  \label{fig:plot}}
\end{figure}

When $n=1/4$, so the converged amplitude takes place only when
\begin{equation}
-i\mathcal{M}^{(1/4)}\bigg(\frac{1}{4},D\bigg)= \frac{\Gamma\big(3-\frac{D}{2} \big)}{2^{D+2}\sqrt{2 \pi^D} m_{\mathrm{H}}^{2-D} v_{\frac{1}{4}}^2}\bigg[\frac{2}{D-4} \bigg(\frac{2}{D-2} -1 \bigg)\bigg] \,.
\end{equation}

The amplitude against spacetime dimension is plotted as follow:
\begin{figure}[H]
\centering
\includegraphics[trim=0cm 0cm 0cm 0cm, clip, scale=0.9]{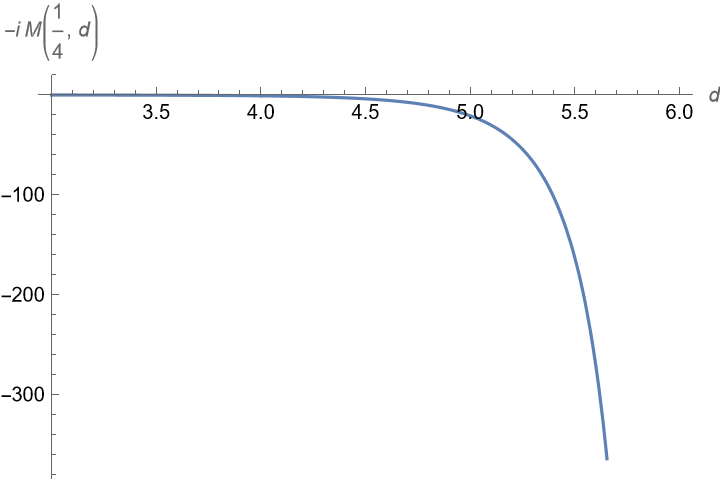}
\caption[]
{The plot of $-i\mathcal{M}\Big(\frac{1}{4},D\Big)$ against spacetime dimension $D$.  \label{fig:plot2}}
\end{figure}

In particular, in our living spacetime dimension $D=4$, using the L-Hospital rule,
\begin{equation}
J\bigg(\frac{1}{4} ,4\bigg) = \lim_{D\rightarrow 4} \frac{2}{D-4} \bigg(\frac{2}{D-2}-1 \bigg) = -1 \,.
\end{equation}
We thus obtain a convergent amplitude for $n=1/4$, $D=4$
\begin{equation}
i\mathcal{M}^{(1/4)}\bigg(\frac{1}{4},4\bigg)= \frac{m_{\mathrm{H}}^2}{64 \sqrt{2} \pi^2  v_{\frac{1}{4}}^2} = \frac{m_{\mathrm{H}}^2}{128\pi^2  v^2} = 2.044 \times 10^{-4} \,,
\end{equation}
where we have used the fact that $v_n^2= 4^n v^2$ in equation (\ref{eq:124}) and $v = 246 \,\mathrm{GeV}$ is the vacuum expectation value. The amplitude is inversely proportional to the square of the Higgs boson mass. It is noted that the amplitude is ill-defined when $D=2$, for which it diverges. 

Now, we investigate the physical $n,D \in \mathbb{Z}^+$ cases. We find that for each integer $n$, the minimum dimension $D$ that gives convergent $J$ integral and amplitude $i\mathcal{M}^{\mathrm{(n)}}$ is given by the formula
\begin{equation}
D_{\mathrm{min}}(n) = 8n +1 \,.
\end{equation}
For example, for $n=1$ case, $D_{\mathrm{min}}(1)$ = 9. That means the lowest spacetime dimension for which the rotor mechanism to give finite result is nine. For $n=1$, if $D<9$, the result will be all infinite. For example,  for the $n=2$ case, $D_{\mathrm{min}}(2)$ = 17. That means the lowest spacetime dimension for which the rotor mechanism to give finite result is seventeen. For $n=2$, if $D<17$, this will be all infinite. This applies to higher positive integer $n$ values. And we conclude that for the Higgs boson self-correction by 3rd order Higgs-self interaction, the minimum physical dimension for taming the divergences (for both IR and UV) of the lowest-order rotor model $(n=1)$ is 9. 

\begin{figure}[H]
\centering
\includegraphics[trim=0cm 0cm 0cm 0cm, clip, scale=1.0]{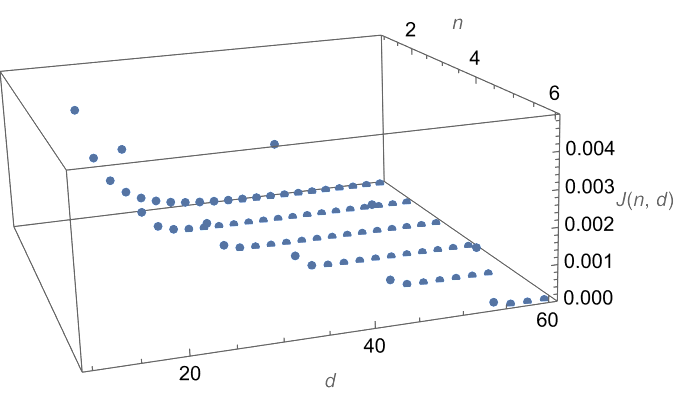}
\caption[]
{The plot of $J(n,D)$ against rotor number $n$ and spacetime dimension $D$. \label{fig:3Dplot}}
\end{figure}

Now we numerically work out the pair of  $(n,d)$ which contributes the finite convergent $J(n,D)$ integral. The result is plotted in figure (\ref{fig:3Dplot}).  

The full amplitude is given by
\begin{equation}
-i\mathcal{M}^{(n)}(n, D) =\frac{\Gamma\big( 4n+2-\frac{D}{2}\big)}{2^{D+4n+2} \Gamma(4n)\sqrt{\pi^D} m_{\mathrm{H}}^{8n-D} v^2} J(n,D)\,.
\end{equation}
In general, in the nominator, the Gamma function $\Gamma\big(4n +2 -\frac{D}{2}\big)$ imposes constraints such that the amplitude is finite. First notice that $\Gamma(-m)$ is infinity for $m$ equal to positive integers. Therefore, such condition occurs when $D= 2k$ and $4n+2 -k \leq 0$, so it takes place when $k \geq 4n +2$. Together with $D\geq 8n+1$ and $D < 8n +4 $, the conditions for finite amplitude are
\begin{equation}
\begin{cases}
D & < 8n+4 \\
D &\geq 8n + 1\\
k &\ngeq 4n+2 \\
D &\neq 2k\\
\end{cases} \,.
\end{equation}
For example $n=1$, we have $k\geq 6$ , $D \geq 12 $, this will give $\Gamma(-m)$, which is divergent. So for $n=1$, the required $(n,D)$ pair for finite amplitude is $(1,9),(1,10), (1,11)  $. For example $n=2$, the required $(n,D)$ pair for finite amplitude is $(2, 17),(2,18),(2,19)$. For example,
\begin{equation}
-i\mathcal{M}^{(1)}(1,9) = \frac{\Gamma\big( \frac{3}{2}\big)}{2^{15}\Gamma(4) \pi^{9/2} m_{\mathrm{H}}^{-1}v^2} J(1,9) = 4.931 \times 10^{-11}\,\mathrm{GeV}^{-1} \,,
\end{equation}
which is a very small value. Therefore, under the rotor mechanism with rotor index $n=1$, at $D=9$, the divergence is removed in high dimension. Thus, this demonstrates how high-order-derivative field theory under rotor mechanism can remove both UV and IR divergences in high spacetime dimension. This shows that the Hierarchy problem can be solved by rotor mechanism in high spacetime dimension. 

\subsection{Correction by the $W$ boson}
Next, we will study the self-energy correction of Higgs boson by massive $W$ boson. Using the result in (\ref{eq:132}), we have the amplitude as 
\begin{equation}
\begin{aligned}
\mathcal{M}^{(n)} &=\frac{g_W^2  M_W^2}{4^n (2\pi)^D} \bigg( \int d^D k \frac{D }{ k^{4n} (p-k)^{4n} (k^2-M_W^2 )( (p-k)^2 -M_W^2  )} \\ 
& \quad\quad\quad\quad\quad - \frac{1}{M_W^2}\int d^D k \frac{(p-k)^2}{k^{4n} (p-k)^{4n} (k^2-M_W^2 )( (p-k)^2 -M_W^2  )} \\
&\quad\quad\quad\quad\quad - \frac{1}{M_W^2}\int d^D k \frac{k^2 }{k^{4n} (p-k)^{4n} (k^2-M_W^2 )( (p-k)^2 -M_W^2  )}\\
&\quad\quad\quad\quad\quad + \frac{1}{M_W^4}\int d^D k \frac{[k\cdot (p-k)]^2} {k^{4n} (p-k)^{4n} (k^2-M_W^2 )( (p-k)^2 -M_W^2  )} \bigg) \,.
\end{aligned}
\end{equation} 
From this we define the following integrals. First, from the first line, we have
\begin{equation}
I^{(n)}_1 = \int d^D k \frac{D }{ (k^{2})^{2n} [(p-k)^{2}]^{2n} (k^2-M_W^2 )( (p-k)^2 -M_W^2  )} \,.
\end{equation}
From the second line we have
\begin{equation}
I^{(n)}_2 = \frac{1}{M_W^2} \int d^D k \frac{1}{ (k^{2})^{2n} [(p-k)^{2}]^{2n-1} (k^2-M_W^2 )( (p-k)^2 -M_W^2  )} \,.
\end{equation}
From the third line we have
\begin{equation}
I^{(n)}_3 = \frac{1}{M_W^2} \int d^D k \frac{1}{ (k^{2})^{2n-1} [(p-k)^{2}]^{2n} (k^2-M_W^2 )( (p-k)^2 -M_W^2  )} \,.
\end{equation}
For the last line, we use the following standard trick to convert the inner product of $k\cdot p$ to squared values. The following identity is used:
\begin{equation}
a\cdot b = \frac{1}{2}[ a^2 + b^2 - (a-b)^2     ] \,.
\end{equation} 
Therefore,
\begin{equation}
\begin{aligned}
I^{(n)}_4  &= \frac{1}{M_W^4}\int d^D k \frac{(k\cdot p - k^2)^2 }{ (k^{2})^{2n} [(p-k)^{2}]^{2n} (k^2-M_W^2 )( (p-k)^2 -M_W^2  )} \\
&= \frac{1}{M_W^4}\int d^D k \frac{ [\frac{1}{2} \big( (k^2 + p^2) - (k-p)^2 \big) - k^2 ]^2 }{ (k^{2})^{2n} [(p-k)^{2}]^{2n} (k^2-M_W^2 )( (p-k)^2 -M_W^2  )}\\
&=\frac{1}{4M_W^4} \int d^D k \frac{(p^2 - k^2 )^2 - 2(p^2 -k ^2)(p-k)^2 + (p-k)^4}{(k^{2})^{2n} [(p-k)^{2}]^{2n} (k^2-M_W^2 )( (p-k)^2 -M_W^2  )} \\
&= \frac{1}{4M_W^4} \int d^D k \frac{(p^4 - 2p^2 k^2 + k^4) - 2(p^2 -k ^2)(p-k)^2 + (p-k)^4}{(k^{2})^{2n} [(p-k)^{2}]^{2n} (k^2-M_W^2 )( (p-k)^2 -M_W^2  )} \,.
\end{aligned}
\end{equation}
For each term in the last line, we define six more integrals:
\begin{equation}
I^{(n)}_{4a} = \frac{p^4}{4 M_W^4} \int d^D k \frac{1}{(k^{2})^{2n} [(p-k)^{2}]^{2n} (k^2-M_W^2 )( (p-k)^2 -M_W^2  )} \,,
\end{equation}
\begin{equation}
I_{4b}^{(n)}=\frac{-2p^2}{4M_W^4}\int d^D k \frac{1}{(k^{2})^{2n-1} [(p-k)^{2}]^{2n} (k^2-M_W^2 )( (p-k)^2 -M_W^2  )} \,,
\end{equation}
\begin{equation}
I_{4c}^{(n)}=\frac{1}{4M_W^4}\int d^D k \frac{1}{(k^{2})^{2n-2} [(p-k)^{2}]^{2n} (k^2-M_W^2 )( (p-k)^2 -M_W^2  )} \,,
\end{equation}
\begin{equation}
I_{4d}^{(n)} = \frac{-2p^2}{4M_W^4}\int d^D k \frac{1}{(k^{2})^{2n} [(p-k)^{2}]^{2n-1} (k^2-M_W^2 )( (p-k)^2 -M_W^2  )} \,,
\end{equation}
\begin{equation}
I_{4e}^{(n)} = \frac{2}{4M_W^4}\int d^D k \frac{1}{(k^{2})^{2n-1} [(p-k)^{2}]^{2n-1} (k^2-M_W^2 )( (p-k)^2 -M_W^2  )} \,,
\end{equation}
\begin{equation}
I_{4f}^{(n)} = \frac{1}{4M_W^4}\int d^D k \frac{1}{(k^{2})^{2n} [(p-k)^{2}]^{2n-2} (k^2-M_W^2 )( (p-k)^2 -M_W^2  )} \,,
\end{equation}
Then now the computation becomes straight forward. The computation is similar to that of the 3rd order Higgs vertex case, therefore we will just show the final result in case some special issues are noted. First of all,
\begin{equation}
I^{(n)}_1 = \frac{iD \pi^{D/2} \Gamma\big( 4n+2-\frac{D}{2}\big)}{\Gamma(4n) m_W^{8n+4-D}} \int_0^1 B\bigg( 1-\frac{1}{r} , 4n , 2-\frac{D}{2}  \bigg) \,.
\end{equation}
if $ 8n +1 \leq  D < 8n+4$. Secondly,
\begin{equation}
I_2^{(n)} = \frac{i\pi^{D/2} \Gamma\big( 4n+1-\frac{D}{2}\big)}{\Gamma(4n-1) m_W^{8n+4-D}}\int_0^1 B\bigg( 1-\frac{1}{r} , 4n-1 , 2-\frac{D}{2}  \bigg) \,.
\end{equation}
if $ 8n -1 \leq  D < 8n+2$. Thirdly,
\begin{equation}
I_3^{(n)} =\frac{i\pi^{D/2}\sqrt{\pi} \Gamma\big( 4n + 1 -\frac{D}{2} \big)}{4^{2n-1}\Gamma(2n)\Gamma\big( 2n -\frac{1}{2} \big)m_W^{8n+4-D} } \int_0^1 B\bigg( 1-\frac{1}{r} , 4n-1 , 2-\frac{D}{2}  \bigg) \,.
\end{equation}
if $ 8n -1 \leq  D < 8n+2$. Then
\begin{equation}
I^{(n)}_{4a} =\frac{p^4}{4M_W^4}\times \frac{I^{(n)}_{1}}{D}  = \frac{i \pi^{D/2} \Gamma\big( 4n+2-\frac{D}{2}\big) p^4}{4\Gamma(4n) m_W^{8n+8-D}} \int_0^1 B\bigg( 1-\frac{1}{r} , 4n , 2-\frac{D}{2}  \bigg) \,.
\end{equation}
if $ 8n +1 \leq  D < 8n+4$. Then
\begin{equation}
I^{(n)}_{4b} = -\frac{2 p^2}{4M_W^2} I_3^{(n)} = -\frac{2i \pi^{D/2} \sqrt{\pi}\Gamma\big(4n+1 -\frac{D}{2} \big)p^2}{4^{2n} \Gamma(2n) \Gamma\big(2n -\frac{1}{2} \big)  m_W^{8n+6-D}}\int_0^1 B\bigg( 1-\frac{1}{r} , 4n-1 , 2-\frac{D}{2}  \bigg) \,.
\end{equation}
For the $I_{4c}^{(n)}$ integral, the situation is more complicated. It it is worth to go through the whole computation process. 
\begin{equation} \label{eq:206}
\begin{aligned}
I_{4c}^{(n)} &= \frac{1}{4M_W^4} \int d^D l \int_0^1 \int_0^1 \int_0^1 \int_0^1 dxdydrds \delta(x+y+r+s-1) \frac{x^{2n-3}y^{2n-1}}{(l^2 -\Delta)^{4n}} \frac{\Gamma(4n)}{\Gamma(2n-2)\Gamma(2n)} \\
&= \frac{i(-1)^{4n} \Gamma(4n) (2\pi^{D/2})}{4M_W^4 \Gamma(2n-2) \Gamma(2n) \Gamma\big(\frac{D}{2}\big)}\int_0^1 \int_0^1 \int_0^1 \int_0^1 dxdydrds \delta(x+y+r+s-1) x^{2n-3} y^{2n-1}\\
&\quad\quad\quad\quad\quad\quad\quad\quad\quad\quad\quad\quad\quad\quad\quad\quad\quad\quad\quad\quad\quad\quad\quad\quad\quad\quad\quad \times \int dl_E \frac{l_E^{D-1}}{(l_E^2 + \Delta)^{4n}} \\
&=\frac{i(-1)^{4n} \Gamma(4n) (2\pi^{D/2})}{4M_W^4 \Gamma(2n-2) \Gamma(2n) \Gamma\big(\frac{D}{2}\big)}\int_0^1 \int_0^1 \int_0^1 \int_0^1 dxdydrds \delta(x+y+r+s-1) x^{2n-3} y^{2n-1}\\
&\quad\quad\quad\quad\quad\quad\quad\quad\quad\quad\quad\quad\quad\quad\quad\quad\quad\quad\quad\quad\quad\quad\quad\quad\quad\quad\quad \times \frac{2n\Gamma\big(\frac{D}{2}\big)\Gamma\big(4n-\frac{D}{2}\big)}{\Delta^{4n-\frac{D}{2}} \Gamma(4n+1)} \\
&= \frac{i (-1)^{4n} n \pi^{D/2}\Gamma(4n)\Gamma\big(4n-\frac{D}{2}\big)}{\Gamma(2n-2)\Gamma(2n)\Gamma(4n+1)M_W^{8n+4-D}}\int_0^1 \int_0^1 \int_0^1 \int_0^1 dxdydrds \delta(x+y+r+s-1) \frac{x^{2n-3} y^{2n-1}}{(r+s)^{4n-\frac{D}{2}}} \\
&=\frac{i (-1)^{4n} n \pi^{D/2}\Gamma(4n)\Gamma\big(4n-\frac{D}{2}\big)}{\Gamma(2n-2)\Gamma(2n)\Gamma(4n+1)M_W^{8n+4-D}}\int_0^1 dr \int_0^{1-r}ds \int_0^{1-r-s}  dy\frac{(1-y-r-s)^{2n-3} y^{2n-1}}{(r+s)^{4n-\frac{D}{2}}} \\
&=\frac{i (-1)^{4n} n \pi^{D/2}\Gamma(4n)\Gamma\big(4n-\frac{D}{2}\big)}{\Gamma(2n-2)\Gamma(2n)\Gamma(4n+1)M_W^{8n+4-D}}\int_0^1 dr \int_0^{1-r}ds \frac{(1-r-s)^{4n-3}}{(r+s)^{4n-\frac{D}{2}}}\frac{\Gamma(2n)\Gamma(2n-2)}{\Gamma(4n-2)} \\
&=\frac{i (-1)^{4n} n \pi^{D/2}\Gamma(4n)\Gamma\big(4n-\frac{D}{2}\big)}{\Gamma(4n+1)\Gamma(4n-2) M_W^{8n+4-D}} \int_0^1 dr \int_0^{1-r} ds\frac{(1-r-s)^{4n-3}}{(r+s)^{4n-\frac{D}{2}}}
\end{aligned}
\end{equation}
if $0< D < 8$. Now we evaluate the last integral. This gives
\begin{equation} \label{eq:207}
\int_0^{1-r} ds\frac{(1-r-s)^{4n-3}}{(r+s)^{4n-\frac{D}{2}}} =\frac{\Gamma(4n-2)}{(1-r)^{4n-2} r^{4n-\frac{D}{2}}} \,_2\tilde{F}_1 \bigg(1 , 4n-\frac{D}{2} , 4n-1 , 1-\frac{1}{r}    \bigg) \,,
\end{equation}
where $\,_2\tilde{F}_1$ is the regularized generalized hypergeometric $(2,1)$ function. We will give the definition of generalized hypergeometric function and regularized generalized hypergeometric $(p,q)$ function here, which is formally denoted as $\,_p F_q$. It is given that
\begin{equation}
\,_p F_q (a_1 , \cdots a_p ; b_1 , \cdots b_q ; z) = \sum_{n=0}^{\infty} \frac{\prod_{k=1}^p (a_k)^{(n)} }{\prod_{k=1}^q (b_k)^{(n)} } \frac{z^n}{n!} \,,
\end{equation}
where the Pochhammer symbol for rising factorial is used,
\begin{equation}
(a)^{(0)} = 1 , \cdots, (a)^{(n)} = a(a+1)(a+2)\cdots (a+n-1) = \prod_{l=0}^{n-1}(a+l) \,. 
\end{equation}
The corresponding regularized generalized hypergeometric $(p,q)$ function is defined by
\begin{equation}
\,_p \tilde{F}_q (a_1 , \cdots a_p ; b_1 , \cdots b_q : z) = \frac{\,_p F_q (a_1 , \cdots a_p ; b_1 , \cdots b_q ; z)}{\Gamma(b_1)\cdots \Gamma(b_q)} \,.
\end{equation}
For our case, we are concerned with the regularized generalized hypergeometric $(2,1)$ function, which is
\begin{equation}
\,_2 \tilde{F}_1 (a_1 , a_2 ; b_1 ; z) =\frac{\,_2 F_1 (a_1 , a_2 ; b_1 ; z)}{\Gamma(b_1)} = \frac{1}{\Gamma(b_1)} \sum_{n=0}^{\infty}\frac{(a_1 )^{(n)} (a_2)^{(n)} }{(b_1)^{(n)}} \frac{z^n}{n!}\,. 
\end{equation}
Therefore, for our case in (\ref{eq:207}), we have
\begin{equation}
\begin{aligned}
\,_2\tilde{F}_1 \bigg(1 , 4n-\frac{D}{2} , 4n-1 , 1-\frac{1}{r}    \bigg) &= \frac{1}{\Gamma(4n-1)} \sum_{m=0}^{\infty} \frac{1^{(m)} \big(4n-\frac{D}{2}\big)^{(m)}}{m!(4n-1)^{(m)}} \bigg(1-\frac{1}{r} \bigg)^m \\
 &= \frac{1}{\Gamma(4n-1)} \sum_{m=0}^{\infty} \frac{ \big(4n-\frac{D}{2}\big)^{(m)}}{(4n-1)^{(m)}} \bigg(1-\frac{1}{r} \bigg)^m \,.
\end{aligned}
\end{equation}
Therefore, the last line of (\ref{eq:206}) yields
\begin{equation} \label{eq:213}
\begin{aligned}
I^{(n)}_{4c}&= \frac{i (-1)^{4n} n \pi^{D/2}\Gamma(4n)\Gamma\big(4n-\frac{D}{2}\big)}{\Gamma(4n+1)\Gamma(4n-1) M_W^{8n+4-D}} \int_0^1 dr \frac{1}{(1-r)^{4n-2} r^{4n-\frac{D}{2}}} \,_2\tilde{F}_1 \bigg(1 , 4n-\frac{D}{2} , 4n-1 , 1-\frac{1}{r}    \bigg)\\
&= \frac{i (-1)^{4n} n \pi^{D/2}\Gamma(4n)\Gamma\big(4n-\frac{D}{2}\big)}{\Gamma(4n+1)\Gamma(4n-1) M_W^{8n+4-D}} \int_0^1 dr \frac{1}{(1-r)^{4n-2} r^{4n-\frac{D}{2}}}\sum_{m=0}^{\infty} \frac{ \big(4n-\frac{D}{2}\big)^{(m)}}{(4n-1)^{(m)}} \bigg(1-\frac{1}{r} \bigg)^m \,.
\end{aligned}
\end{equation}
Thus this completes the mathematical calculation of $I_{4c}^{(n)}$.

Then we continue to calculate the integral $I_{4d}^{(n)}$. This is easy,
\begin{equation}
I_{4d}^{(n)} = \frac{-2p^2}{4M_W^2}I_2^{(n)} = \frac{-i\pi^{D/2}\Gamma\big(4n+1-\frac{D}{2}\big)p^2}{2\Gamma(4n-1)m_W^{8n+6-D}} \int_0^1 B\bigg( 1-\frac{1}{r} , 4n-1 , 2-\frac{D}{2}  \bigg) \,.
\end{equation}
Next,
\begin{equation}
I_{4e}^{(n)}=\frac{i(-1)^{4n} \pi^{D/2}\sqrt{\pi}\Gamma(4n)\Gamma(4n-2)\Gamma\big(4n-\frac{D}{2} \big)}{2^{4n-4}\Gamma(2n-1)\Gamma(4n+1)\Gamma\big(2n-\frac{1}{2}\big)M_W^{8n+4-D}} \int_0^1 dr \frac{\,_2\tilde{F}_1 \big(1 , 4n-\frac{D}{2} , 4n-1 , 1-\frac{1}{r}    \big)}{(1-r)^{4n-2} r^{4n-\frac{D}{2}}} \,.
\end{equation}
Finally,
\begin{equation}
I_{4f}^{(n)} = \frac{i(-1)^{4n}\Gamma(4n)\Gamma\big(4n-\frac{D}{2}\big)}{4\Gamma(4n+1)M_{W}^{8n+4-D}}\int_0^1 dr \frac{\,_2\tilde{F}_1 \big(1 , 4n-\frac{D}{2} , 4n-1 , 1-\frac{1}{r}    \big)}{(1-r)^{4n-2} r^{4n-\frac{D}{2}}} \,.
\end{equation}
The amplitude of the Higgs-self energy correction by $W$ boson is hence
\begin{equation}
\mathcal{M}^{(n)}= \frac{g^2 M_W^2}{4^n (2\pi)^D} \big( I_1^{(n)} - I_2^{(n)}   - I_3^{(n)} + I_{4a}^{(n)} + I_{4b}^{(n)} + I_{4c}^{(n)} + I_{4d}^{(n)} + I_{4e}^{(n)} + I_{4f}^{(n)} \big) \,.
\end{equation}
It is worth to analyse the properties of these integrals. We can see that unlike the case of the  self-energy correction for the Higgs-boson from the 3rd order Higgs interaction, here the amplitude is ill-defined for the $n=1/4$ case, as we have the denominator of $\Gamma(4n-1)$ in $I_2^{(n)}, I_{4c}^{(n)},I_{4d}^{(n)}, I_{4e}^{(n)} ,I_{4f}^{(n)} $, which diverges when $n = 1/4$. Now, in order to get an idea of what pairs of $(n,D)$ give us convergent result, first we would like to analyse the integral in (\ref{eq:213}). Define
\begin{equation}
p(n,D,r)=\frac{1}{(1-r)^{4n-2} r^{4n-\frac{D}{2}}} \,_2\tilde{F}_1 \bigg(1 , 4n-\frac{D}{2} , 4n-1 , 1-\frac{1}{r}    \bigg) \,.
\end{equation}
Then we plot $p(n,D,r)$ with different $n$ and $D$ values. 

\begin{figure}[H]
\centering
\includegraphics[trim=0cm 0cm 0cm 0cm, clip, scale=1.0]{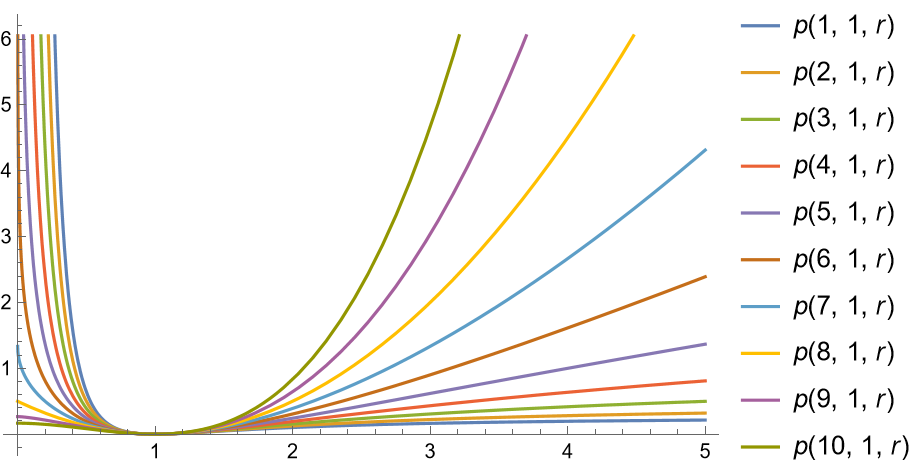}
\caption[]
{The plot of $p(n,D,r)$ with  of $n=1$ and different values of $D$. \label{fig:p1}}
\end{figure}
\begin{figure}[H]
\centering
\includegraphics[trim=0cm 0cm 0cm 0cm, clip, scale=1.0]{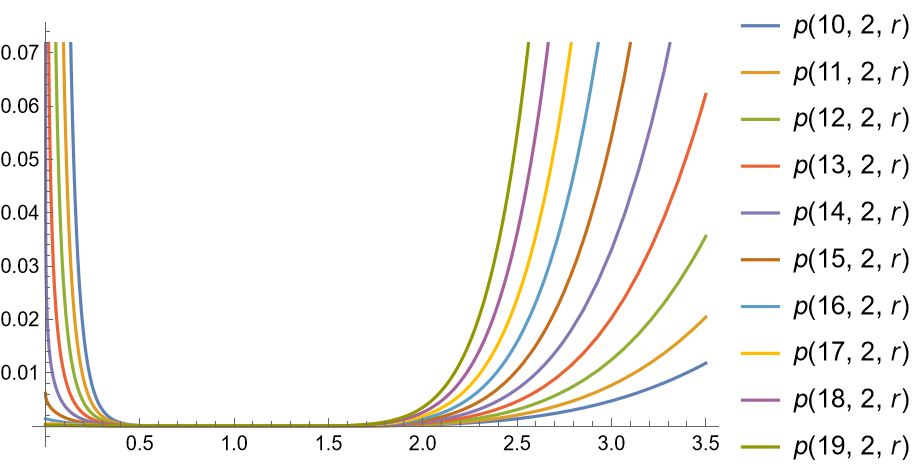}
\caption[]
{The plot of $p(n,D,r)$ with  of $n=2$ and different values of $D$. \label{fig:p2}}
\end{figure}
\begin{figure}[H]
\centering
\includegraphics[trim=0cm 0cm 0cm 0cm, clip, scale=1.0]{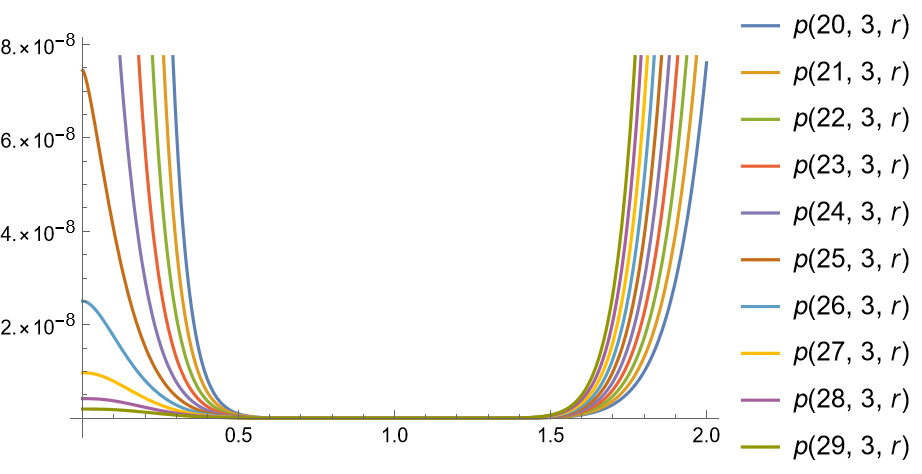}
\caption[]
{The plot of $p(n,D,r)$ with  of $n=3$ and different values of $D$. \label{fig:p3}}
\end{figure}
\begin{figure}[H]
\centering
\includegraphics[trim=0cm 0cm 0cm 0cm, clip, scale=1.0]{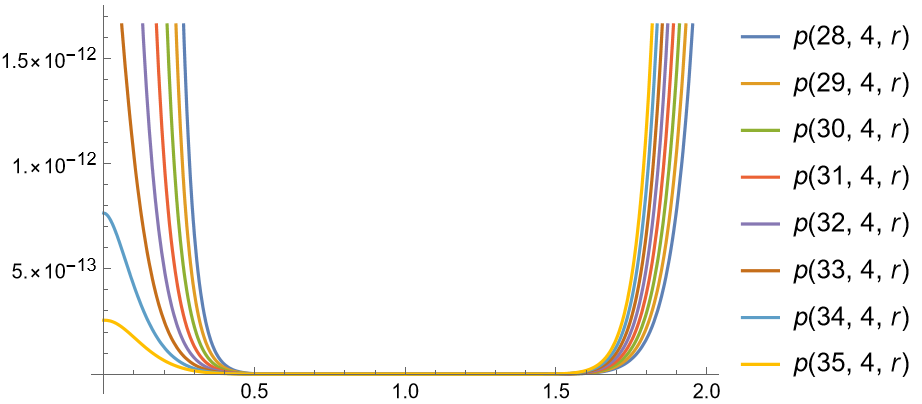}
\caption[]
{The plot of $p(n,D,r)$ with  of $n=4$ and different values of $D$. \label{fig:p4}}
\end{figure}
\begin{figure}[H]
\centering
\includegraphics[trim=0cm 0cm 0cm 0cm, clip, scale=1.0]{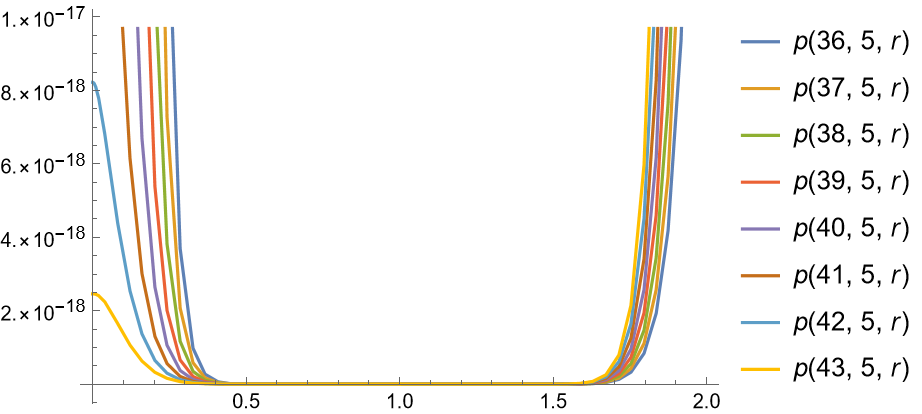}
\caption[]
{The plot of $p(n,D,r)$ with  of $n=5$ and different values of $D$. \label{fig:p5}}
\end{figure}

We study the general properties of the $p(n,D,r)$ function and the integration of it from 0 to 1. Let's first study the $(1,D)$ case in detail. For each set of $(1,D)$, we compute $p(1,D,r)$ and $\int_0^1 p(1,D,r) dr$.
\begin{table}[H]
\begin{center}
\begin{tabular}{ c | c | c }
 $(n,D)$ & $p(n,D,r)$ & $\int_0^1 p(1,D,r) dr$ \\
\hline
  $(1,1)$ & $-\frac{2}{3}r^{-3/2} + \frac{2}{5}r^{-5/2} + \frac{4}{15} $ & $\infty$ \\
  $(1,2)$ & $ \frac{(r-1)^2}{2r^2} $ & $\infty$ \\
  $(1,3)$ & $-\frac{2}{\sqrt{r}} + \frac{2}{3}r^{-3/2} +4 $ & $\infty$ \\
  $(1,4)$ & $\frac{1}{r} + \ln r -1$ & $\infty$ \\
  $(1,5)$ & $2\sqrt{r} \bigg(\sqrt{\frac{1}{r}} -1 \bigg)^2 $ & $1.333$ \\
  $(1,6)$ & $r-1 - \ln r$ & $0.5$ \\
  $(1,7)$ & $\frac{2}{3}\sqrt{r} \bigg( r+2 \sqrt{\frac{1}{r}} -3\bigg)$ & $0.267$ \\
  $(1,8)$ & $\frac{1}{2}(r-1)^2$ & $0.167$ \\
  $(1,9)$ & $\frac{2}{15}\bigg( 2\sqrt{\frac{1}{r}} -5r + 3r^2 \bigg)$ & $0.1143$ \\
  $(1,10)$ & $\frac{1}{6}(r-1)^2 (2r+1)$ & $0.0833$ \\
  \hline
\end{tabular}
\end{center}
\end{table}
We can see that for the $n=1$ case, the integral only converges when $D> 4$. And in general, for higher dimension $D$ the integral decreases. This pattern also applies for other $n$. For each $n$, there exists a threshold dimension $D_{\mathrm{min}}(n)$ such that below this dimension the integral diverges, upon computation we find that 
\begin{equation}
D_{\mathrm{min}}(n) = 8n-3 \,.
\end{equation}
When $n=1$, the threshold is $D_{\mathrm{min}}=5$, which is confirmed by the above table. And when $n=2$, $D_{\mathrm{min}} =13$, and so on. From all the plots above, we can see that larger $n$ with larger $D$ results in smaller values of the integral. This shows that we will obtain smaller amplitudes with increasing $(n,D)$ values.

The full amplitude is now given by
\begin{equation} \label{eq:215}
\begin{aligned}
\mathcal{M}^{(n)} &= \frac{ig_W^2 }{2^{2n+D} \sqrt{\pi^D} M_W^{8n+2-D}}\bigg\{ \frac{\Gamma\big( 4n+2-\frac{D}{2}\big) }{\Gamma(4n)} \bigg( D+ \frac{p^4}{4M_W^4} \bigg) \int_0^1 B\bigg( 1-\frac{1}{r} , 4n , 2-\frac{D}{2}  \bigg) \\
&\,\,-\Gamma\bigg(4n+1-\frac{D}{2} \bigg)\bigg[\frac{1}{\Gamma(4n-1)}\bigg(1+\frac{p^2}{2M_W^2} \bigg)  +\frac{\sqrt{\pi}}{4^{2n} \Gamma(2n)\Gamma\big(2n-\frac{1}{2} \big)}\bigg(4+\frac{2p^2}{M_W^2}\bigg) \bigg] \\
&\quad\quad\quad\quad\quad\quad\quad\quad\quad\quad\quad\quad\quad\quad\quad\quad\quad\quad\quad\quad\quad\times \int_0^1 B\bigg( 1-\frac{1}{r} , 4n-1 , 2-\frac{D}{2}  \bigg) \\
&+\frac{(-1)^{4n}\Gamma(4n)\Gamma\big(4n-\frac{D}{2} \big)}{\Gamma(4n+1)}\bigg[\frac{n}{\Gamma(4n-1)} + \frac{\sqrt{\pi}\Gamma(4n-2)}{\Gamma(2n-1)\Gamma\big(2n-\frac{1}{2} \big)} + \frac{1}{4}  \bigg] \\
&\quad\quad\quad\quad\quad\quad\quad\quad\quad\quad\quad\quad\quad\quad\quad\quad\quad\quad\times\int_0^1 dr \frac{\,_2\tilde{F}_1 \big(1 , 4n-\frac{D}{2} , 4n-1 , 1-\frac{1}{r}    \big)}{(1-r)^{4n-2} r^{4n-\frac{D}{2}}} \bigg\} \,.
\end{aligned}
\end{equation}
The full amplitude is a function of $n$,$D$ and incoming momentum $p$, $\mathcal{M}^{(n)}\equiv \mathcal{M}^{(n)}(n,D,p)$. It is found that for the unrotored case $(n=0)$ in our $D=4$ spacetime, this amplitude diverges to infinity. This is obvious due to the term $\Gamma(4n)$ appearing in the denominator. This result is expected as in the Standard Model, one-loop correction of the Higgs boson by $W$ boson diverges. 

However, as by the above results, we see that the integrals converge only in different set of $(n,D)$ values. We want to find the common $(n,D)$ values for all the integrals, which amount to solve the three inequalities:
\begin{equation}
\begin{cases}
& 8n+1 \leq D < 4n+4\\
& 8n -1 \leq  D< 8n+2\\
& 0 \leq D < 8 \\
& D \geq 8n-3 \\
\end{cases} \,.
\end{equation}
However, there is no solution for this set of inequality. Therefore, the amplitude in (\ref{eq:215}) means to be diverged in any particular set of $(n,D)$ values.   

\subsection{Correction by the fermion loop}
Finally, we remain to calculate the self-energy correction of Higgs Boson by fermion loop under rotor mechanism. This is essentially to compute the amplitude calculation in (\ref{eq:143}). We need to develop some Dirac algebra and trace identities involving $\gamma^{\mu\dagger}$, as well as $\gamma^{\mu}$ and $\gamma^{\mu\dagger}$ together in the most general $D$ spacetime dimension. First notice that the complex conjugate of the $\gamma^\mu$ matrix is defined by
\begin{equation} \label{eq:217}
\gamma^{\mu\dagger} = \gamma^0 \gamma^{\mu}\gamma^{0\dagger}= \gamma^0 \gamma^{\mu}\gamma^{0}\,,
\end{equation}
where we have used the fact that $\gamma^{0\dagger}= \gamma^0$. Then we immediately obtain our first identity :
\begin{equation}
\mathrm{Tr}(\gamma^{\mu^\dagger}) =\mathrm{Tr}(\gamma^0 \gamma^{\mu}\gamma^{0}) = \mathrm{Tr}(\gamma^0\gamma^0\gamma^\mu ) = \mathrm{Tr}(\gamma^\mu )=0 \,,
\end{equation}
where we have used the trace identity $\mathrm{Tr}(ABC)=\mathrm{Tr}(CAB)$ and the fact that $(\gamma^0)^2 =I$. Next, we prove the Dirac algebra for hermitian gamma matrices
\begin{equation}
\{\gamma^{\mu\dagger} , \gamma^{\nu\dagger}   \} = 2\eta^{\mu\nu}I_{D} \,.
\end{equation}
The proof is straight forward,
\begin{equation}
\begin{aligned}
\{\gamma^{\mu\dagger} , \gamma^{\nu\dagger}   \} &= \gamma^{\mu\dagger}\gamma^{\nu\dagger} + \gamma^{\nu\dagger}\gamma^{\mu\dagger} \\
&= \gamma^0 \gamma^{\mu}\gamma^0 \gamma^0 \gamma^{\nu}\gamma^0 + \gamma^0 \gamma^{\nu}\gamma^0 \gamma^0 \gamma^{\mu}\gamma^0
&=\gamma^0( \gamma^\mu\gamma^\nu +\gamma^\nu\gamma^\mu )\gamma^0 \\
&=\gamma^0 (2\eta^{\mu\nu}I_{D\times D})\gamma^0\\
&=2\eta^{\mu\nu}\gamma^0 I_{D\times D} \gamma^0\\
&=2\eta^{\mu\nu}I_{D\times D}\,,
\end{aligned}
\end{equation}
where we have used the fact that $(\gamma^0)^2 = I$. This result implies
\begin{equation}
\slashed{k}\slashed{k} = \slashed{k}^\dagger\slashed{k}^\dagger = k^2 I_{D\times D} \quad \text{and} \quad \slashed{k}\slashed{p} = \slashed{k}^\dagger\slashed{p}^\dagger = (k\cdot p)I_{D \times D} \,.
\end{equation}
Next we will prove the following identity:
\begin{equation}
\mathrm{Tr}(\gamma^{\mu}\gamma^{\rho\dagger}) =D(2\eta^{0\mu} \eta^{0\rho} - \eta^{\mu\rho})\,.
\end{equation} 
By the definition (\ref{eq:217}), it is noticed that $\mathrm{Tr}(\gamma^{\mu}\gamma^{\rho\dagger}) = \mathrm{Tr}(\gamma^{\mu}\gamma^0 \gamma^\rho \gamma^0 )$.
Then using the identity of $\mathrm{Tr}(\gamma^\mu \gamma^\nu \gamma^\rho \gamma^\sigma) = D(\eta^{\rho\sigma}\eta^{\mu\nu} -\eta^{\nu\sigma}\eta^{\mu\rho} +\eta^{\mu\sigma}\eta^{\nu\rho}  )$, by putting $\nu$ and $\sigma$ equal to 0, then we get
\begin{equation}
\begin{aligned}
\mathrm{Tr}(\gamma^{\mu}\gamma^{\rho\dagger}) &= \mathrm{Tr}(\gamma^{\mu}\gamma^0 \gamma^\rho \gamma^0 ) \\
&=D(\eta^{\rho 0} \eta^{\mu 0} -\eta^{00}\eta^{\mu\rho} + \eta^{\mu 0}\eta^{0\rho} ) \\
&=D(2\eta^{0\mu} \eta^{0\rho} -\eta^{\mu\rho})\,, 
\end{aligned}
\end{equation}
where $\eta^{00} =1 $ in our $\mathrm{diag}(+---)$ metric convention. Next we verify the following computationally,
\begin{equation} \label{eq:224}
\mathrm{Tr}(\gamma^{\mu\dagger}\gamma^\nu \gamma^{\rho\dagger} ) = \mathrm{Tr}(\gamma^0\gamma^\mu\gamma^0\gamma^\nu\gamma^0 \gamma^{\rho}\gamma^0 )=\mathrm{Tr}(\gamma^\mu \gamma^{\nu\dagger}\gamma^\rho) =0
\end{equation}
that the trace of a mixture of odd number of $\gamma^\mu$ and $\gamma^{\mu\dagger}$ matrices is zero. Next, another useful identity we need to use later is 
\begin{equation} \label{eq:225}
\begin{aligned}
&\quad\,\,\mathrm{Tr}(\gamma^\mu \gamma^{\nu\dagger} \gamma^\rho \gamma^{\sigma^\dagger}) =\mathrm{Tr}(\gamma^\mu\gamma^0\gamma^\nu \gamma^0 \gamma^\rho \gamma^0\gamma^\sigma \gamma^0 ) \\
&=D(8\eta^{0\mu}\eta^{0\nu}\eta^{0\rho}\eta^{0\sigma} -2\eta^{0\rho}\eta^{0\sigma}\eta^{\mu\nu} -2 \eta^{0\mu} \eta^{0\sigma}\eta^{\nu\rho} -2\eta^{0\nu}\eta^{0\rho}\eta^{\mu\sigma} -2\eta^{0\mu} \eta^{0\nu}\eta^{\rho\sigma}        \\
&\quad\quad\quad\quad\quad\quad\quad\quad\quad\quad\quad\quad\quad\quad\quad\quad\quad\quad\quad+\eta^{\mu\nu}\eta^{\rho\sigma} -\eta^{\mu\rho}\eta^{\nu\sigma}+ \eta^{\mu\sigma} \eta^{\nu\rho} )\,.        
\end{aligned}
\end{equation}
For simplicity, we will consider the $n=1$ case first in (\ref{eq:143}). Now we can evaluate the nominator of (\ref{eq:143})
\begin{equation}
\begin{aligned}
&\quad(\slashed{k}) (\slashed{k}+m_f )(\slashed{k}^\dagger )  (\slashed{p}-\slashed{k} ) (\slashed{p}-\slashed{k} + m_f ) (\slashed{p}^\dagger-\slashed{k}^\dagger ) \\
&=(k^2 \slashed{k}^\dagger + m_f \slashed{k}\slashed{k}^\dagger)\,[(p-k)^2 (\slashed{p}^\dagger - \slashed{k}^\dagger) + m_f (\slashed{p} - \slashed{k})(\slashed{p}^\dagger - \slashed{k}^\dagger)  ] \\
&=k^2 (p-k)^2 (k\cdot p -k^2)I_{D \times D} + m_f (p-k)^2 \slashed{k}\slashed{k}^\dagger (\slashed{p}^\dagger -\slashed{k}^\dagger  )\\
&\quad +m_f k^2 \slashed{k}^\dagger (\slashed{p} - \slashed{k})(\slashed{p}^\dagger - \slashed{k}^\dagger) + m_f^2 \slashed{k}\slashed{k}^\dagger (\slashed{p}-\slashed{k})(\slashed{p}^\dagger - \slashed{k}^\dagger) \\
&=k^2 (p-k)^2 (k\cdot p -k^2)I_{D \times D} +m_f (p-k)^2 \slashed{k}(k\cdot p - k^2) \\
&\quad + m_f k^2 \slashed{k}^\dagger (\slashed{p}\slashed{p}^\dagger - \slashed{p}^\dagger \slashed{k}^\dagger - \slashed{k}^\dagger\slashed{p}^\dagger + \slashed{k}\slashed{k}^\dagger  ) +m_f^2 \slashed{k}\slashed{k}^\dagger(\slashed{p}\slashed{p}^\dagger - \slashed{p}^\dagger \slashed{k}^\dagger - \slashed{k}^\dagger\slashed{p}^\dagger + \slashed{k}\slashed{k}^\dagger  )
\end{aligned}
\end{equation}
Now we need to take the trace of the above expression. First notice that,
\begin{equation}
\mathrm{Tr}(\slashed{k}) = k_{\mu}\mathrm{Tr}(\gamma^\mu) = 0 \,.
\end{equation}
Then using the identity of \ref{eq:224},
\begin{equation}
\mathrm{Tr}(\slashed{k}^\dagger \slashed{p} \slashed{q}^\dagger  ) = k_{\mu}p_{\nu}q_{\rho} \mathrm{Tr}(\gamma^{\mu\dagger}\gamma^\nu \gamma^{\rho\dagger} ) =0 \,.
\end{equation}
Therefore using these two results and 
\begin{equation}
\begin{aligned}
&\quad \mathrm{Tr} \Big( k^2 (p-k)^2 (k\cdot p -k^2)I_{D \times D} +m_f (p-k)^2 \slashed{k}(k\cdot p - k^2) \\
&\quad + m_f k^2 \slashed{k}^\dagger (\slashed{p}\slashed{p}^\dagger - \slashed{p}^\dagger \slashed{k}^\dagger - \slashed{k}^\dagger\slashed{p}^\dagger + \slashed{k}\slashed{k}^\dagger  ) +m_f^2 \slashed{k}\slashed{k}^\dagger(\slashed{p}\slashed{p}^\dagger - \slashed{p}^\dagger \slashed{k}^\dagger - \slashed{k}^\dagger\slashed{p}^\dagger + \slashed{k}\slashed{k}^\dagger  ) \Big) \\
&=D k^2 (p-k)^2 (k\cdot p -k^2) + m_f^2 \mathrm{Tr}( \slashed{k}\slashed{k}^\dagger \slashed{p}\slashed{p}^\dagger       -  \slashed{k}\slashed{k}^\dagger \slashed{p}\slashed{k}^\dagger  - \slashed{k}\slashed{k}^\dagger \slashed{k}\slashed{p}^\dagger) +  \slashed{k}\slashed{k}^\dagger \slashed{k}\slashed{k}^\dagger)\\
&=D k^2 (p-k)^2 (k\cdot p -k^2) + m_f^2 (k_{\mu}k_{\nu}p_{\rho}p_{\sigma}-k_{\mu}k_{\nu}p_{\rho}k_{\sigma}-k_{\mu}k_{\nu}k_{\rho}p_{\sigma}+ k_{\mu}k_{\nu}k_{\rho}k_{\sigma})\mathrm{Tr}(\gamma^\mu \gamma^{\nu\dagger}\gamma^{\rho}\gamma^{\sigma\dagger})\\
&=D k^2 (p-k)^2 (k\cdot p -k^2) + m_f^2 (k_{\mu}k_{\nu}p_{\rho}p_{\sigma}-k_{\mu}k_{\nu}p_{\rho}k_{\sigma}-k_{\mu}k_{\nu}k_{\rho}p_{\sigma}+ k_{\mu}k_{\nu}k_{\rho}k_{\sigma}) \\
&\quad\times D(8\eta^{0\mu}\eta^{0\nu}\eta^{0\rho}\eta^{0\sigma} -2\eta^{0\rho}\eta^{0\sigma}\eta^{\mu\nu} -2 \eta^{0\mu} \eta^{0\sigma}\eta^{\nu\rho} -2\eta^{0\nu}\eta^{0\rho}\eta^{\mu\sigma} -2\eta^{0\mu} \eta^{0\nu}\eta^{\rho\sigma}        \\
&\quad\quad\quad\quad\quad\quad\quad\quad\quad\quad\quad\quad\quad\quad\quad\quad\quad\quad\quad+\eta^{\mu\nu}\eta^{\rho\sigma} -\eta^{\mu\rho}\eta^{\nu\sigma}+ \eta^{\mu\sigma} \eta^{\nu\rho} )\\
&= D k^2 (p-k)^2 (k\cdot p -k^2)\\ 
&\quad-2D (k^0)^2 p^2 +8D k^0 p^0 k^2 -2D(p^0)^2 k^2 + Dk^2 p^2 +8D(k^0)^2 (k\cdot p) -4Dk^0 p^0 (k\cdot p)\\
&\quad -2Dk^2(k\cdot p) -8D(k^0)k^2 + Dk^4 -16D(k^0)^3 p^0 + 8D(k^0)^2 (p^0)^2 + 8D(k^0)^4 \,.
\end{aligned}
\end{equation}
Therefore, we have at least 13 integrals. But some of them vanish due to the odd parity, we will look into details when we come across them. The integrals are much more technically difficult than the previous ones, as now here we have to also integrate terms regarding $k^0$. With some observation, first we define the generic integral
\begin{equation}
\begin{aligned}
I(a,b,D) &= \int d^D k \frac{1}{(k^2)^a [(p-k)^2]^b (k^2 -m_f^2) ( (p-k)^2 - m_f^2  )   }\\
&=\frac{2i\pi^{D/2}\Gamma\big(a+b+2-\frac{D}{2} \big)}{\Gamma(a+b) m_f^{2(a+b+2)-D}} \int_0^1 dr B\bigg( 1-\frac{1}{r} ; a+b , 2-\frac{D}{2} \bigg)
\end{aligned}
\end{equation}
if $0 < D < 2a+2b+4 $. In addition, the integral of the beta function is also convergent for some threshold $D$ values.
Define the first integral as
\begin{equation}
\begin{aligned}
I^{(1)}_1 &= \int d^D k\frac{D k^2 (p-k)^2 (k\cdot p -k^2)}{k^4 (p-k)^4 (k^2 -m_f^2)((p-k)^2 -m_f^2)} \\
&= \frac{D}{2} \int d^D k \frac{(p^2 -k^2) - (p-k)^2}{k^2 (p-k)^2 (k^2 -m_f^2)((p-k)^2 -m_f^2) }\\
&=\frac{D}{2}\Big( p^2 I(1,1,D) - I(0,1,D) -I(1,0,D)                \Big)\,. \\
\end{aligned}
\end{equation}
The range for each integral is: for $I(1,1,D), 4< D< 8$, for $I(1,0,D)$ and $I(0,1,D)$, $ 2<D< 6$. Therefore the overlap $D$ for the first integral is $D=5$. 

The second integral is defined to be
\begin{equation}
I^{(1)}_2 = -2Dp^2 \int d^D k \frac{(k^0)^2}{k^4 (p-k)^4 (k^2 -m_f^2)((p-k)^2 -m_f^2)} \,.
\end{equation}
Then using the old trick of Feynman parameters, we have
\begin{equation}
I^{(1)}_2 = \frac{-2Dp^2 \Gamma(6)}{\Gamma^2 (2)\Gamma^2 (1)} \int d^D k \int_0^1 \int_0^1 \int_0^1 \int_0^1 dxdydrds \delta(x+y+r+s-1) \frac{(k^0)^2 xy}{[(k-(y+s)p)^2 - (r+s)m_f^2 ]^6} \,.
\end{equation}
Note that again we have $l = k-(y+s)p$,\,$\Delta = (r+s)m_f^2$. And we have $k^0 =l^0+(y+s)p^0$, therefore the integral becomes,
\begin{equation}
\begin{aligned}
I^{(1)}_2 &=-240Dp^2 \int d^D l \int_0^1 \int_0^1 \int_0^1 \int_0^1 dxdydrds \delta(x+y+r+s-1) \frac{(k^0)^2 xy}{(l^2 - \Delta )^6} \\
&= -240Dp^2 \int d^D l \int_0^1 \int_0^1 \int_0^1 \int_0^1 dxdydrds \delta(x+y+r+s-1)\frac{xy (l^0 + (y+s)p)^2  }{(l^2 -\Delta)^6} \\
&= -240Dp^2 \int d^D l \int_0^1 \int_0^1 \int_0^1 \int_0^1 dxdydrds \delta(x+y+r+s-1)\\
&\quad\quad\quad\quad\quad\quad\quad\quad\quad\quad\quad\quad\quad\quad\quad\quad\quad\quad\quad\quad\times\frac{xy ( (l^0)^2 + 2l^0 (y+s)p +(y+s)^2 p^2  )}{(l^2 -\Delta)^6} \,.
\end{aligned}
\end{equation}
Next we carry out Wick's rotation and perform the substitution of $l^0 = il_E^0$,
\begin{equation}
\begin{aligned} \label{eq:234}
I^{(1)}_2 &= -(-1)^6 240iDp^2 \int_0^\infty dl_E \int d\Omega_D \int_0^1 \int_0^1 \int_0^1 \int_0^1 dxdydrds\delta(x+y+r+s-1) \\
&\quad\quad\quad\quad\quad\quad\quad\quad\quad\quad\quad\quad\quad\quad\quad\quad\quad\times\frac{xy[l_E^{D-1}\big(-(l_E^0)^2 + 2il_E^0 (y+s)p + (y+s)^2 p^2)]}{(l_E^2 + \Delta)^6} \,.
\end{aligned}
\end{equation}
Now this integral involve the integrating on $l_E^0$. To tackle this integral, we concern integration on $D-$sphere. We set the spherical coordinates $(l_E^0,l_E^1, \cdots l_E^{D-1} )$ in the Euclidean $l$ momentum space as follows:
\begin{equation}
\begin{cases}
l_E^0 &=l_E \sin\varphi_1 \sin\varphi_2 \cdots\sin\varphi_{D-2} \cos\varphi_{D-1} \\ 
l_E^1 &= l_E \sin\varphi_1 \sin\varphi_2 \cdots\sin\varphi_{D-2} \sin\varphi_{D-1} \\
\vdots & \vdots \\
l_E^{n-3} &= l_E \sin\varphi_1 \sin\varphi_2 \cos \phi_3 \\
l_E^{n-2} &= l_E \sin\varphi_1 \cos \varphi_2\\
l_E^{n-1} &= l_E  \cos \varphi_1 \,, \\
\end{cases}
\end{equation}
where
\begin{equation}
l_E^2 = (l_E^0)^2+(l_E^1)^2 + \cdots + (l_E^{D-1})^2 \,.
\end{equation}
The differential volume is given by
\begin{equation}
d^D l_E = l_E^{D-1} \sin^{D-2} \varphi_1 \sin^{D-3} \varphi_2 \cdots  \sin \varphi_{D-2} dl_E d\varphi_1 d\varphi_2 \cdots d\varphi_{D-1} \,.
\end{equation}
And the differential solid angle is given by
\begin{equation}
d\Omega_D = \sin^{D-2} \varphi_1 \sin^{D-3} \varphi_2 \cdots  \sin \varphi_{D-2}  d\varphi_1 d\varphi_2 \cdots d\varphi_{D-1} \,,
\end{equation}
where $\varphi_1 ,\varphi_2 ,\cdots , \varphi_{D-2} \in [0,\pi] $ and $\varphi_{D-1} \in [0,2\pi]$.
We first carry out the integral for the first term involving $(l_E^0)^2$, we have
\begin{equation}
\begin{aligned}
I^{(1)}_{2a} & =
240iDp^2\int_0^\infty dl_E \int d\Omega_D \int_0^1 \int_0^1 \int_0^1 \int_0^1 dxdydrds \delta(x+y+r+s-1)\frac{xyl_E^{D-1}(l_E^0)^2}{(l_E^2 + \Delta )^6} \\
&=240iDp^2\int_0^\infty dl_E \int\cdots\int d\varphi_1 d\varphi_2 \cdots d\varphi_{D-1} \sin^{D-2} \varphi_1 \sin^{D-3} \varphi_2 \cdots  \sin \varphi_{D-2}   \\
&\quad\times\int_0^1 \int_0^1 \int_0^1 \int_0^1 dxdydrds \delta(x+y+r+s-1) \frac{xyl_E^{D-1}(l_E \sin\varphi_1 \sin\varphi_2 \cdots\sin\varphi_{D-2} \cos\varphi_{D-1})^2  }{(l_E^2 + \Delta)^6} \\
&=240iDp^2 \int_0^\pi \sin^D \varphi_1 d\varphi_1 \int_0^\pi\sin^{D-1} \varphi_2 d\varphi_2 \int_0^\pi \sin^{D-2} \varphi_3 d\varphi_3 \cdots \int_{0}^{\pi}\sin^{3} \varphi_{D-2} \int_{0}^{2\pi}\cos^2 \varphi_{D-1} d\varphi_{D-1}\\
&\quad\times \int_0^1 \int_0^1 \int_0^1 \int_0^1 dxdydrds \delta(x+y+r+s-1)\int_0^\infty dl_E \frac{xy l_E^{D+1}}{(l_E^2 + \Delta)^6} \,.
\end{aligned}
\end{equation}
In general,
\begin{equation}
\int_0^\pi \sin^{m}x\, dx=\frac{\sqrt{\pi}\Gamma\big( \frac{m+1}{2}\big) }{\Gamma\big(\frac{m}{2}+1 \big)} \quad \text{if}\,\, m > -1.
\end{equation}
Also,
\begin{equation}
\int_0^\infty dl_E \frac{l_E^{D+1}}{(l_E^2 +\Delta)^6} = \frac{\pi (D-8)(D-6)(D-4)(D-2)D\csc \big(\frac{\pi D}{2} \big)}{7680\Delta^{5-\frac{D}{2}}} \,.
\end{equation}
Then we get
\begin{equation}
\begin{aligned}
I^{(1)}_{2a} & = \frac{iDp^2 \pi^2}{32 m_f^{10-D}} \bigg(\prod_{k=1}^{D-2}\int_0^\pi \sin^{D-k+1}\varphi_k d\varphi_k \bigg) \int_0^1 \int_0^1 \int_0^1 \int_0^1 dxdydrds \delta(x+y+r+s-1) \frac{xy}{(r+s)^{5-\frac{D}{2}}} \\
&\quad\quad\quad\quad\quad\quad\quad\quad\quad\quad\quad\quad\quad\quad\quad\quad\times(D-8)(D-6)(D-4)(D-2)D\csc \bigg(\frac{\pi D}{2}\bigg)\\
&=\frac{D^2 (D-8)(D-6)(D-4)(D-2) p^2 \pi^{D/2 + 1}}{32 \sin\big(\frac{\pi D}{2}\big)m_f^{10-D}} \bigg(\prod_{k=1}^{D-2} \frac{\Gamma\big( \frac{D-k+2}{2}\big)}{\Gamma\big( \frac{D-k+3}{2}\big)} \bigg)\\
&\quad\quad\quad\quad\quad\quad\quad\quad\quad\quad\quad\quad\quad\quad\quad\quad\times\int_0^{1}dr\int_0^{1-r}ds\int_0^{1-r-s}dy \frac{(1-y-r-s) y}{(r+s)^{5-\frac{D}{2}}} \\
&=\frac{D^2 (D-8)(D-6)(D-4)(D-2) p^2 \pi^{D/2 + 1}}{32 \sin\big(\frac{\pi D}{2}\big)m_f^{10-D}} \frac{\Gamma(2)}{\Gamma\big(\frac{D+2}{2}  \big)} \frac{1}{6}\int_0^{1}dr\int_0^{1-r}ds\frac{(1-r-s)^3}{(r+s)^{5-\frac{D}{2}}}
\end{aligned}
\end{equation}
if $-2<D <10$. Out of this range the integral will diverge. Now we calculate the last integral
\begin{equation}
\begin{aligned}
\int_0^{1-r} ds \frac{(1-r-s)^3}{(r+s)^{5-\frac{D}{2}}} &= 2\bigg(\frac{r^3}{D-2} - \frac{3r^2}{D-4}+\frac{3r}{D-6}+ \frac{1}{8-D}\bigg)r^{\frac{D}{2}-4}\\
&\quad\quad\quad\quad + \frac{96}{(D-8)(D-6)(D-4)(D-2)} \,.
\end{aligned}
\end{equation}
And finally 
\begin{equation}
\int_0^{1}dr\int_0^{1-r}ds\frac{(1-r-s)^3}{(r+s)^{5-\frac{D}{2}}} = \frac{96}{(D-6)(D-4)(D-2)D}
\end{equation}
if $D>6$. It is noted that if $D \leq 6,$ the integral does not converge. Therefore we get
\begin{equation}
I_{2a}^{(1)}= \frac{4ip^2 \pi^{D/2+1} D(D-8)}{ \sin\big(\frac{\pi D}{2} \big)\Gamma\big(\frac{D+2}{2}  \big) m_f^{10-D}}
\end{equation}
in the overall range of $ 6 < D < 10$.

The next term to integrate will be 
\begin{equation}
I_{2b}^{(1)} = (-2i)(240)Dp^2 \int_0^\infty dl_E \int d\Omega_D \int_0^1 \int_0^1\int_0^1\int_0^1 dxdydrds\delta(x+y+r+s-1)\frac{l_E^0 (y+s)}{(l_E^2 + \Delta)^6} \,.
\end{equation}
Since $l_E^0 =l_E \sin\varphi_1 \sin\varphi_2 \cdots\sin\varphi_{D-2} \cos\varphi_{D-1}$, when we integrate over $d\Omega_D$ where will be a vanishing term of
\begin{equation}
\int_0^{2\pi} \cos \varphi_{D-1} d\varphi_{D-1} =0 \,.
\end{equation}
Therefore the whole integral vanishes. In general, since if the integral is odd
\begin{equation}
\int\frac{d^D l}{(2\pi)^D} l^m (l^0 ) =0
\end{equation}
for any integer $m$. Thus when we have odd order of $k^0$ in the integral, the integral must vanish. Next we have the $I_{2c}^{(1)}$ integral, which is
\begin{equation}
\begin{aligned}
I_{2c}^{(1)}& =-240iDp^4 \int_0^\infty dl_E \int d\Omega_D \int_0^1 \int_0^1 \int_0^1 \int_0^1 dxdydrds\delta(x+y+r+s-1) \frac{l_E^{D-1} xy(y+s)^2}{(l_E^2 + \Delta)^6} \\
&=-240iDp^2\int_0^\infty dl_E \int\cdots\int d\varphi_1 d\varphi_2 \cdots d\varphi_{D-1} \sin^{D-2} \varphi_1 \sin^{D-3} \varphi_2 \cdots  \sin \varphi_{D-2} \\
&\quad\times\int_0^1 \int_0^1 \int_0^1 \int_0^1 dxdydrds \delta(x+y+r+s-1) \frac{xy (y+s)^2 l_E^{D-1} }{(l_E^2 + \Delta)^6} \,.
\end{aligned}
\end{equation}
And as
\begin{equation}
\int_0^\infty dl_E \frac{l_E^{D-1}}{(l_E^2 +\Delta)^6} = -\frac{\pi (D-10)(D-8)(D-6)(D-4)(D-2)\csc \big(\frac{\pi D}{2} \big)}{7680\Delta^{6-\frac{D}{2}}}
\end{equation}
if $0<D<12$. Then
\begin{equation}
\begin{aligned}
I_{2c}^{(1)} &=\frac{ip^4 (D-10)(D-8)(D-6)(D-4)(D-2)D\pi^{D/2+1}}{32 \sin\big(\frac{\pi D}{2}\big) m_f^{12-D} } \bigg( \prod_{k=1}^{D-2} \frac{\Gamma\big(\frac{D-k}{2} \big)}{\Gamma\big(\frac{D-k+1}{2} \big)} \bigg)\\
&\quad\quad\quad\quad\quad\quad\quad\quad\quad\quad\quad\quad\times\int_0^{1}dr\int_0^{1-r}ds\int_0^{1-r-s}dy \frac{(1-y-r-s) y(y+s)^2}{(r+s)^{6-\frac{D}{2}}} \\
&=\frac{ip^4 (D-10)(D-8)(D-6)(D-4)(D-2)D \cdot 2\pi^{D/2+1}}{32 \sin\big(\frac{\pi D}{2}\big) m_f^{12-D} } \frac{\Gamma(1)}{\Gamma\big(\frac{D}{2} \big)}\\
&\quad\quad\quad\quad\quad\quad\quad\quad\quad\times\frac{1}{60}\int_0^{1}dr\int_0^{1-r}ds\frac{(1-r-s)^3 (3(r-1)^2 -4(r-1)s + 3s^2 )}{(r+s)^{6-\frac{D}{2}}}\\
&=\frac{ip^4 \pi^{\frac{D}{2}+1} (D-10)(D^2 -2D + 24)}{6(D+2)\sin\big(\frac{\pi D}{2}\big) \Gamma\big(\frac{D}{2} \big) m_f^{12-D}}
\end{aligned}
\end{equation}
if $D>8$. When $D<8$ the integral does not converge. So the $I_{2c}^{(1)}$ integral is converged in $8< D <12$. Therefore, for the $I^{(1)}$ integral to be convergent, the spacetime dimension constraint is $6<D<10$ and $8<D<12$. The solution is $D=9$ which is unique for the convergence. 

The third integral is defined to be
\begin{equation}
I^{(1)}_3 = 8Dp^0 \int d^D k \frac{(k^0) k^2}{k^4 (p-k)^4 (k^2 -m_f^2)((p-k)^2 -m_f^2)} =0 
\end{equation}
as the integral is odd.

The fourth integral is defined to be
\begin{equation}
I_4^{(1)} = -2D(p^0)^2 \int d^D k \frac{ k^2}{k^4 (p-k)^4 (k^2 -m_f^2)((p-k)^2 -m_f^2)} =-2D(p^0)^2 I(1,2)
\end{equation}
which only converges when $6<D<10$. 

The fifth integral is defined to be
\begin{equation}
I_5^{(1)} = Dp^2 \int d^D k \frac{ k^2}{k^4 (p-k)^4 (k^2 -m_f^2)((p-k)^2 -m_f^2)} =Dp^2 I(1,2)
\end{equation}
which only converges when $6<D<10$.

The sixth integral is defined to be
\begin{equation}
\begin{aligned}
I_6^{(1)} &=8D \int d^D k \frac{ (k^0)^2 (k\cdot p) }{k^4 (p-k)^4 (k^2 -m_f^2)((p-k)^2 -m_f^2)} \\
&=4D\int d^D k  \frac{(k^0)^2[(p^2 -k^2) - (p-k)^2]}{k^4 (p-k)^4 (k^2 -m_f^2)((p-k)^2 -m_f^2)} \,.
\end{aligned}
\end{equation}
Then we further define three more integrals,
\begin{equation}
I^{(1)}_{6a} = 4Dp^2 \int d^D k \frac{(k^0)^2}{k^4 (p-k)^4 (k^2 -m_f^2)((p-k)^2 -m_f^2)}
\end{equation}
which takes the same form as $I_{2a}^{(1)}$ except the front constant. We have $I^{(1)}_{6a} = -2I_{2a}^{(1)}$, which is defined in the overall range of $ 6 < D < 10$. Next we have,
\begin{equation}
I^{(1)}_{6b} = -4D\int d^D k\frac{(k^0)^2}{k^2 (p-k)^4 (k^2 -m_f^2)((p-k)^2 -m_f^2)} \,.
\end{equation}
Using the similar technique as above, this amounts to give
\begin{equation}
\begin{aligned} 
I^{(1)}_{6b} &= -(-1)^5 96iD \int_0^\infty dl_E \int d\Omega_D \int_0^1 \int_0^1 \int_0^1 \int_0^1 dxdydrds\delta(x+y+r+s-1) \\
&\quad\quad\quad\quad\quad\quad\quad\quad\quad\quad\quad\quad\quad\quad\quad\quad\quad\times\frac{y[l_E^{D-1}\big(-(l_E^0)^2 + 2il_E^0 (y+s)p + (y+s)^2 p^2)]}{(l_E^2 + \Delta)^5}
\\
&=\frac{i\pi^{D/2 +1} D(D-6)}{8\sin\big(\frac{\pi D}{2}\big)\Gamma\big(\frac{D+2}{2} \big) m_f^{8-D}} +\frac{2i\pi^{D/2+1} p^2 (D-8)(D^2+2D+24) }{3(D+2)\sin\big(\frac{\pi D}{2}\big) \Gamma\big(\frac{D}{2} \big) m_f^{10-D}}  
\end{aligned}
\end{equation}
for $6< D < 8$. Hence only when $D=7$ this integral is finite.

Finally, we have the integral of $I^{(1)}_{6c}$,
\begin{equation}
I^{(1)}_{6c} = -4D \int d^D k \frac{(k^0)^2}{k^4 (p-k)^2 (k^2 -m_f^2)((p-k)^2 - m_f^2  )} \,.
\end{equation}
Using the similar technique as above, this amounts to give,
\begin{equation}
\begin{aligned}
I^{(1)}_{6c} &= -(-1)^5 24iD\int_0^\infty dl_E \int d\Omega_D \int_0^1 \int_0^1 \int_0^1 \int_0^1 dxdydrds\delta(x+y+r+s-1) \\
&\quad\quad\quad\quad\quad\quad\quad\quad\quad\quad\quad\quad\quad\quad\quad\quad\quad\times\frac{x[l_E^{D-1}\big(-(l_E^0)^2 + 2il_E^0 (y+s)p + (y+s)^2 p^2)]}{(l_E^2 + \Delta)^5} \\
&= \frac{2i\pi^{D/2+1}(D-8)(D-6)}{\sin\big( \frac{\pi D}{2}\big)\Gamma\big(\frac{D+2}{2} \big)m_f^{8-D}}   -\frac{2i\pi^{D/2+1}(D-6)(D-8)(D^2+8)}{(D+2)(D+4) \sin\big( \frac{\pi D}{2}\big) \Gamma\big(\frac{D}{2}\big)  m_f^{8-D}}
\end{aligned}
\end{equation}
if $4<D<8$.

The seventh integral vanishes as it is an odd integral,
\begin{equation}
I_7^{(1)} = -4D p^0 \int d^D k \frac{k^0(k\cdot p)}{k^4 (p-k)^4 (k^2 -m_f^2)((p-k)^2 -m_f^2)} =0 \,.
\end{equation}
The eighth integral is defined to be
\begin{equation}
\begin{aligned}
I_8^{(1)} &= -2D \int d^D k \frac{k^2 (k\cdot p)}{k^4 (p-k)^4 (k^2 -m_f^2)((p-k)^2 -m_f^2)} \\
&=-D\int d^D k \frac{(p^2 - k^2)-(p-k)^2}{k^2 (p-k)^4 (k^2 -m_f^2)((p-k)^2 -m_f^2)} \\
&=-D(p^2 I (1,2 ,D) - I(0,2,D) - I(1,1 ,D)        ) \,.
\end{aligned}
\end{equation}
The range for each integral to be finite is: for $I(1,2,D)$, $6<D<10$. For $I(0,2,D)$, $4<D<8$. For $I(1,1,D)$, $4<D<8$. Therefore the overall range is $6<D<8$, which is $D=7$.

The ninth integral vanishes as it is an odd integral,
\begin{equation}
I_9^{(1)} = -8D \int d^D k \frac{(k^0)k^2}{k^4 (p-k)^4 (k^2 -m_f^2)((p-k)^2 -m_f^2)} =0 \,.
\end{equation}
The tenth integral is defined to be
\begin{equation}
I_{10}^{(1)} = D\int d^Dk \frac{k^4}{k^4 (p-k)^4 (k^2 -m_f^2)((p-k)^2 -m_f^2)} = DI(0,2,D)\,,
\end{equation}
which is in the range of $4<D<8$.

The eleventh integral vanishes as it is an odd integral,
\begin{equation}
I_{11}^{(1)} = -16D p^0 \int d^D k \frac{(k^0)^3}{k^4 (p-k)^4 (k^2 -m_f^2)((p-k)^2 -m_f^2)} =0 \,.
\end{equation}

The twelfth integral is defined to be
\begin{equation}
I_{12}^{(1)} = 8D (p^0)^2 \int d^D k \frac{(k^0)^2}{k^4 (p-k)^4 (k^2 -m_f^2)((p-k)^2 -m_f^2)} =-\frac{4I_2^{(1)} (p^0)^2}{p^2}
\end{equation}
with $D=9$.

The thirteenth integral is defined to be
\begin{equation}
\begin{aligned}
I_{13}^{(1)} &= 8D \int d^D k \frac{(k^0)^4}{k^4 (p-k)^4 (k^2 -m_f^2)((p-k)^2 -m_f^2)}\\
&= 960D\int d^D l \int_0^1 \int_0^1 \int_0^1 \int_0^1 dxdydrds\delta(x+y+r+s-1)\frac{xy((l^0)+(y+s)p)^4 }{(l^2 - \Delta)^6}\\
&= 960D\int d^D l \int_0^1 \int_0^1 \int_0^1 \int_0^1 dxdydrds\delta(x+y+r+s-1) \\
&\quad\quad\quad
\times \frac{xy((l^0)^4 +4(l^0)^3 (y+s)p + 6(l^0)^2 (y+s)^2 p^2 + 6(l^0)(y+s)^3 p^3 + (y+s)^4 p^4 )}{(l^2 -\Delta)^6} \,.
\end{aligned}
\end{equation}
The integrals involving $(l^0)^3$ and $(l^0)$ vanish as they are odd integrals. We subdivide $I_{13}^{(1)}$ into three more integrals.
\begin{equation}
\begin{aligned}
I_{13a}^{(1)} &=960D\int d^D l \int_0^1 \int_0^1 \int_0^1 \int_0^1 dxdydrds \delta(x+y+r+s-1) \frac{xy (l^0)^4}{(l^2 - \Delta)^6}\\
&=960iD\int_0^\infty dl_E \int\cdots\int d\varphi_1 d\varphi_2\cdots d\varphi_{D-1}\sin^{D-2}\varphi_1 \sin^{D-3}\varphi_2\cdots \sin \varphi_{D-2} \\
&\quad \times\int_0^1 \int_0^1 \int_0^1 \int_0^1 dxdydrds \delta(x+y+r+s-1) \frac{xyl_E^{D-1}(l_E \sin\varphi_1 \sin\varphi_2 \cdots\sin\varphi_{D-2} \cos\varphi_{D-1})^4}{(l_E^2 + \Delta)^6} \\
&=960iD \int_0^\pi \sin^{D+2} \varphi_1 d\varphi_1 \int_0^\pi\sin^{D+1} \varphi_2 d\varphi_2 \int_0^\pi \sin^{D} \varphi_3 d\varphi_3 \cdots \int_{0}^{\pi}\sin^{5} \varphi_{D-2} \int_{0}^{2\pi}\cos^4 \varphi_{D-1} d\varphi_{D-1}\\
&\quad\times \int_0^1 \int_0^1 \int_0^1 \int_0^1 dxdydrds \delta(x+y+r+s-1)\int_0^\infty dl_E \frac{xy l_E^{D+3}}{(l_E^2 + \Delta)^6} \\
&-\frac{3i(D-6)(D-4)(D-2)D^2 (D+2)\pi^{D/2+1}}{32 \sin\big(\frac{\pi D}{2} \big) m_f^{8-D}}\bigg(\prod_{k=1}^{D-2}\frac{\Gamma\big( \frac{D-k+4}{2}\big)}{\Gamma\big( \frac{D-k+5}{2} \big)}  \bigg) \times \\
&\quad\quad\quad\quad\quad\quad\quad\quad\quad\quad\quad\quad\quad\quad \int_0^1 dr \int_{0}^{1-r}ds \int_{0}^{1-r-s}dy \frac{(1-y-r-s)y}{(r+s)^{4-\frac{D}{2}}} \\
&=-\frac{3i\pi^{D/2 +1}D(D-6)}{\sin\big(\frac{\pi D}{2}\big)\Gamma\big( \frac{D+4}{2}  \big) m_f^{8-D}}
\end{aligned}
\end{equation}
for $4<D<8$. Next, we have
\begin{equation}
\begin{aligned}
I_{13b}^{(1)} &= 5760iDp^2 \int_0^\pi \sin^D \varphi_1 \int_0^\pi \sin^{D-1} \varphi_2 \cdots \int_0^\pi \sin^3 \varphi_{D-2}        d\varphi_{D-2} \int_0^{2\pi} \cos\varphi_{D-1} d \varphi_{D-1}\\
&\quad\times \int_0^1 \int_0^1 \int_0^1 \int_0^1 dxdydrds \delta(x+y+r+s-1)\int_0^\infty dl_E \frac{xy(y+s)^2 l_E^{D+1}}{(l_E^2 + \Delta)^6} \\
&=\frac{72 i \pi^{D/2+1} D(D-8)p^2}{\sin\big(\frac{\pi D}{2}\big)\Gamma\big( \frac{D+2}{2}\big) m_f^{10-D}  }
\end{aligned}
\end{equation}
for $6<D<10$. 

Finally, we have
\begin{equation}
\begin{aligned}
I_{13c}^{(1)} &= 960iDp^4 \int\cdots\int d\varphi_1 d\varphi_2 \cdots d\varphi_{D-1} \sin^{D-2} \varphi_1 \sin^{D-3} \varphi_2 \cdots  \sin \varphi_{D-2}\\
&\times \int_0^1 \int_0^1 \int_0^1 \int_0^1 dxdydrds\delta(x+y+r+s-1) \int dl_E \frac{xy(y+s)^4 l_E^{D-1}}{(l^2_E +\Delta)^6}\\
&=-\frac{4i\pi^{D/2+1}p^4(D-10)(D^4+20D^2+ 240D+2304)}{5(D+2)(D+4)(D+6)\sin\big(\frac{\pi D}{2}\big)\Gamma\big( \frac{D}{2}\big) m_f^{12-D} } \,.
\end{aligned}
\end{equation}
for $8<D<12$. 

Therefore, we have completed all the calculations of the integrals. The total amplitude would be \begin{equation}
\mathcal{M}^{(n)} = -\frac{m_f^2}{v_n^2 (2\pi)^D} \sum \mathrm{Integrals}\,.
\end{equation} 

Since from the above calculations, each integral corresponds to a specific range of $D$ or specific value of $D$ such that the integral converges, we see that there is no common $D$ for all the integrals to be convergent. The rotor mechanism can only save the divergence for each term of the calculation with higher dimension $D$ one by one in the $W$-boson self-energy correction diagram.

\section{Discussion on non-Unitarity of Quantum Field Theory under Rotor Mechanism }
In normal quantum field theory, the S-matrix satisfies unitarity, subsequently this leads to the optical theorem. However, high-order derivative quantum field theory is infamous for dynamic instability, where the Hamiltonian is unbounded below \cite{d1,d2,d3,d4,BW2}. This causes the non-unitarity of S-matrix of the theory. The non-unitarity issue also appears in quantizing high-order derivative gravitational theories. 
High-order derivative field theory has a great advantage of renormalizability, but so often if suffers from the unitary problem. The quantum field theory of rotor mechanism is not exceptional too.

Under rotor mechanism, the S-matrix is given by the time-ordered exponential of interaction Hamiltonian of the $n$-th rotored fields in the interaction picture,
\begin{equation}
S = \mathrm{T}\exp\bigg(i\int d^{D}x \hat{H}_{\mathrm{int}}[\Box^n \text{fields}(x)] \bigg) \,,
\end{equation}
where $\mathrm{T}$ is the time-ordered operator. 
The generic quantum amplitude is given by
\begin{equation}
\mathcal{M} = \langle f |S | i\rangle = \langle \pmb{\mathrm{p}}_1 \pmb{\mathrm{p}}_2 \cdots \pmb{\mathrm{p}}_m | S | \pmb{\mathrm{k}}_1 \pmb{\mathrm{k}}_2 \cdots \pmb{\mathrm{k}}_m \rangle \,,
\end{equation}
where $|i\rangle =| \pmb{\mathrm{k}}_1 \pmb{\mathrm{k}}_2 \cdots \pmb{\mathrm{k}}_m \rangle  $ is the initial incoming states and $|f\rangle =| \pmb{\mathrm{p}}_1 \pmb{\mathrm{p}}_2 \cdots \pmb{\mathrm{p}}_m \rangle$ is the final outgoing states. Explicitly, these states can be expressed as the excitation of the vacuum, giving
\begin{equation}
| \pmb{\mathrm{k}}_1 \pmb{\mathrm{k}}_2 \cdots \pmb{\mathrm{k}}_m  \rangle = \prod_{i=1}^m \sqrt{2E_{i}}\, \hat{a}_{\pmb{\mathrm{k}}_i}^\dagger |0\rangle \quad\text{and}\quad | \pmb{\mathrm{p}}_1 \pmb{\mathrm{p}}_2 \cdots \pmb{\mathrm{p}}_m\rangle = \prod_{j=1}^m \sqrt{2E_{j}}\, \hat{a}_{\pmb{\mathrm{p}}_j}^\dagger |0\rangle \,.
\end{equation} 
The normalization is given by
\begin{equation}
\langle \pmb{\mathrm{k}}_i | \pmb{\mathrm{k}}_j \rangle = 2E_i (2\pi)^{D-1} \delta^{D-1} (\pmb{\mathrm{k}}_i - \pmb{\mathrm{k}}_j) \,.
\end{equation}
Now the quantum amplitude reads
\begin{equation}
\mathcal{M} = \prod_{i=1}^m 2E_i (2\pi)^{D-1} \delta^{D-1} (\pmb{\mathrm{p}}_i - \pmb{\mathrm{k}}_i) \, + \, \langle \pmb{\mathrm{p}}_1 \pmb{\mathrm{p}}_2 \cdots \pmb{\mathrm{p}}_m | \sum_{n=1}^\infty \frac{i^n}{n!} \bigg(\int d^{D}x \hat{H}_{\mathrm{int}}[\Box^n \text{fields}(x)] \bigg)^n | \pmb{\mathrm{k}}_1 \pmb{\mathrm{k}}_2 \cdots \pmb{\mathrm{k}}_m \rangle \,,
\end{equation}
The S matrix can be written as 
\begin{equation}
S = 1 + iT \,, 
\end{equation}
where $T$ is the transfer matrix.
If the unitarity condition of the S-matrix is satisfied, it demands
\begin{equation}
S^\dagger S  =1 \,,
\end{equation}
giving
\begin{equation} \label{eq:59}
-i(T-T^\dagger) = T^\dagger T \,.
\end{equation}
Then, this would follow
\begin{equation} \label{eq:60}
\begin{aligned}
&\quad  \langle \pmb{\mathrm{p}}_1 \pmb{\mathrm{p}}_2 \cdots \pmb{\mathrm{p}}_m | T^\dagger T | \pmb{\mathrm{k}}_1 \pmb{\mathrm{k}}_2 \cdots \pmb{\mathrm{k}}_m \rangle \\
&= \sum_{n}\prod_{i=1}^n \int \frac{d^{D-1} \pmb{\mathrm{q}}_i}{(2\pi)^{D-1}} \frac{1}{2E_i} \langle \pmb{\mathrm{p}}_1 \pmb{\mathrm{p}}_2 \cdots \pmb{\mathrm{p}}_m |T^\dagger | \pmb{\mathrm{q}}_1 \pmb{\mathrm{q}}_2 \cdots \pmb{\mathrm{q}}_n  \rangle \langle \pmb{\mathrm{q}}_1 \pmb{\mathrm{q}}_2 \cdots \pmb{\mathrm{q}}_n | T | \pmb{\mathrm{k}}_1 \pmb{\mathrm{k}}_2 \cdots \pmb{\mathrm{k}}_m \rangle  \,,
\end{aligned}
\end{equation}
where we have imposed the completeness relation of
\begin{equation} \label{eq:complete}
\int \frac{d^{D-1} \pmb{\mathrm{q}}_i}{(2\pi)^{D-1}} \frac{1}{2E_i}| \pmb{\mathrm{q}}_1 \pmb{\mathrm{q}}_2 \cdots \pmb{\mathrm{q}}_n  \rangle \langle \pmb{\mathrm{q}}_1 \pmb{\mathrm{q}}_2 \cdots \pmb{\mathrm{q}}_n | = \pmb{1} \,.
\end{equation}
As 
\begin{equation}
\langle \pmb{\mathrm{p}}_1 \pmb{\mathrm{p}}_2 \cdots \pmb{\mathrm{p}}_m |T^\dagger | \pmb{\mathrm{q}}_1 \pmb{\mathrm{q}}_2 \cdots \pmb{\mathrm{q}}_n  \rangle = (2\pi)^D \delta^D \bigg(\sum_{i=1}^m p_i - \sum_{i=1}^n q_i \bigg) \mathcal{M}^\dagger(
\pmb{\mathrm{p}}_1 \pmb{\mathrm{p}}_2 \cdots \pmb{\mathrm{p}}_m \rightarrow    \pmb{\mathrm{q}}_1 \pmb{\mathrm{q}}_2 \cdots \pmb{\mathrm{q}}_n ) \,.
\end{equation}
Thus (\ref{eq:60}) is evaluated to be
\begin{equation}
\begin{aligned}
&\quad\langle \pmb{\mathrm{p}}_1 \pmb{\mathrm{p}}_2 \cdots \pmb{\mathrm{p}}_m | T^\dagger T | \pmb{\mathrm{k}}_1 \pmb{\mathrm{k}}_2 \cdots \pmb{\mathrm{k}}_m \rangle  \\
&=  \sum_{n}\prod_{i=1}^n \int \frac{d^{D-1} \pmb{\mathrm{q}}_i}{(2\pi)^{D-1}} \frac{1}{2E_i} (2\pi)^{2D} \delta^D \bigg(\sum_{i=1}^m p_i - \sum_{i=1}^m q_i \bigg)\\
&\quad \times \mathcal{M}^\dagger( \pmb{\mathrm{p}}_1\pmb{\mathrm{p}}_2 \cdots \pmb{\mathrm{p}}_m \rightarrow    \pmb{\mathrm{q}}_1 \pmb{\mathrm{q}}_2 \cdots \pmb{\mathrm{q}}_n )\mathcal{M}(
\pmb{\mathrm{k}}_1 \pmb{\mathrm{k}}_2 \cdots \pmb{\mathrm{k}}_m \rightarrow    \pmb{\mathrm{q}}_1 \pmb{\mathrm{q}}_2 \cdots \pmb{\mathrm{q}}_n ) \,.
\end{aligned}
\end{equation}
Using (\ref{eq:59}), take the bra states of $\langle \pmb{\mathrm{p}}_1 \pmb{\mathrm{p}}_2 \cdots \pmb{\mathrm{p}}_m |$ and ket states of $|\pmb{\mathrm{k}}_1 \pmb{\mathrm{k}}_2 \cdots \pmb{\mathrm{k}}_m \rangle$ it follows that
\begin{equation}
\begin{aligned}
&\quad -i\big(\mathcal{M}(
\pmb{\mathrm{k}}_1 \pmb{\mathrm{k}}_2 \cdots \pmb{\mathrm{k}}_m \rightarrow    \pmb{\mathrm{p}}_1 \pmb{\mathrm{p}}_2 \cdots \pmb{\mathrm{p}}_n ) -\mathcal{M}^\dagger(
\pmb{\mathrm{k}}_1 \pmb{\mathrm{k}}_2 \cdots \pmb{\mathrm{k}}_m \rightarrow    \pmb{\mathrm{p}}_1 \pmb{\mathrm{p}}_2 \cdots \pmb{\mathrm{p}}_n )  \big) \\
&=  \sum_{n}\prod_{i=1}^n \int \frac{d^{D-1} \pmb{\mathrm{q}}_i}{(2\pi)^{D-1}} \frac{1}{2E_i} (2\pi)^{D}   \mathcal{M}^\dagger( \pmb{\mathrm{p}}_1\pmb{\mathrm{p}}_2 \cdots \pmb{\mathrm{p}}_m \rightarrow    \pmb{\mathrm{q}}_1 \pmb{\mathrm{q}}_2 \cdots \pmb{\mathrm{q}}_n )\mathcal{M}(
\pmb{\mathrm{k}}_1 \pmb{\mathrm{k}}_2 \cdots \pmb{\mathrm{k}}_m \rightarrow    \pmb{\mathrm{q}}_1 \pmb{\mathrm{q}}_2 \cdots \pmb{\mathrm{q}}_n ) \,.
\end{aligned}
\end{equation}
Therefore, the optical theorem is obtained if unitary is satisfied
\begin{equation} \label{eq:optical}
\begin{aligned}
&\quad 2\mathrm{Im}\mathcal{M}(
\pmb{\mathrm{k}}_1 \pmb{\mathrm{k}}_2 \cdots \pmb{\mathrm{k}}_m \rightarrow    \pmb{\mathrm{p}}_1 \pmb{\mathrm{p}}_2 \cdots \pmb{\mathrm{p}}_n )\\
 &=\sum_{n}\prod_{i=1}^n \int \frac{d^{D-1} \pmb{\mathrm{q}}_i}{(2\pi)^{D-1}} \frac{1}{2E_i} (2\pi)^{D}  \mathcal{M}^\dagger( \pmb{\mathrm{p}}_1\pmb{\mathrm{p}}_2 \cdots \pmb{\mathrm{p}}_m \rightarrow    \pmb{\mathrm{q}}_1 \pmb{\mathrm{q}}_2 \cdots \pmb{\mathrm{q}}_n )\mathcal{M}(
\pmb{\mathrm{k}}_1 \pmb{\mathrm{k}}_2 \cdots \pmb{\mathrm{k}}_m \rightarrow    \pmb{\mathrm{q}}_1 \pmb{\mathrm{q}}_2 \cdots \pmb{\mathrm{q}}_n ) \,.
\end{aligned}
\end{equation}
However, in the following, we will show that in fact the quantum field theory under rotor mechanism resonates a similar problem with higher-derivative quantum gravity. For example, in Stelle theory, where second-order derivative is concerned, the action reads \cite{d5}
\begin{equation}
S_{\mathrm{QG}} = \int d^D x \sqrt{-g} \bigg( 2\Lambda_C + \zeta R + \alpha\bigg(R_{\mu\nu}R^{\mu\nu} -\frac{1}{3}R^2 \bigg) -\frac{\xi}{6}R^2 \bigg) \,,
\end{equation}
where $R_{\mu\nu}$ is the Ricci tensor, $R = g^{\mu\nu}R_{\mu\nu}$ is the Ricci scalar and $\Lambda_C$ is the cosmological constant. Upon quantization, this action will give Feynman propagators in form of 
\begin{equation}
\frac{1}{p^2 - m_1^2}\frac{1}{p^2 - m_2^2} = \frac{1}{m_1^2 - m_2^2}\bigg(\frac{1}{p^2 -m_1^2} -\frac{1}{p^2 -m_2^2}  \bigg) \,.
\end{equation}
The dangerous minus sign in front of the second term causes the issue of non-unitarity, for which the completeness relation in (\ref{eq:complete}) cannot be satisfied. Similarly for our rotor model case, we have 
\begin{equation}
\frac{i}{p^{4n}(p^2 - m^2)} \,.
\end{equation}
For example for $n=1$, we have
\begin{equation}
\frac{i}{p^4 (p^2 - m^2)} =\frac{i}{m^4 (p^2 -m^2) } -\frac{i}{m^4 p^2} -\frac{i}{m^2 p^4} \,,
\end{equation}
where the minus sign for the second and the third term causes the problem. The is the origin of non-unitarity such that the optical theorem in (\ref{eq:optical}) cannot be satisfied . To rescue the unitary problem, one can follow the approach by introducing `fakeons', which is a fake particle that provides a fake degree of freedom  that propagates inside the Feynman diagrams \cite{fake1,fake2}. Fakeons can make a high-order derivative quantum field theory unitary \cite{fake1,fake2}. However, the full analysis would be beyond the scope of this paper and requires further studies in future.

\section{Conclusion}
In conclusion, we have applied the rotor mechanism and quantization to the standard model of particle physics, which naturally generates high-order derivative quantum fields in the standard model's Lagrangian. Upon path integral quantization, we develop feynman propagators of scalar particles, gauge bosons and Dirac fermions under rotor model. When the rotor index is $n=0$, this restores to the original standard model case. Then we explicitly calculate the quantum amplitudes of the one-loop self-energy correction diagrams of the Higgs Boson under rotor mechanism. This includes the correction by the Higgs-self interaction, $W$-boson and fermion respectively. We find that the rotor mechanism can generally remove the UV divergences, however, IR divergence is arisen at the same time. We discover that the rotor model is able to remove infinities (both UV and IR), or suppress divergences arise from the case of the Higgs-self interaction. We find that the minimum spacetime dimension $D$ for $n=1$ rotor index is $9$, and for $n=2$ is $17$, and so on with a general formal of $D_{\mathrm{min}} = 8n+1$. This suggests that the rotor mechanism can remove infinities (both UV and IR) arise from simple integral calculation, thus giving a new way to partially solve the  Hierarchy problem.  However, for diagrams with more complicated integrals, such as the $W$-boson loop correction and fermion loop correction, due to specific dimension range arise from each integral term, there does not exist a general $D$ that can cure the divergence all at once for specific $n$. More future work has to be done. Finally, it is understood that a renormalizable high-order derivative quantum field theory breaks unitarity, and yet can be rescued by the the `fakeons' approach, but detailed analysis requires rigorous work in the future.

\end{document}